\documentclass[twocolumn]{aastex631}

\newcommand{\feh}{[Fe/H]}
\newcommand{\meh}{[M/H]}
\newcommand{\xh}[1]{[#1/H]}
\newcommand{\xm}[1]{[#1/M]}
\newcommand{\xfe}[1]{[#1/Fe]}
\newcommand{\teff}{T$_{\rm eff}$}
\newcommand{\logg}{$\log g$}
\newcommand{\kms}{km s$^{-1}$}

\begin{document}

\title{The Apache Point Observatory extra-Galactic Evolution Experiment (APOeGEE): Chemical Abundance Trends for Seven Dwarf Spheroidal Galaxies in the APOGEE Survey}

\author[0000-0003-0509-2656]{Matthew Shetrone}
\affiliation{University of California Observatories, University of California Santa Cruz, Santa Cruz, CA, 95064}

\author[0000-0002-1691-8217]{Rachael L. Beaton}
    \affiliation{Space Telescope Science Institute, 3700 San Martin Drive, Baltimore, MD 21218, USA}

\author[0000-0003-2969-2445]{Christian R. Hayes}
\affiliation{Space Telescope Science Institute, 3700 San Martin Drive, Baltimore, MD 21218, USA}
\author[0000-0001-5388-0994]{Sten Hasselquist}
\affiliation{Space Telescope Science Institute, 3700 San Martin Drive, Baltimore, MD 21218, USA}
    
\author[0000-0002-4733-4994]{Joshua~D.~Simon}
\affiliation{Observatories of the Carnegie Institution for Science, 813 Santa Barbara St., Pasadena, CA 91101, USA}    

\author[0000-0002-9771-9622]{Jon A. Holtzman}
\affiliation{Department of Astronomy, New Mexico State University, P.O.Box 30001, MSC 4500, Las Cruces, NM, 88003, USA}

\author[0000-0001-6476-0576]{Katia Cunha}
\affil{Observat\'orio Nacional/MCTI, R. Gen. Jos\'e Cristino, 77,  20921-400, Rio de Janeiro, Brazil}
\affiliation{Steward Observatory, University of Arizona Tucson AZ 85719}

\author[0000-0003-2025-3147]{Steven R. Majewski}
\affiliation{Department of Astronomy, University of Virginia, Charlottesville, VA 22904, USA}

\author[0000-0002-4989-0353]{Jennifer Sobeck}
\affiliation{Department of Astronomy, University of Washington, Box 351580, Seattle, WA 98195, USA}

\author[0000-0002-2244-0897]{Ricardo Schiavon}
\affiliation{Liverpool John Moores University: Liverpool, Merseyside, GB}

\author[0000-0002-6939-0831]{Thomas Masseron}
\affil{Instituto de Astrofísica de Canarias (IAC), E-38205 La Laguna, Tenerife, Spain}
\affil{Universidad de La Laguna (ULL), Departamento de Astrofísica, 38206 La Laguna, Tenerife, Spain}

\author[0000-0002-0134-2024]{Verne V. Smith}
\affiliation{NOIRLab, Tucson AZ 85719}
\affil{Institut d'Astrophysique de Paris, UMR 7095 CNRS \& Sorbonne Universit\'e, 98 bis Blvd Arago, Paris  75014  France}

\author[0000-0002-1793-3689]{David L. Nidever}
\affil{Department of Physics, Montana State University, P.O. Box 173840, Bozeman, MT 59717-3840}


\begin{abstract}
    In addition to comprehensive surveys of the Milky Way bulge, disk, and halo, the Apache Point Galactic Evolution Experiment (APOGEE) project observed seven dwarf spheroidal satellites (dSphs) of the Milky Way: Carina, Sextans, Sculptor, Draco, Ursa Minor, Bootes 1, and Fornax. APOGEE radial velocities, stellar parameters, and Gaia EDR3 proper motions are used to identify member stars in the vicinity of each dwarf. To properly analyze the abundance patterns of these galaxies, a novel procedure was developed to determine the measurable upper limits of the APOGEE chemical abundances as a function of the effective temperature and the spectral signal-to-noise ratio. In general, the APOGEE abundance patterns of these galaxies (limited to [Fe/H] $>$ -2.5) agree with those found in high-resolution optical studies after abundance offsets are applied. Most of the galaxies studied have abundance patterns that are distinctly different from the majority of stars found in the MW halo, suggesting that these galaxies contributed little to the MW halo above [Fe/H] $>$ -2.0. From these abundance patterns, we find that these dSphs tend to follow two types of chemical evolution paths: episodic and continuous star formation, a result that is consistent with previous photometric studies of their star formation histories. We explore whether mass and/or environment have an impact on whether a galaxy has an episodic or continuous star formation history, finding evidence that, in addition to the galaxy's mass, proximity to a larger galaxy and the cessation of star formation may drive the overall shape of the chemical evolution.

\end{abstract}

\section{Introduction} \label{sec:intro}


Detailed chemical abundances determined from high resolution spectroscopy of even small numbers of stars in dwarf Spheroidal (dSph) galaxies \citep[e.g., ][]{Shetrone_2001,Shetrone_2003} have long demonstrated that these objects are predominantly metal-poor ([Fe/H] $<$ -0.5 {for most dSphs}) and often have significant enhancements in their $\alpha$-elements ([$\alpha$/Fe] $>$ +0.4) {at [Fe/H] $<$ -2.0.} 
Decades later, the number of stars in individual dwarf galaxies with measured chemical patterns remains small \citep[notable exceptions being ][with as many as $\sim$100 stars]{Hill_2019,Letarte_2010,Lemasle_2014,theler_2020} {with the largest literature compilations consisting of 1058 unique stars across 25 galaxies}  \citep{Saga_1,Saga_2,saga_3,saga_4}. 
In addition, many of the objects were targeted for having extraordinary properties, such as being extremely metal-poor \citep[e.g.,][among others]{Norris_2010b,Norris_2010c,Simon_2015,Jablonka_2015,Tafelmeyer_2010,Cohen_2010,Skuladottir_2021} or exhibiting peculiar element enhancements \citep[e.g.,][]{Ji2023}. 
Even when ensembles of stars in the literature can be compiled, these compilations consist of observations using different instrumental setups and often contain a wide range of different abundance analysis procedures, all of which can lead to significant systematic uncertainties in a compiled chemical dataset.
Lastly, relatively few studies of dSphs also study significant samples of Milky Way stars, making it difficult to compare and contrast the abundance-inferred star formation histories (SFHs) of the dSphs to each other and to the MW, particularly to the halo of the MW. Therefore, open questions remain about how these galaxies may have formed stars differently, both early on in their formation and at more recent times, when their interactions with the MW (and in some cases each other) either stalled or restarted star formation.

An ideal chemical abundance study of these dSph galaxies would then consist of observations from a single instrument with stellar parameters and abundances derived in a homogeneous way. Such a survey at high-resolution is nearly time-prohibitive without dedicated multi-year surveys, but such a dataset was obtained by \citet{Kirby_2010} using low resolution spectroscopy (R $\simeq$ 6,000) of several hundred stars across Sculptor, Fornax, Leo I, Sextans, Leo II, Canes Venatici I, Ursa Minor, and Draco. From the metallicity distribution functions (MDFs) of these galaxies, \citet{Kirby2011_iii} concluded that the more luminous dwarf galaxies tended to enrich to higher metallicity, suggesting that more massive galaxies better retain gas that can then be used to continue star formation. From the [$\alpha$/Fe]-[Fe/H] abundance patterns of these dSphs, they found that all galaxies had Type Ia SNe contributing to the chemical abundances of most stars at [Fe/H] $>$ -2.5 \citep{Kirby2011_iv}, in stark contrast to the MW. They also used chemical evolution models to investigate whether some of the dSph galaxies showed chemical signatures of star bursts. Although the Kirby et al. studies have been monumental in providing a more universal picture of how these dwarf galaxies form and evolve over time, additional details can be inferred from the measurements of individual chemical abundances, which are typically accessible to higher precision at higher spectral resolution (although Kirby et al. are able to measure some individual elements, e.g., \citealt{Kirby_2009,Kirby2011_iii}).


In this work, we extend the work on local group dwarf galaxies of \citealt{Hasselquist_2017,Nidever_2020,Hasselquist2021_sats} by presenting detailed chemical abundances of up to 12 elements for 537 stars across seven dSphs: Carina, Sextans, Sculptor, Draco, Ursa Minor, Bootes 1, and Fornax. These abundances were measured from spectra obtained by the Apache Point Galactic Evolution Experiment (APOGEE), which was a high resolution (R $\simeq$ 20,000) near-infrared spectroscopic \citep{majewski_2017} survey of stars across the MW and its satellite galaxies. Although fainter than the typical APOGEE target, deep observations were able to target the upper red-giant branch of these systems. However, the varying S/N and metal-poor nature of these observations require careful analysis and discriminating between true measurements and upper limits. In this study, we derive a prescription for determining upper limits on abundances based on the S/N of the observation and stellar parameters of the star. We used these abundance upper limits to better analyze the chemical abundance patterns of the dSph galaxies at [Fe/H] $>$ -2.5, but these upper limits could, in principle, be applied to all APOGEE results. The data used and the target selection are described in Section \ref{sec:sample}. Stellar parameters and abundance determination, along with the prescription for upper limits, are discussed in Section \ref{sec:metalpoor}. The chemical abundance patterns for the 7 dSphs are presented in Section \ref{sec:chem_patterns}, and comparisons with optical literature samples are shown in Section \ref{sec:litcomp}. We discuss what these chemical abundance patterns of the dSphs mean for their detailed SFHs in Section \ref{sec:discussion}, dividing the dSphs into two groups based on whether they have had continuous or episodic SFHs as indicated by their chemical abundance patterns.

\section{Data Source and Target Selection} \label{sec:sample}

For our dSph sample, we start with data from the 17th Data Release (DR17, \citealt{SDSS_DR17}) of the fourth iteration of the Sloan Digital Sky survey (SDSS-IV, \citealt{blanton_2017}), making use of the spectroscopic observations from the twin APOGEE spectrographs \citep{wilson_2019} onboard the Apache Point Observatory SDSS 2.5m \citep{gunn_2006} and Las Campanas Observatory 2.5m \citep{Bowen_1973} telescopes, referred to as APOGEE-N and APOGEE-S, respectively. These spectrographs obtained $H$-band spectra of $\sim$ 750,000 stars across the MW and its nearby satellite galaxies. We first describe how the dSphs were initially targeted in \autoref{sec:targ}, and describe how we selected the most likely members of the dSphs from these observations in \autoref{sec:dsph_targ}.  


\subsection{Target Selection for APOGEE Survey} \label{sec:targ}


Targeting dSph stars in any project is complex because of their low stellar density, subjecting them to MW foreground contamination, and given their large distances, their stars are quite faint. 
Both concerns are relevant for APOGEE. With only a 2.5m aperture, even stars at the tip of the red giant branch require more than 20 hours of integration to reach a sufficient signal-to-noise ratio for spectral synthesis analysis \citep[typically around signal-to-noise of 70, e.g.,][]{Jonsson_2020}. 
The long integration time required means that only a single dedicated field would be designed for each dSph, 
Moreover, the fiber-collision limits for APOGEE are larger than other dSph spectroscopic surveys; two fibers in APOGEE can only be placed as close as 71.5$''$ for the APOGEE-N instrument and 56$''$ on the APOGEE-S instrument. 
The field of view for APOGEE is sufficiently large to span nearly the full sizes of many dSphs with a 1.5$^{\circ}$ radius for APOGEE-N and 0.95$^{\circ}$ radius for APOGEE-S\footnote{We note that APOGEE-S also has two large obscured regions for the positioning and guiding optics that block parts of the field-of-view. See \citet{Santana_2021} for a detailed description.}. These targeting constraints mean that there were fibers that could not be assigned to dSph targets, and some bright dSph stars could not be observed. All of these factors meant that target selection for these fields differed substantially from typical APOGEE survey target selection described in \citet{Zasowski_2013} and \citet{Zasowski_2017}.

With all of these constraints, only seven of the many Milky Way dSphs could be readily targeted by APOGEE-2: Draco, Ursa Minor, and Bootes I with APOGEE-N and Fornax, Sculptor, Carina, and Sextans with APOGEE-S.
We note that, as described in \citet{Santana_2021}, Fornax was added to the dSph program later in the design of the survey and could therefore make use of \emph{Gaia} proper motion measurements; the other dSph target selection was completed before \emph{Gaia} DR2.
All plates were originally scheduled for 24 visits, roughly 24 hours, of on-sky time with an additional 12 hours granted as part of the Bright Time Extension for dSphs observed at APO \citep{Beaton_2021}. 
We note that some targets in Ursa Minor and Draco received up to 48 visits over the course of the APOGEE survey. 

The targeting occurred in three primary phases:
    (1) selection of previously confirmed radial velocity members,
    (2) selection of high-likelihood red-giant members selected using the Washington$+$DDO51 pre-selection \citep{majewski_2000},
    (3) selection of other candidate members using other external data.
The selection of members relied heavily on prior spectroscopic work, in particular the radial velocities, and these stars have target flags of \texttt{APOGEE\_DSPH\_MEMBER}.  

The selection of candidates beyond those targets already identified in the literature relied most heavily on Washington photometry, some of which was acquired for APOGEE-2 (J.~Munn, private communication; Bootes I, Sextans) and some of which was available in the existing literature \cite[for Carina, Ursa Major, Draco, and Sculptor:][]{Majewski_2000_carina,Palma_2003,Munoz_2005,Westfall_2006}; in lieu of Washington photometry, \emph{Gaia} DR2 proper motions \citep{gaia_dr2} were used to select Fornax candidates.
The Washington photometry allows for the selection of giants from foreground dwarfs of the same spectral type using the $DDO51$ intermediate band filter centered on the surface-gravity sensitive Mgb spectral feature \citep{majewski_2000}. 
Targets selected as candidate members based on external data are flagged as \texttt{APOGEE\_DSPH\_CANDIDATE}. Note that the targeting strategy described above means that our sample presented in this work cannot contain any potential young ($\sim$ 1 Gyr old) stars that have been found in some of these galaxies (e.g., \citealt{Yang2024}).

\subsection{Selecting dSph members in APOGEE DR17}\label{sec:dsph_targ}
Although the targeting strategy described in the previous section results in a high targeting fraction of available dSph members (described in detail in \autoref{sec:MEMBER}), we make use of APOGEE radial velocities and \emph{Gaia} proper motions to ensure that the initial APOGEE targets are indeed members of these dSphs. APOGEE radial velocities are obtained for each visit using the methods described in \citet{nidever_2015}. Much of this process was revised for DR17, partly inspired by the challenge of the low per-visit signal-to-noise ratio of the dSph stars. These revisions are discussed in detail in  Holtzman et al. (in prep.) and used the \texttt{Doppler}\footnote{\url{https://github.com/dnidever/doppler}} code \citep{2021zndo...4906681N}. This code determines RVs through convolution followed by a series of cross correlations to spectral grids of increasing complexity generated by the Cannon \citep{Casey2016}. These RVs are combined with \emph{Gaia} EDR3 proper motions and sky positions to create a ``6-sigma Membership Sample'', described in \autoref{sec:6sigma}. The completeness of this sample is discussed in \autoref{sec:ages}. 



\begin{figure*} 
    \centering
    \includegraphics[width=0.85\textwidth]{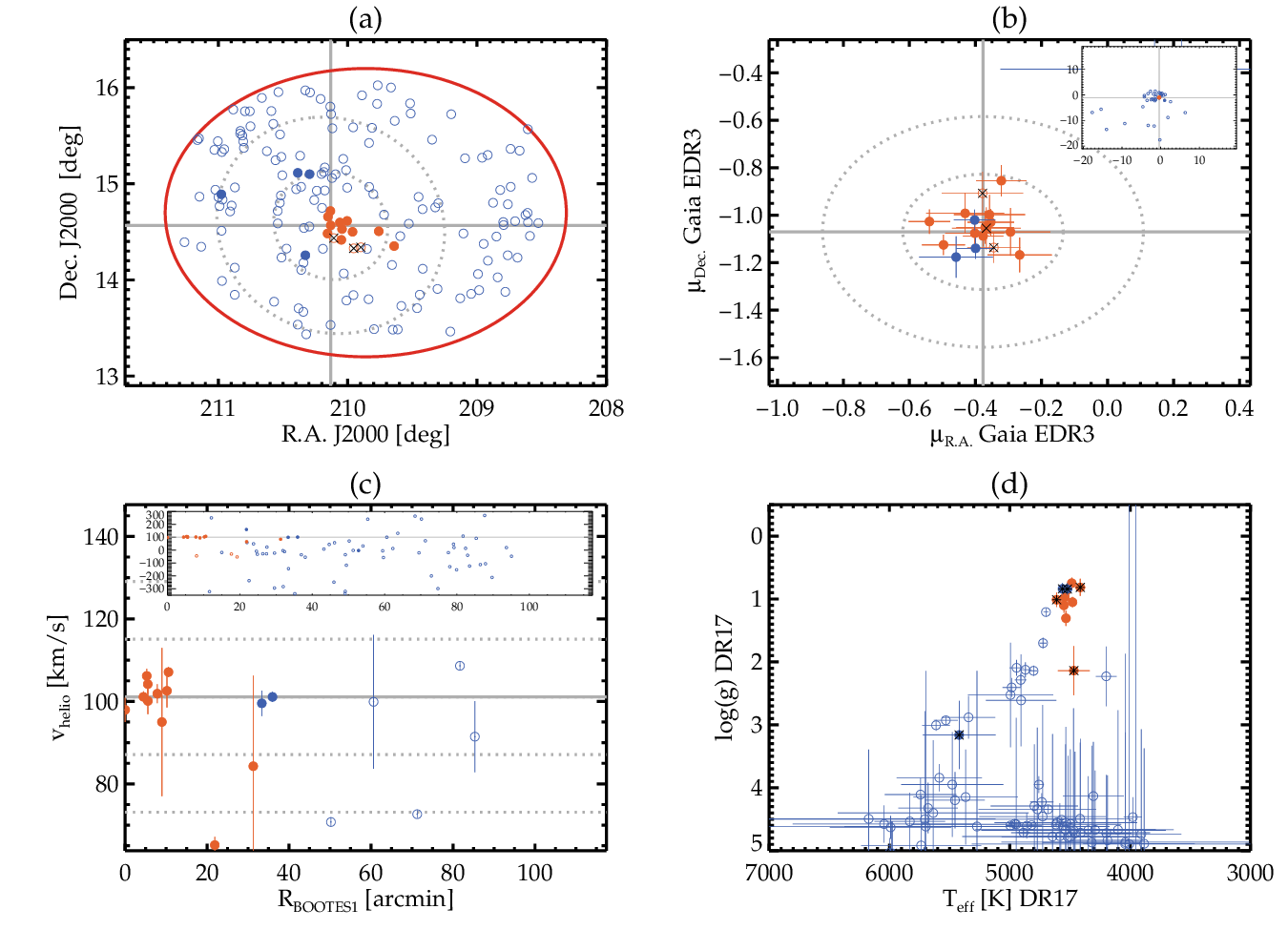}
    \caption{ 
    Four panel targeting and membership demonstrations for all seven dSphs, in the following order: Bootes 1, Carina, Draco, Fornax, Sculptor, Sextans, and Ursa Minor. 
    In all panels, dSph members are filled circles and non-members are open circles, stars targeted as literature members are in orange and candidates in blue, and targets with overall too low spectral quality to be evaluated confidently are indicated with a black ``x.''
    For each dSph the panels are as follows:
     (a) Plate-centric sky distribution with the plate FOV in red, the central coordinates of the dSph are shown with the solid grey lines, and the 3 and 6 times the half-light radius is indicated with dashed lines \citep[structural parameters taken from the 2021 update to][see our \autoref{tab:dsph_prop}]{McConnachie_2012}. 
     (b) Proper motion distribution of targets, the central PM is given by the solid lines, and the circles represent 3 and 6 sigma from the central value. The inset panel expands to a much larger PM range.
     (c) Radial velocity versus the radial distance from the dwarf center with the central velocity from APOGEE DR17 RVs as solid lines and 3 and 6 sigma given as dashed lines. The error bars plotted in this panel are \textsc{VSCATTER} that measures RV variability. 
     The inset panel expands to a much larger RV range.
     (d) Kiel diagram for stars with calibrated ASPCAP results.
     Asterisks in this panel flag those stars with generally unreliable abundances. The complete figure set (7 images) is available in the online journal.
    }
    \label{fig:targ}
\end{figure*} 

\figsetstart
\figsetnum{1}
\figsettitle{dSph Targeting and Membership}
\figsetgrpstart
\figsetgrpnum{1.1}
\figsetgrptitle{Bootes 1 Targeting}
\figsetplot{TargMemb_BOOTES1_v2}
\figsetgrpnote{Targeting and membership demonstration for Bootes 1. The four panels are as follows: (a) Plate-centric sky distribution; (b) Proper motion distribution of targets; (c) Radial velocity versus the radial distance from the dwarf center; (d) Kiel diagram for stars with calibrated ASPCAP results.}
\figsetgrpend

\figsetgrpstart
\figsetgrpnum{1.2}
\figsetgrptitle{Carina Targeting}
\figsetplot{TargMemb_CARINA_v2}
\figsetgrpnote{Targeting and membership demonstration for Carina. The four panels are as follows: (a) Plate-centric sky distribution; (b) Proper motion distribution of targets; (c) Radial velocity versus the radial distance from the dwarf center; (d) Kiel diagram for stars with calibrated ASPCAP results.}
\figsetgrpend

\figsetgrpstart
\figsetgrpnum{1.3}
\figsetgrptitle{Draco Targeting}
\figsetplot{TargMemb_DRACO_v2}
\figsetgrpnote{Targeting and membership demonstration for Draco. The four panels are as follows: (a) Plate-centric sky distribution; (b) Proper motion distribution of targets; (c) Radial velocity versus the radial distance from the dwarf center; (d) Kiel diagram for stars with calibrated ASPCAP results.}
\figsetgrpend

\figsetgrpstart
\figsetgrpnum{1.4}
\figsetgrptitle{Fornax Targeting}
\figsetplot{TargMemb_FORNAX_v2}
\figsetgrpnote{Targeting and membership demonstration for Fornax. The four panels are as follows: (a) Plate-centric sky distribution; (b) Proper motion distribution of targets; (c) Radial velocity versus the radial distance from the dwarf center; (d) Kiel diagram for stars with calibrated ASPCAP results.}
\figsetgrpend

\figsetgrpstart
\figsetgrpnum{1.5}
\figsetgrptitle{Sculptor Targeting}
\figsetplot{TargMemb_SCULPTOR_v2}
\figsetgrpnote{Targeting and membership demonstration for Sculptor. The four panels are as follows: (a) Plate-centric sky distribution; (b) Proper motion distribution of targets; (c) Radial velocity versus the radial distance from the dwarf center; (d) Kiel diagram for stars with calibrated ASPCAP results.}
\figsetgrpend

\figsetgrpstart
\figsetgrpnum{1.6}
\figsetgrptitle{Sextans Targeting}
\figsetplot{TargMemb_SEXTANS_v2}
\figsetgrpnote{Targeting and membership demonstration for Sextans. The four panels are as follows: (a) Plate-centric sky distribution; (b) Proper motion distribution of targets; (c) Radial velocity versus the radial distance from the dwarf center; (d) Kiel diagram for stars with calibrated ASPCAP results.}
\figsetgrpend

\figsetgrpstart
\figsetgrpnum{1.7}
\figsetgrptitle{Ursa Minor Targeting}
\figsetplot{TargMemb_URMINOR_v2}
\figsetgrpnote{Targeting and membership demonstration for Ursa Minor. The four panels are as follows: (a) Plate-centric sky distribution; (b) Proper motion distribution of targets; (c) Radial velocity versus the radial distance from the dwarf center; (d) Kiel diagram for stars with calibrated ASPCAP results.}
\figsetgrpend

\figsetend

\subsubsection{DR17 Membership Classifications} \label{sec:MEMBER}

Our starting point for identifying members in the dSphs is the \textsc{MEMBER} tag included in the DR17 summary files.
This tag combined a number of membership criteria to identify a set of the most likely members:
    (i) targets within 90 arcmin of the dSph center, 
    (ii) comparing the APOGEE \texttt{vhelio\_avg} to the known systemic velocity of the dSph within 6 sigma of the cluster median, and
    (iii) comparing the target \emph{Gaia} EDR3 proper motions to the known systemic proper motions of the dSph within 6 sigma of the cluster median values. 
The data used for these classifications are given in the top portion of \autoref{tab:dr17_member} and a given APOGEE target has to meet all of the stated criteria to be flagged as a member in the \texttt{MEMBER} tag.\footnote{Diagnostic figures affiliated with the \texttt{MEMBER} tag from the DR17 release can be found here: \url{https://data.sdss.org/sas/dr17/apogee/spectro/redux/dr17/stars/dsph/}} 

\begin{table*}[h]
    \centering
    \caption{Membership Criteria} \label{tab:dr17_member}
    \begin{tabular}{c| cc c c | cc | cc  }
    \hline \hline
    \multicolumn{6}{l}{DR17 \texttt{MEMBER} dSph Membership Criteria} \\
    dSph        & \multicolumn{2}{c}{Central Coord.}    & Tele. & Radius   & $v_{sys}$ & $\Delta_{RV}$ & \multicolumn{2}{c}{Proper Motion$^{b}$} \\
                &  deg  & deg       &       & arcmin    &  km/s     &   km/s        &  mas/yr  & mas/yr    \\ 
    \hline 
    Bootes I    & 210.024994 &  14.500000  & N & 90. &   99.0 & 50. & -0.46$\pm$0.16 & -1.06 $\pm$0.16  \\
    Carina      & 100.835228 & -51.099789  & S & 90. &  229.0 & 50. &  0.50$\pm$0.10 & -0.14 $\pm$0.10  \\
    Draco       & 260.049988 &  57.915001  & N & 90. & -291.0 & 50. & -0.02$\pm$0.14 & -0.14 $\pm$0.14  \\
    Fornax      & 39.995800  & -34.449001  & S & 90. &   55.3 & 50. &  0.38$\pm$0.07 & -0.41 $\pm$0.07  \\ 
    Sculptor    &  15.220700 & -33.688961  & S & 90. &  111.4 & 50. &  0.08$\pm$0.12 & -0.13 $\pm$0.12  \\
    Sextans     & 153.262497 &  -1.614700  & S & 90. &  224.0 & 50. & -0.50$\pm$0.12 &  0.08 $\pm$0.12  \\
    Ursa Minor  & 227.283325 &  67.222504  & N & 90. & -247.0 & 50. & -0.18$\pm$0.14 &  0.07 $\pm$0.14  \\
    \hline \hline 
    \multicolumn{6}{l}{Final Criteria Adopted in this Work} \\
        dSph    & \multicolumn{2}{c}{Median Coord.}     & Tele. & Radius$^{a}$   & $RV_{Median}$ & $RV_{SMAD}$ & \multicolumn{2}{c}{Proper Motion$^{b}$}  \\
                &  deg                  &  deg &     & arcmin    &  km/s     &   km/s        &  mas/yr  & mas/yr \\
    \hline 
    Bootes I    & 210.131  &  14.598     & N &  \nodata &   101.11 &  4.57 & -0.37$\pm$0.047  & -1.05  $\pm$0.054  \\
    Carina      & 100.359  & -50.9635	 & S &  \nodata &   224.5  &  5.53 &  0.558$\pm$0.062 &  0.128 $\pm$0.079  \\
    Draco       & 259.978  &  57.9379    & N &  \nodata & -293.07  & 10.47 &  0.032$\pm$0.073 & -0.181 $\pm$0.089  \\
    Fornax      & 39.9322  & -34.5463	 & S &  \nodata &   54.07  & 10.05 &  0.384$\pm$0.066 & -0.381 $\pm$0.104  \\ 
    Sculptor    & 15.0002  & -33.7103	 & S &  \nodata &  110.53  &  9.43 &  0.098$\pm$0.072 & -0.167 $\pm$0.056  \\
    Sextans     & 153.294  &  -1.5986    & S &  \nodata &  225.43  &  8.56 & -0.405$\pm$0.125 &  0.013 $\pm$0.135  \\
    Ursa Minor  & 227.415  &  67.2545	 & N &  \nodata & -246.85   & 7.43 & -0.105$\pm$0.096 &  0.08 $\pm$0.075   \\ 
    \hline \hline
    \multicolumn{8}{l}{$^{a}$ \footnotesize{No spatial restrictions were used, but the plate radius formally limits this value.}}\\
    \multicolumn{8}{l}{$^{b}$ \footnotesize{Proper motion of our sample, there are significantly more accurate proper motions in the literature \citep[e.g.,][]{Pace_2022}.}}
    \end{tabular}
\end{table*}


\subsubsection{The 6-sigma Membership Sample} \label{sec:6sigma}

We provide an update to the \texttt{MEMBER} flag included in DR17 by extending our membership criteria to 6-sigma based on the median kinematic properties determined from the members: sky position, $v_{sys}$, and proper motion; the parameters are given in \autoref{tab:dr17_member}.
In particular, all 6-sigma members have original targeting flags of either \texttt{APOGEE\_DSPH\_MEMBER}, meaning it was an RV member in prior work, or \texttt{APOGEE\_DSPH\_CANDIDATE} meaning that it had photometry properties consistent with being a metal-poor giant at the distance of the dSph, with the exception of Fornax whose candidate selection also included proper motions \citep[details on Fornax selection are given in][]{Santana_2021}.  Thus, the membership used in this work does not exactly match the DR17 \texttt{MEMBER} flag.   More details of these differences can be found in \autoref{app:memb}.



\begin{table*}
    \centering
    \caption{APOGEE dSph Sample}
    \begin{tabular}{l c | c c c | c c c c | c}
     \hline \hline
     dSph &               &  \multicolumn{3}{c}{Kinematics$^{a}$} & \multicolumn{4}{c}{ASPCAP$^{b}$} & Measurements$^{c}$ \\ 
          & $N_{Targets}$ 
          & $N_{PM}$ & $N_{RV}$ & $N_{PM+RV}$ 
          & $N_{ASPCAP}$ & $N_{Cool Giant}$ & $N_{M/H}$ &  $N_{Good ASPCAP}$ 
          & $N_{Fe}$ \\
     \hline 
     Bootes I   &  161 &  17 &  15 &  12 &  10 &  10 &   7 &   6 &   5 \\
     Carina     &  215 &  64 &  64 &  62 &  56 &  51 &  52 &  52 &  50 \\
     Draco      &  155 &  70 &  68 &  63 &  51 &  47 &  44 &  41 &  42 \\
     Fornax     &  248 & 239 & 247 & 239 & 212 & 203 & 211 & 209 & 208 \\
     Sculptor   &  238 & 208 & 204 & 201 & 181 & 173 & 168 & 161 & 149 \\     
     Sextans    &  131 &  92 &  83 &  72 &  68 &  56 &  45 &  40 &  36 \\
     Ursa Minor &  167 &  84 &  83 &  82 &  70 &  68 &  55 &  54 &  47 \\
     \hline      
     Total      & 1315 & 774 & 764 & 731 & 648 & 608 & 582 & 563 & 537 \\   
     \hline \hline 
     \multicolumn{9}{l}{$^{a}$ \footnotesize{Using a 6-sigma window around the final criteria given in \autoref{tab:dr17_member}.}}\\    
     \multicolumn{9}{l}{$^{b}$ \footnotesize{See discussion in \autoref{ssec:dq}.}} \\        
     \multicolumn{9}{l}{$^{c}$ \footnotesize{Stars with detectable measurements in [Fe/H] as described in \autoref{ssec:detection}.}} \\ 
    \end{tabular}
    \label{tab:memb_accounting}
\end{table*}


As summarized in \autoref{tab:memb_accounting}, there were 1315 stars targeted in the dSph program; of these, 530 (40\%) were targeted as members of a dSph based on the results of the literature (primarily RVs) and 785 (60\%) were candidate members using photometry or, for Fornax, proper motions. 
Applying the 6-sigma criteria in \autoref{tab:dr17_member}, 774 (59\%) pass the proper motion criteria and 764 (58\%) pass the RV criteria; combining the two, we have 731 stars meeting the kinematic membership cut (56\%). 
No members were serendipitously identified, e.g., through the criteria of the ``Main Red Star Sample'' for a 24-visit field \citep{Zasowski_2013,Zasowski_2017,Santana_2021,Beaton_2021}; this is approximately consistent with the magnitude limits and the tip of the red giant branch (TRGB) for the dSphs. 
In detailed inspection of the stellar spectra, 181 members (24\%) are excluded from detailed chemical analyzes due to the poor quality of the spectra (see \autoref{ssec:dq}). We do find that some of our dSphs members exhibit RV variability consistent with having a binary companion, but an analysis of binaries in these dSphs is beyond the scope of this work.

\autoref{fig:targ} provides a visual assessment of the final membership samples for each of the seven dSphs.
\autoref{fig:targ} has four panels:
 (a) spatial location compared to the plate radius (red) and the structural parameters of the dSph,
 (b) proper motion from Gaia EDR3 (\autoref{tab:shetronemembers}),
 (c) radial velocity relative to the systemic velocity and the velocity dispersion, here plotted against projected distance from the dSph center, and 
 (d) the Kiel diagram from APOGEE calibrated stellar parameters. 
The velocity dispersion and structural parameters for each dwarf were obtained from the updated compilation of \cite{McConnachie_2012}\footnote{The 2021 updated compilation is available: \url{https://www.cadc-ccda.hia-iha.nrc-cnrc.gc.ca/en/community/nearby/}}; the values and original references are provided in \autoref{tab:dsph_prop}. 
These values of \cite{McConnachie_2012} are used only for visualization purposes, and these values did not affect the selection of members as described in \autoref{sec:MEMBER} and \autoref{sec:6sigma}.
As such, we note that there are more recent values for structural parameters such as \citealt{Jenkins2021} for Bootes I, \citealt{Pace2020} for Ursa Minor, and \citealt{munoz_2018} for several dwarfs, among others. 
In each panel of \autoref{fig:targ}, the open symbols do not meet the 6-sigma membership criteria whereas the filled symbols do; orange symbols are those stars targeted as known members, and the blue symbols were selected as candidates.

\subsubsection{Completeness} \label{sec:ages}
\begin{figure*}
    \centering
    \includegraphics[width=0.8\textwidth]{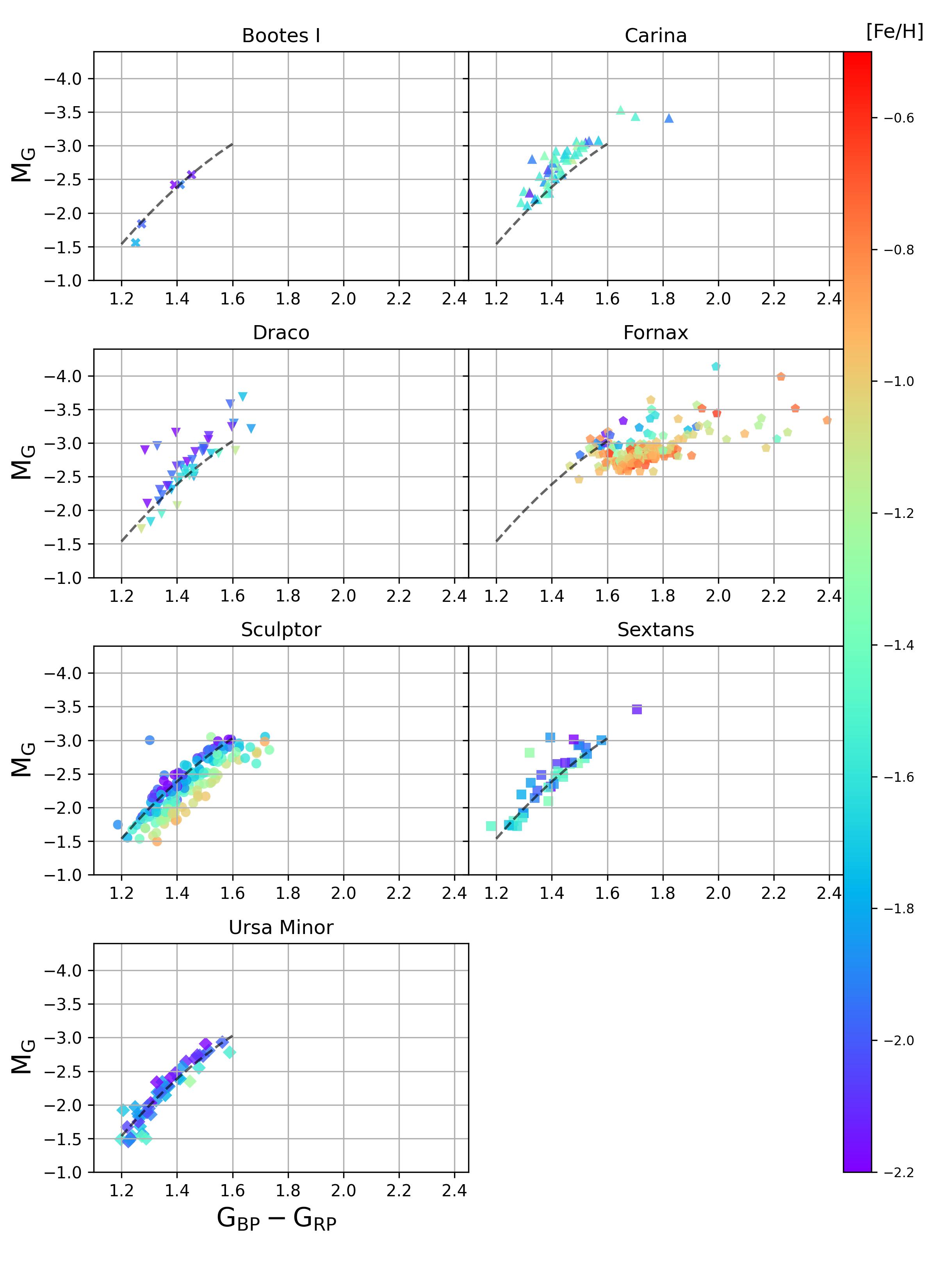}  
    \caption{
    \emph{Gaia} absolute color-magnitude diagrams for each dSph. Points are colored by [Fe/H], as indicated by the color bar at right. In each panel, a fit to Ursa Minor is reproduced (black dashed line).}
    \label{fig:sample}
\end{figure*}

To understand the extent to which we are sampling the red giant branches of these dSphs, we show in \autoref{fig:sample} a \emph{Gaia} dereddened-color to absolute-magnitudes ($M_{G} = m_{G}-\mu_{gal}-A_{G}$) diagram in the \emph{Gaia} photometric bands for EDR3 for each dSph. The distances and reddening $E(B-V)$ adopted for \autoref{fig:sample} are given in \autoref{tab:dsph_prop}; these come from the compilation of \citet{McConnachie_2012} where $E(B-V)$ is from \citet{Schlegel_1998}.
To place $E(B-V)$ in the \emph{Gaia} filter system, we assume $R_V$ = 3.1 and then follow \citet{Wang_2019}; more specifically, $A_{G} = 0.789 A_{V}$ and $E(BP-RP) = A_{V}/2.394$.
The $A_{G}$ ranges from 0.04 (Bootes I) to 0.15 mag (Carina) and the $E(BP-RP)$ varies from 0.02 (Bootes I) to 0.08 mag (Carina). 


\begin{table*}[t]
    \centering
    \caption{dSph Properties}
    \begin{tabular}{c | ccc | cccc | cc }
    \hline \hline
    dSph        & \multicolumn{2}{c}{Distance$^{a}$} & $E(B-V)^{b}$ & PA$^{c}$  & ellip$^{c}$ & $r_h^{c}$  & $\sigma^{d}$      &  $M_{*}^{e}$             &  $\tau_{70}^{e}$ \\
                & kpc      & mag     & mag      & deg &       & arcmin &   km s$^{-1}$ & $10^{6}$ $M_{\odot}$ &  $\log_{10}$ (yrs)       \\
    \hline \hline
    Bootes I    &  66 & 19.11$^{0.08}_{0.08}$ & 0.017 &  7$^{3}_{3}$ & 0.25$^{0.02}_{0.02}$ & 11.3$^{0.27}_{0.27}$ &  2.4$^{0.9}_{0.5}$ & 0.029 & 10.12 \\
    Carina      & 105 & 20.11$^{0.13}_{0.13}$ & 0.061 & 60$^{1}_{1}$ & 0.37$^{0.01}_{0.01}$ & 11.4$^{0.12}_{0.12}$ &  6.6$^{1.2}_{1.2}$ & 0.38  &  9.56    \\
    Draco       &  76 & 19.4$^{0.17}_{0.17}$ & 0.027 & 97$^{1}_{1}$ & 0.30$^{0.01}_{0.01}$ &  9.9$^{0.09}_{0.09}$ &  9.1$^{1.2}_{1.2}$ & 0.29  & 10.03   \\
    Fornax      & 147 & 20.84$^{0.18}_{0.18}$ & 0.021 & 46$^{1}_{1}$ & 0.28$^{0.01}_{0.01}$ & 18.4$^{0.02}_{0.02}$ & 11.7$^{0.9}_{0.9}$ & 20.   &  9.67   \\ 
    Sculptor    &  86  & 19.67$^{0.14}_{0.14}$ & 0.018 & 94$^{1}_{1}$ & 0.37$^{0.01}_{0.01}$ & 12.3$^{0.05}_{0.05}$ &  9.2$^{1.4}_{1.4}$ & 2.3   & 10.06    \\
    Sextans     &  86 & 19.67$^{0.1}_{0.1}$   & 0.047 & 56$^{5}_{5}$ & 0.35$^{0.05}_{0.05}$ & 27.8$^{1.2}_{1.2}$   &  7.9$^{1.3}_{1.3}$ & 0.44  & 10.08  \\
    Ursa Minor  & 76 & 19.4$^{0.1}_{0.1}$    & 0.032 & 50$^{1}_{1}$ & 0.55$^{0.01}_{0.01}$ & 17.3$^{0.11}_{0.11}$ &  9.5$^{1.2}_{1.2}$ & 0.29  &  9.98 \\ 
    \hline \hline
    \multicolumn{10}{l}{\footnotesize $^{a}$ Distance Ref.       -- Boo: \citet{DallOra_2006}; Car: \citet{Pietrzynski_2009b}; Dra: \citet{Bonanos_2004}; } \\
    \multicolumn{10}{l}{\footnotesize ~~~For: \citet{Pietrzynski_2009b}; Scl: \citet{Pietrzynski_2009}; Sex: \citet{Lee_2009}; UMi: \citet{Carrera_2002}.} \\
    \multicolumn{10}{l}{\footnotesize $^{b}$ Reddening Ref. -- Taken from the position of the dSph in \citet{Schlegel_1998} maps by \citet{McConnachie_2012}.} \\
    \multicolumn{10}{l}{\footnotesize $^{c}$ Struct. Param. Ref. -- Boo:  \citet{Martin_2008}; Car, For, Scl, Sex, UMi: \citet{irwin_1995};} \\
    \multicolumn{10}{l}{\footnotesize ~~~  Dra: \citet{Martin_2008}.} \\
    \multicolumn{10}{l}{\footnotesize $^{d}$ $\sigma$ Ref. -- Boo: \citet{Koposov_2011}; Dra, UMi: \citet{Wilkinson_2004}; Car, For, Scl, Sex: \citet{Walker_2008}.} \\
    %
    \multicolumn{10}{l}{\footnotesize $^{e}$ Mass \& $\tau_{70}$ Ref. -- Car, Dra, For, Scl, UMi: \citet{Weisz_2014}; Sex: \citet{Bettinelli_2018}} \\       
    \end{tabular}
    \label{tab:dsph_prop}
\end{table*}

From Figure \ref{fig:sample}, we find that most dSphs, with the exception of Bootes I and Fornax, have $\sim$ 1.5 mags of their upper giant branch sampled. Sculptor exhibits the most ``traditional'' CMD consistent with simple chemical evolution, with a smooth metallicity gradient across the giant branch. Carina, despite being more metal-rich than Ursa Minor, has an RGB that is bluer than that of Ursa Minor. This suggests either that the adopted distance for one of the galaxies is incorrect or that the Carina stars are significantly younger than Ursa Minor. Strong evidence that Carina’s stars are young comes from photometric studies (e.g., \citealt{Weisz2014}). Such apparent age differences are discussed further in \autoref{sec:chem_patterns} after considering the chemical abundance trends.

\section{APOGEE Stellar Parameters and Abundances for dSph Stars}
\label{sec:metalpoor}

Stellar parameters and chemical abundances for APOGEE stars are obtained using the APOGEE Stellar Parameters and Chemical Abundances Pipeline \citep[ASPCAP][]{garciaperez_2016}. ASPCAP uses an 8-dimensional synthetic spectral grid created from MARCS model atmospheres \citep{Jonsson_2020} using the \texttt{synspec}  code \citep{synspec} and the line list described in \citet{Smith_2021}. The chi-squared minimization code, \texttt{FERRE} \citep{AllendePrieto2006}, is used to return a best-fit spectrum from this synthetic grid. This synthetic spectral grid only goes as metal-poor as [Fe/H] = -2.5, so our abundance analysis of these galaxies is limited to [Fe/H] $>$ -2.5. More details on the version of ASPCAP used in DR17 can be found in \citet{SDSS_DR17}. 


Unlike many other stellar populations explored in APOGEE, the dSphs are metal-poor, with the most metal-rich stars across the literature being \xh{Fe}$\sim$-0.5 for Fornax and \xh{Fe}$\lesssim$-0.9 across the other dSphs (but see \citealt{Yang2024} for signs of a small handful of more metal-rich stars in e.g., Sculptor).
Although considerable work has been undertaken to understand the precision and accuracy of the APOGEE measurements, these works typically focus on stars in the solar neighborhood for which the abundances are within $\pm0.5$ dex of solar and stretching to \xh{Fe}$\sim -1.0$ at the extremes \citep[see e.g., ][among others]{Jonsson_2018,Jonsson_2020}.  An exception to this is the work on globular clusters, e.g. \citet{Geisler_2021} and \citet{Nidever_2020}.
Moreover, these works generally focus on the most robust results from high S/N spectra {(S/N $\sim$100 per pixel)}; for context, the mean magnitude of the dSph members is $H$=14.8, which is 1 magnitude fainter than the S/N$\sim$100 magnitude limit for a 24 hour plate.
Thus, before we use the measurements for the dSphs, we must explore the APOGEE spectral quality at lower S/N and understand the ASPCAP performance for low-metallicity stars. In \autoref{ssec:dq} we show some dSph spectra and discuss the obvious limitations of the data, and in \autoref{ssec:detection} we describe our prescription for upper limits on the APOGEE abundance measurements of these low-metallicity, low-S/N spectra.

\subsection{APOGEE dSph Data Quality and Abundance Uncertainties} \label{ssec:dq}

\begin{figure*} 
    \centering
    \includegraphics[width=1.0\textwidth]{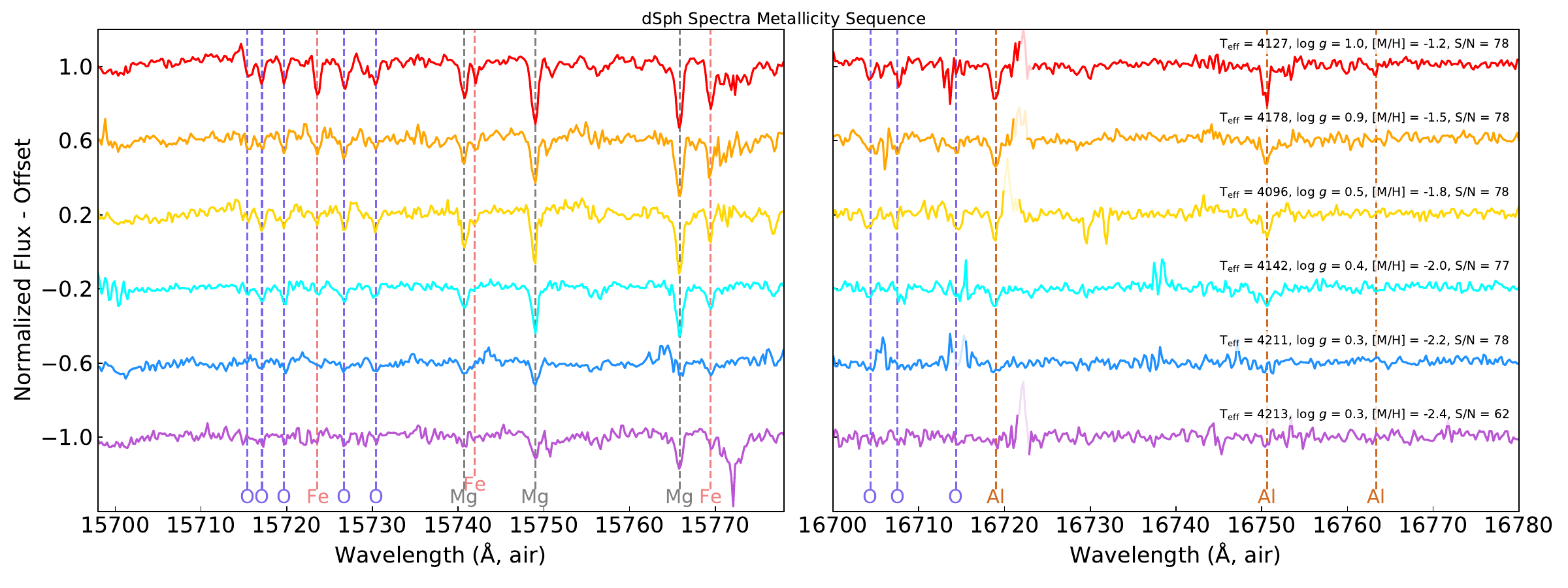}
    \includegraphics[width=1.0\textwidth]{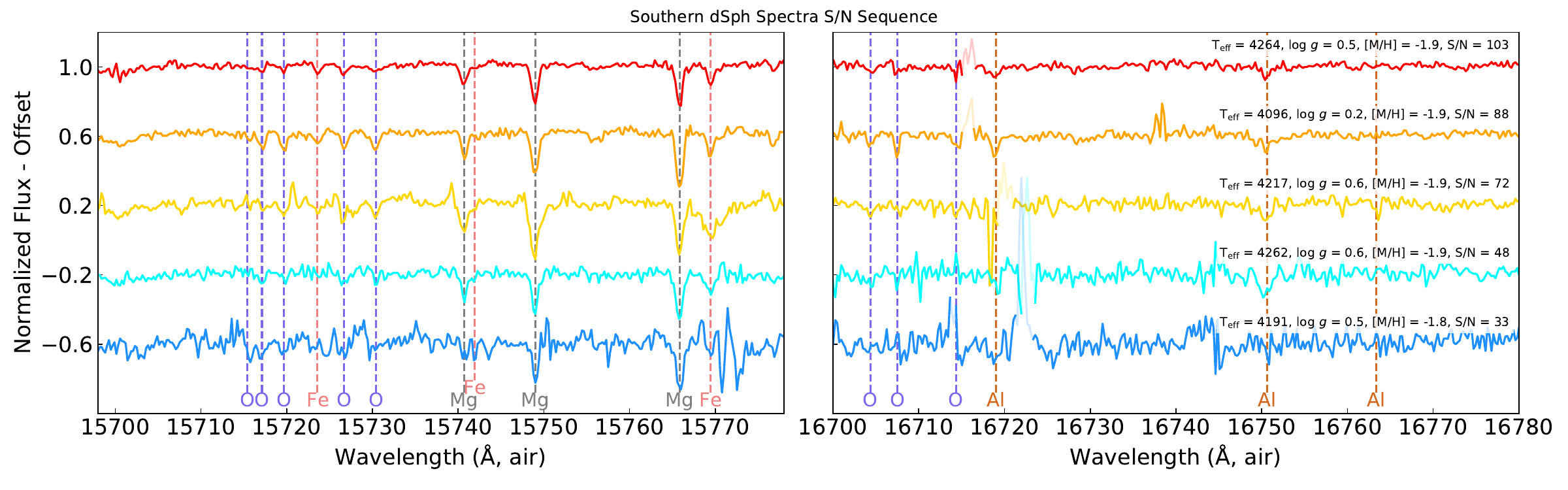}
    \includegraphics[width=1.0\textwidth]{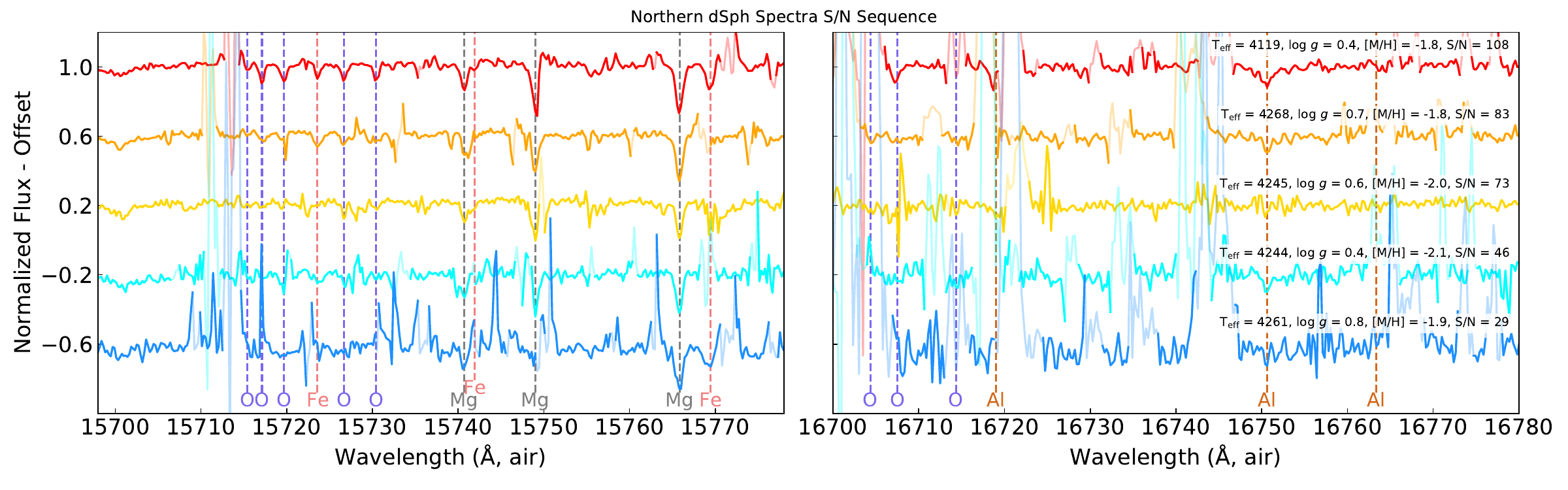}
    \caption{Continuum normalized APOGEE spectra of dSph stars as a function of metallicity (top row), or S/N for stars observed in the south (middle row) and in the north (bottom row).
    Relatively strong lines are identified in a spectral window around the strong Mg I lines (left panel) and Al I lines (right panel). Lines marked as O indicate OH features. Pixels with uncertainties $> 25\%$ are masked as transparent for clarity.
    In the top row, stars with S/N$\sim 70$ with \teff$\sim 4550$ and \logg$\sim 0.5$  are compared from \xh{M} from -1.2 to -2.4 from top to bottom. 
    In the middle and bottom panel, stars with \teff$\sim 4550$ and \logg$\sim 0.5$  are compared with S/N 100 to 30 from top to bottom.
    }
    \label{fig:spec_seq}
\end{figure*} 

Figure \ref{fig:spec_seq} shows example APOGEE spectra for dSph members in spectral windows around the strongest Mg I and Al I lines. 
These spectra were selected to highlight two of the challenges APOGEE faces in measuring stellar parameters and chemical abundances for dSph stars: the low metallicity of the stars, which makes many lines extremely weak, and the low S/N of their spectra, which often makes those weak lines indistinguishable from noise. 

In the top panel of Figure \ref{fig:spec_seq}, we show a sequence of dSph star spectra with \teff$\sim 4550$, \logg$\sim 0.5$, and S/N$\sim$70 for a series of metallicities, \xh{M} from -1.2 (top spectrum, red) to -2.4 (bottom spectrum, purple). 
The vertical lines indicate the wavelength for OH, Fe I, and Mg I (left) and OH and Al I (right). 
The Mg I and Al I features, in particular, are among the strongest lines in the APOGEE spectral range.
This illustrates that at the lowest metallicities that APOGEE can measure (\xh{Fe}$= -2.5$, which is the low metalliticy edge of the synthetic spectral grid) even the strongest spectral features can grow weak or disappear entirely.  This effectively limits the number of elements that APOGEE is able to accurately measure at the lowest metallicities to only the elements with the strongest lines or those species with strong enhancements. As described in detail in the Appendix, dSph spectra taken with APOGEE-N tend to be of lower quality than with APOGEE-S, even at fixed reported S/N ratio, meaning that defining reliable measurements for the full sample means that a global S/N restriction to the dataset of interest, as is commonly used in other works using APOGEE-data, will not result in a uniform quality to any measured quantities, either stellar parameters or abundances.

Figure \ref{fig:spec_seq} shows that there are many measurable features in the APOGEE spectra of dwarf galaxies, but a simple S/N threshold is not enough to select ``reliable'' measurements. Rather than creating criteria that separate reliable from bad measurements, we derive theoretical upper limits, as described in detail in \autoref{ssec:detection}. We first make use of some of the APOGEE quality flags and uncertainties to remove stars with probably bad parameters/abundances from our sample. 

To begin with, we remove any stars that have the \texttt{ASPCAPFLAG} quality flag \texttt{STAR\_BAD} set as this indicates that the outputs from ASPCAP are suspect.  
We also remove any stars that lie near the ASPCAP metallicity grid edge with \xh{M} $< -2.4$ or near the high edge of the carbon grid with \xm{C} $> 0.9$, in either case the stars are likely not well modeled, since they may have parameters beyond these grid edges.

\begin{figure*} 
    \centering
    \includegraphics[width=0.8\textwidth]{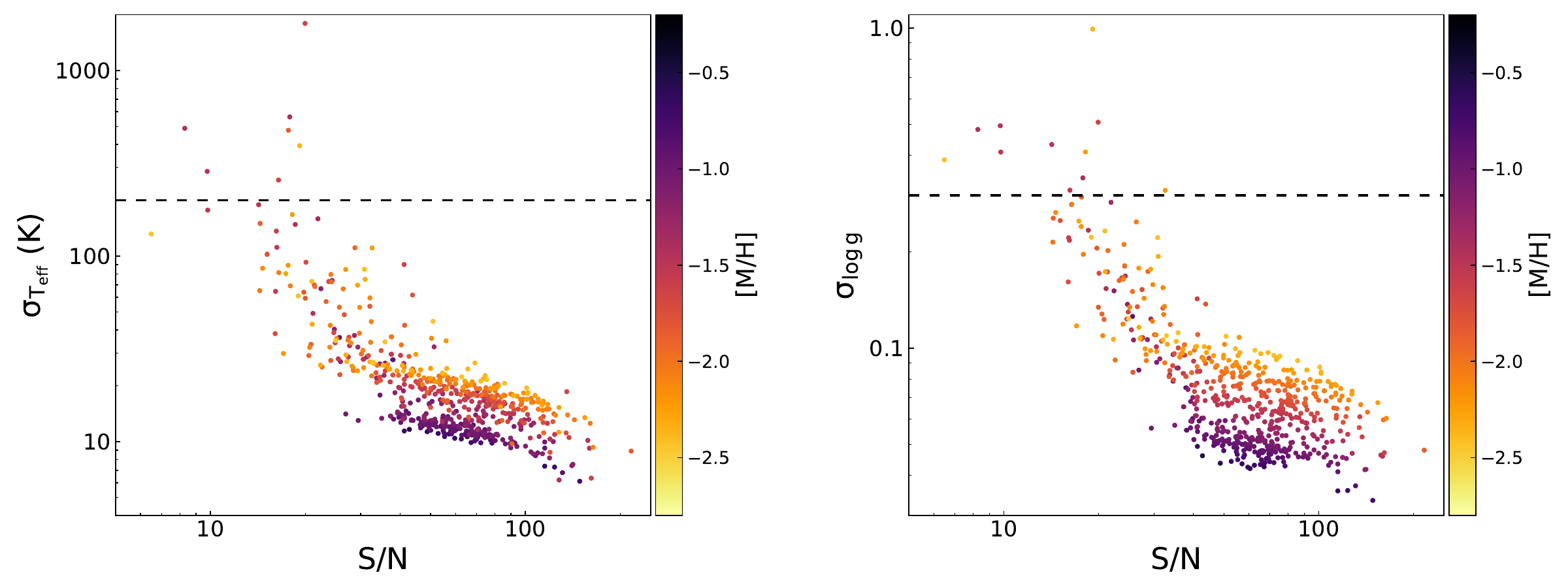}
    \caption{  
    The \teff\ and \logg\ errors from ASPCAP are a complicated function of spectral S/N (as measured by \textsc{snrev}) and the overall metallicity (\meh).  We exclude stars with errors greater than by removing those with errors greater than 200 K and 0.3 in \teff\ and \logg, respectively (indicated by dashed lines), because these stars have parameters that are too uncertain to provide usable information.
    }
    \label{fig:diagnostic_errors}
\end{figure*} 


Second, we remove stars that have large reported ASPCAP uncertainties. \autoref{fig:diagnostic_errors} shows how the ASPCAP uncertainties in \teff ($\sigma_{\rm Teff}$) (left) and in \logg ($\sigma_{logg}$) (right) are complex functions of S/N and overall metallicity (\xh{M}). 
We remove stars with the poorest quality by removing those with errors greater than $\sigma_{Teff} >$ 200 K and $\sigma_{logg} >$ 0.3, which are indicated by dashed lines in \autoref{fig:diagnostic_errors}. 
In addition we remove a sample of stars that have calibrated ASPCAP stellar parameters outside of reasonable ranges for luminous red giants that we would observe in relatively old, metal-poor systems at these distances.
Examples of these stellar parameters are ones that are very hot, putting these stars as B, A, or F type stars (often because their low metallicity and S/N ratios are consistent with featureless spectra), or
stars that ASPCAP assigns as dwarfs. Some of these appear to be variable or carbon stars, which ASPCAP likely has not modeled correctly.
These stars and the rationale for rejection are given in \autoref{tab:bad}.

\subsection{Abundance Upper Limits} \label{ssec:detection}

Much of the vetting of the APOGEE stellar parameters and abundances, to date, has focused on the high signal-to-noise regime, e.g., S/N $\gtrsim$ 100, and/or stars of high to moderate metallicity, which is appropriate for the bulk of the APOGEE main red star sample \citep[e.g.,][]{Zasowski_2013}.
During DR17 calibration, we placed restrictions on which ASPCAP results were considered ``reliable'' based on the effective temperature (\teff) and $\log g$ of the star e.g., limiting the \teff ranges in which abundances of a given element are reported because suspect \teff trends suggest that lines are severely blended or simply not detectable at typical APOGEE  S/N $\gtrsim$ 100.
Although the dSph stars are generally within the normal \teff and $\log g$ range for most species to be reliably measurable at solar abundances (and thus are reported by ASCPAP), the stars are at both significantly lower metallicity and lower S/N than a typical APOGEE star, so APOGEE's lack of abundance upper limit treatment means that unmeasurable abundances may be reported for certain elements. 
In this section, we describe our procedure to determine when a measurement reported in the APOGEE catalog is actually measurable at a given S/N and {\teff}.

Many abundance determinations use a line-by-line approach, but APOGEE spectral fitting is a full-spectrum approach that isolates individual species using windowing functions and fitting all features simultaneously. 
In a line-by-line approach, the abundance detection limit can be determined in its simplest fashion using the depth of the line at a given \teff\  compared to the signal-to-noise ratio of a typical spectrum \citep[see e.g., ][]{Hayes_2022}.
We modify the spirit of this methodology and apply it to the APOGEE methods.  
The method is fully described, with an exemplar element (Ca), in \autoref{app:upper_limits}, but we briefly describe it here.

For a given element, we produce synthetic spectra at a range of temperatures and abundances of that element.  
By comparison of these syntheses to synthetic spectra without the lines of that element, we can measure the line depth of this element alone.  Using the pixel weights from the elemental windows that APOGEE uses to derive abundances, we measure a weighted average line depth, $D_{\rm wavg}$, that is determined across all lines with non-zero window weights for that element at each temperature and abundance.  
In noisy observations, the spectral noise per pixel ($\sigma_{\rm pix}$) averaged over the weighted number of pixels ($N_{\rm wavg}$) is $\sigma_{\rm wavg} = \sigma_{\rm pix} / \sqrt{N_{\rm wavg}}$. 
We can therefore set a noise threshold according to $D_{\rm wavg} > \chi \cdot \sigma_{\rm wavg}$, where we have chosen $\chi = 4.0$, i.e., for an element to be detectable, we require the weighted average line depth to be 4.0 times larger than the noise on that measurement.  This threshold was chosen based on visual inspection of the spectra of stars near the detection limits and comparison with the abundance measurements from \citet{Hill_2019}, when possible.

Because the spectral noise (of normalized spectra) is $\sigma_{\rm pix} = 1/({\rm S/N})$, this is equivalent to setting a threshold signal-to-noise ratio, below which we would not be able to measure that element in the synthetic spectrum:  ${\rm S/N}_{limit} = \chi / (D_{\rm wavg} \, \sqrt{N_{\rm wavg})}$.  
We calculate these $S/N$ limits for our range of synthesized temperatures and abundances.  
This combination of temperatures, abundances, and the $S/N$ thresholds then allows us to parameterize the detection threshold for each element.

Instead of deriving a relation to calculate the S/N threshold as a function of temperature and abundance, however, it is more useful to produce a relationship that defines the upper limit on abundances as a function of temperature and S/N.  By doing so, we can provide abundance upper limits for stars that are below our detection threshold.  
So, we fit a function of the form:


    \begin{equation}\label{eq:upper_limit}
        {\rm [X/H]}_{\rm limit} = \alpha + \beta \bigg(\frac{T_{\rm eff}}{1000\, K}\bigg) + \gamma \log_{10}(S/N) + \delta \bigg[(\log_{10}(S/N))\bigg]^{2}, 
    \end{equation}
    
\noindent to the distributions of $T_{\rm eff}$, [X/H], and S/N threshold values from our synthetic spectra, to determine the optimal coefficients $\alpha$, $\beta$, $\gamma$, and $\delta$ for our upper limit threshold relations.
The coefficients for 17 elements are given in \autoref{tab:detectionlimit}.
A more detailed account of this process is given in \autoref{app:upper_limits}.  


\begin{table}
    \centering
    \caption{Coefficients to Determine Detection Limits for 17 Elements in APOGEE DR17}
    \begin{tabular}{c cccc}
    \hline \hline
    Species & $\alpha$ & $\beta$ & $\gamma$ & $\delta$ \\
    \hline  
     C &  -5.60 &  1.40 & -1.90 & 0.209 \\
     N &  -3.33 &  1.11 & -1.87 & 0.210 \\
     O & -21.61 &  5.75 & -4.76 & 0.601 \\
    Na &  -0.80 &  0.55 & -1.12 & 0.017 \\
    Mg &   0.38 &  0.58 & -5.17 & 1.116 \\
    Al &  -0.18 &  0.64 & -5.09 & 1.226 \\
    Si &   0.85 &  0.33 & -4.68 & 1.078 \\
     S &   4.66 & -0.48 & -2.06 & 0.171 \\
     K &  -0.68 &  0.52 & -1.57 & 0.099 \\
     V &  -4.92 &  1.50 & -1.09 & 0.012 \\
    Ca &  -1.32 &  0.59 & -1.57 & 0.108 \\
    Cr &  -1.45 &  0.67 & -1.31 & 0.050 \\
    Mn &  -1.61 &  0.61 & -1.85 & 0.188 \\
    Co &  -2.69 &  0.89 & -1.32 & 0.058 \\
    Fe &  -1.88 &  0.44 & -1.91 & 0.273 \\
    Ni &  -0.68 &  0.43 & -1.78 & 0.160 \\
    Ce &  -0.93 &  0.51 & -1.84 & 0.174 \\
    \hline \hline 
    \end{tabular}
    \label{tab:detectionlimit}
\end{table}

Using these relations, we then flag any abundances whose \xh{X} measurement from APOGEE is within 1$\sigma$ (using the reported abundance uncertainties) of falling below the upper limit for that star's temperature and S/N.  In the subsequent analysis, we will either not show abundances that are considered upper limits in this analysis, or clearly identify them as having upper limit measurements, where their upper limit is the maximum of the upper limit relation for that star's temperature and S/N or the APOGEE reported abundance (in the case that the star is within 1$\sigma$ of the upper limit relation).

One important caveat about our methodology is that, near the 4-sigma detection threshold, we can potentially get a few remaining false positive detections of stars with very high \xfe{X} abundance ratios, i.e. the reported APOGEE abundance would suggest it should be detectable, but visible inspection shows that these high abundances are merely a result of large, local noise in the spectra, e.g., poor sky-subtraction or telluric corrections. 
Thus, our methodology is sensitive to some high abundance false positives.

Detailed notes on individual elements can be found in \autoref{app:upper_limits}, but based on this upper limit analysis we exclude the Na, S, and V abundances from this work. We also remove 25 stars that should not have measurable [Fe/H] and 12 stars that should not have measurable [Mg/Fe], as suggested by the upper limit analysis.

\subsection{Final Dwarf Galaxy Giant Sample}

Of the 731 stars that we determined to be members of dSphs, 181 are removed for the various reasons listed above and summarized here:

\begin{itemize}
    \item Star Bad
    \item Grid edge in [M/H] and [C/M]
    \item ASPCAP uncertainty 
    \item Unreasonable ASPCAP parameters (e.g., logg values consistent with main sequence stars that would be much too faint to be observed by APOGEE if they were actually in the dSphs)
    \item Beyond theoretical upper limits
\end{itemize}

Approximately 50 stars were targeted as dSph members based on literature results that are not included in our final membership sample. 
Most of the spectra had poor overall data quality (\texttt{SNREV}$<$ 34), but for $\sim$12 we provide further notes in \autoref{app:star_by_star}. The final APOGEE dSph sample with measurable chemical abundances consists of 518 stars across the 7 galaxies that have reliable measurements of at least [Fe/H] and [Mg/Fe].  For the other elements, we show in \autoref{tab:detectionsample} how many stars in each galaxy have measurable abundances after considering the upper limits. Examples of these measurements are presented in Table \ref{tab:data}, and a machine-readable table can be found in the online journal.

\begin{table*}
    \centering
    \caption{Final Chemistry Sample with True Measurements in APOGEE DR17 \label{tab:final_chem_sample}}
    \begin{tabular}{c ccccccc}
    \hline \hline
    Species & Fornax & Sculptor & Carina & Ursa Minor & Draco & Sextans & Bootes \\
    \hline  
     C & 195 &  34 & 20 &  4 &  4 &  1 & 0  \\
     N & 190 &  27 & 12 &  2 &  3 &  0 & 0  \\
     O & 208 & 112 & 41 & 27 & 38 & 19 & 3  \\
    Na &  0 (2) &  0  & 0  &  0 &  0 (1) &  0 & 0  \\
    Mg & 207 & 145 & 49 & 41 & 41 & 32 & 3  \\
    Al & 191 &  99 & 40 & 19 & 21 & 14 & 0  \\
    Si & 208 & 146 & 50 & 43 & 42 & 32 & 5  \\
     S &   0 (32) & 0  (3) &  0  (2) &  0 &  0 &  0 & 0  \\
     K &  79 &   1 & 0 & 0 & 0  & 0  &  0  \\
    Ca & 169 &  19 & 4  &  1 &  2 &  0 & 0   \\
     V & 4(6)&  0  &  0 &  0 &  0 &  0 & 0 \\
    Cr &  33 &   2 &  0 &  0 &  0 &  0 & 0  \\
    Mn & 152 &  31 & 17 &  5 &  2 &  1 & 0 \\
    Fe & 208 & 149 & 50 & 47 & 42 & 36 & 5  \\
    Co &  74 &   3 &  0 &  0 &  0 &  0 & 0   \\
    Ni & 178 &  29 &  8 &  2 &  3 &  1 & 0   \\
    Ce &  68 &   4 &  4 &  2 &  0 &  0 & 0   \\
    \hline \hline 
    \multicolumn{8}{l}{For Entries with two numbers:}\\
    \multicolumn{8}{l}{() number indicates formal formulaic measurements}\\
    \multicolumn{8}{l}{number without () indicates correction for visual inspection}
    \end{tabular}
    \label{tab:detectionsample}
\end{table*}

\begin{table*}
    \centering
    \caption{Stellar Abundances Presented in this Work \label{tab:data}}
    \begin{tabular}{l ccccccccc}
    \hline \hline
    APOGEE ID & Galaxy & [Fe/H] & [C/Fe] & [C/Fe]$_{\rm lim}$ & [Mg/Fe] & [Mg/Fe]$_{\rm lim}$ & [Ni/Fe] & [Ni/Fe]$_{\rm lim}$ & ...  \\
    \hline  
2M13595091+1430029 & Bootes 1 & $-1.92$ &--- &$-0.1$& $-0.1$ & --- &--- &$0.53$ & ...\\ 
2M06405789-5102276 & Carina & $-1.34$ &$-0.55$ & --- &$-0.33$ & --- &$-0.08$ & --- & ...\\
2M17185759+5754142 & Draco & $-2.0$ &--- &$0.14$& $0.13$ & --- &--- & --- & ...\\
2M02393076-3451354 & Fornax & $-0.94$ &$-0.57$ & --- &$-0.26$ & --- &$-0.27$ & --- & ...\\
2M00591209-3346208 & Sculptor & $-1.7$ &$0.2$ & --- &$0.12$ & --- &--- &$0.07$& ...\\ 
2M10135723-0144452 & Sextans & $-1.5$ &--- &$-0.4$& $-0.45$ & --- &--- &$0.14$& ...\\ 
2M15043661+6718538 & Ursa Minor & $-1.8$ &--- &$0.04$& $0.03$ & --- &--- &$0.55$& ...\\ 
... & ... & ... & ... & ... & ... & ... & ... & ... &... \\ 

    \hline \hline 
    \multicolumn{8}{l}{A full machine-readable table including all elements presneted in this work is included in the online journal}\\
    \multicolumn{8}{l}{[X/Fe]$_{\rm lim}$ indicates measurement is an upper limit}\\
    \end{tabular}
\end{table*}

\section{Chemical abundance patterns of dSphs in APOGEE} \label{sec:chem_patterns}

\begin{figure*} 
    \centering
    \includegraphics[width=0.8\textwidth]{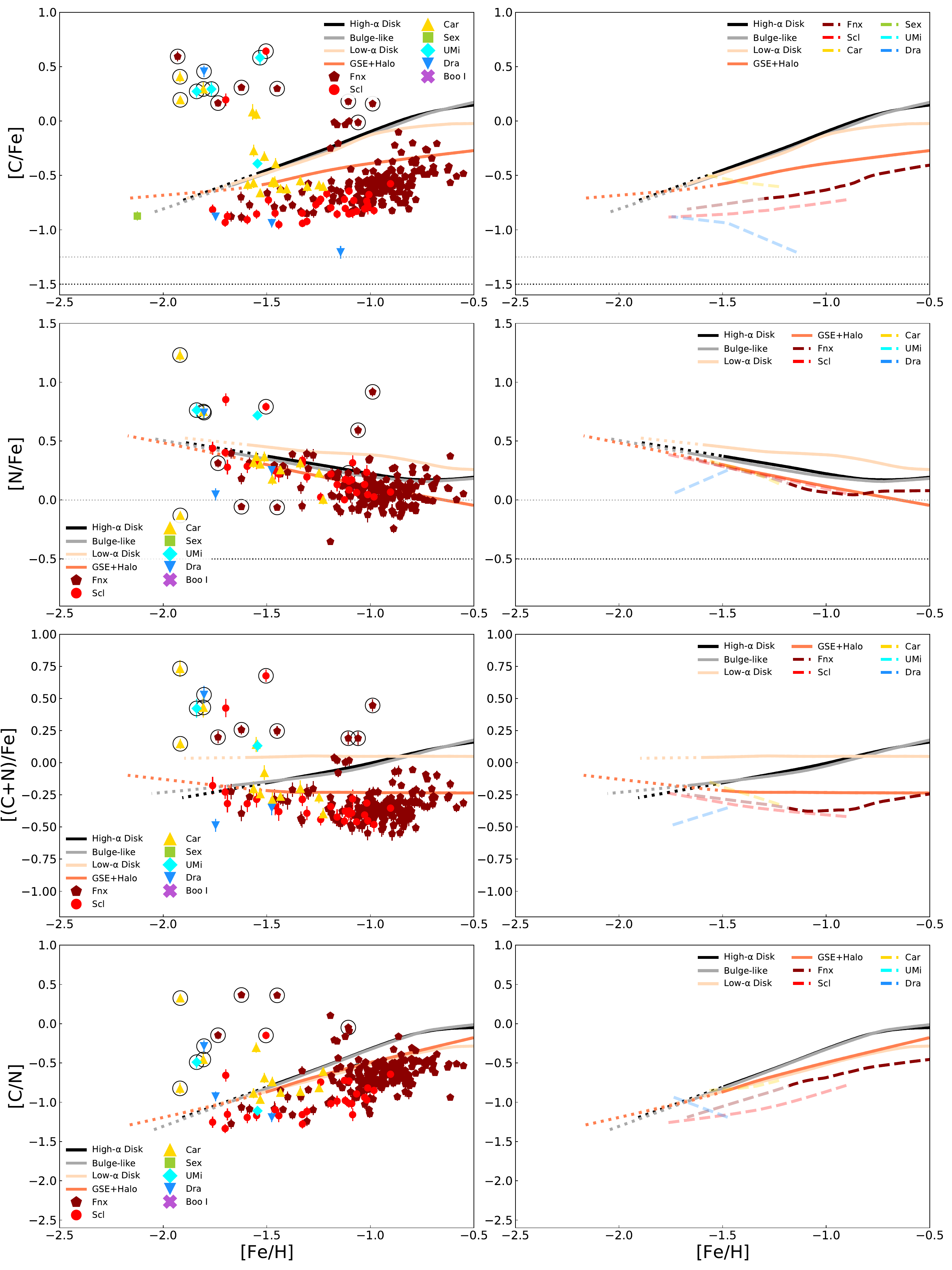}
    \caption{The APOGEE \xfe{C}, \xfe{N}, \xfe{(C+N)}, and [C/N] are shown as a function of metallicity for stars with good abundances in each dSph galaxy (i.e., not identified as upper limits and having uncertainties less than 0.3 dex).   The left panels show the abundances of individual stars (Fornax - dark red pentagons, Sculptor - red circles, Sextans - orange squares, Carina - yellow triangles, Ursa Minor - cyan diamonds, Draco - blue downward triangles, Bootes I - purple crosses) in comparison to LOWESS smoothed distributions of four MW comparison samples (high-$\alpha$ disk - black line, low-$\alpha$ disk - gray, the bulge-like sample - brown, low-$\alpha$ halo - light brown).  In the right panels we show the equivalent LOWESS smoothed abundance distributions for each of the dSphs (dashed lines with the same color-coding), and are transparent at metallicities below which more than half of the abundance measurements are considered bad for that galaxy.  Potential carbon stars are also identified (points surrounded by a black circle).
    }
    \label{fig:cn_grid}
\end{figure*}

\begin{figure*} 
    \centering
    \includegraphics[width=0.8\textwidth]{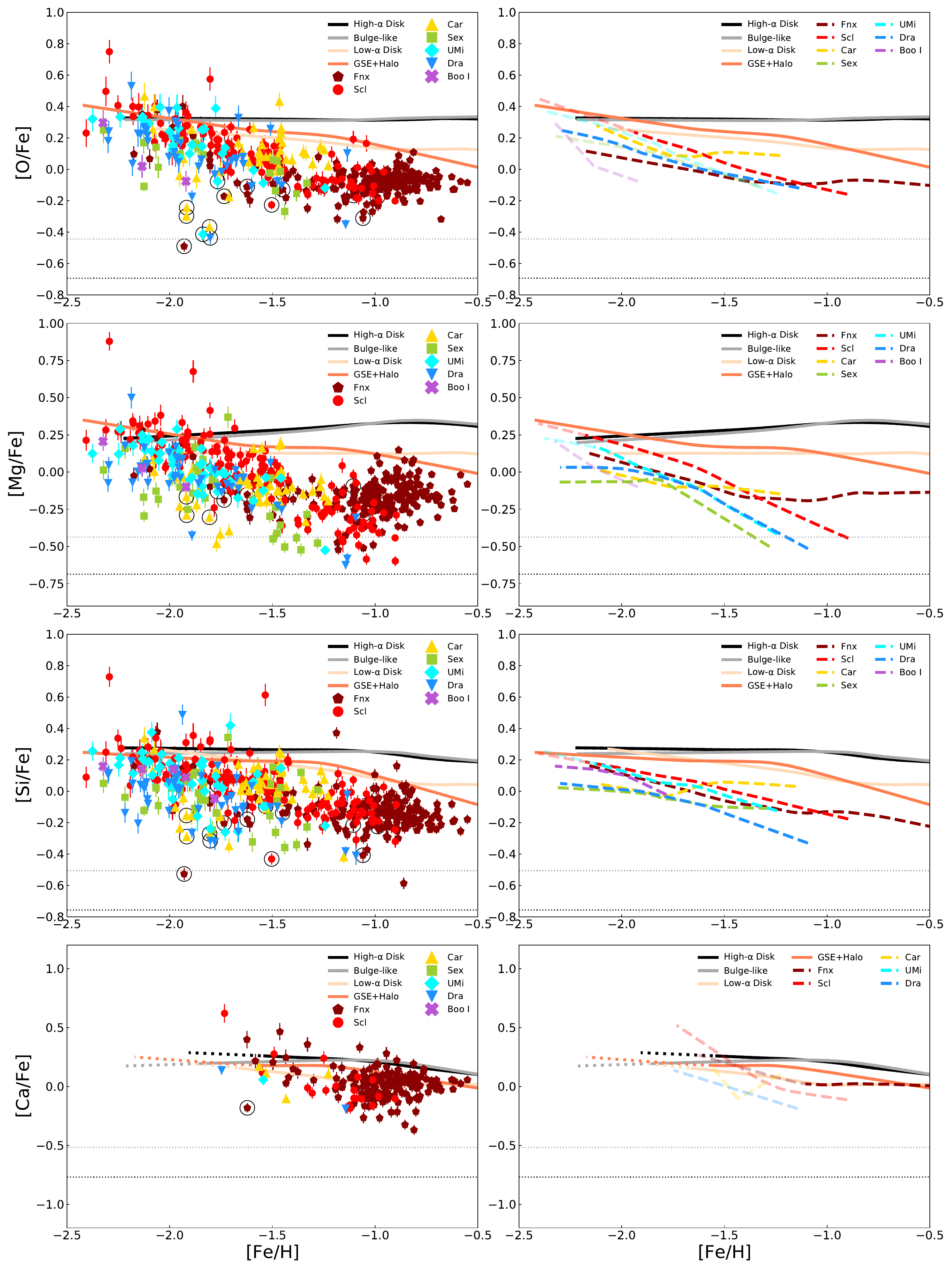}
    \caption{The APOGEE \xfe{X} abundance ratios for O, Mg, Si, and Ca versus metallicity for stars (left panels) with good abundance measurements in each dSph galaxy and the abundance trends (right panels) of each dSph (as shown in Figure \ref{fig:cn_grid}). 
    }
    \label{fig:alpha_grid}
\end{figure*}

\begin{figure*} 
    \centering
    \includegraphics[width=0.8\textwidth]{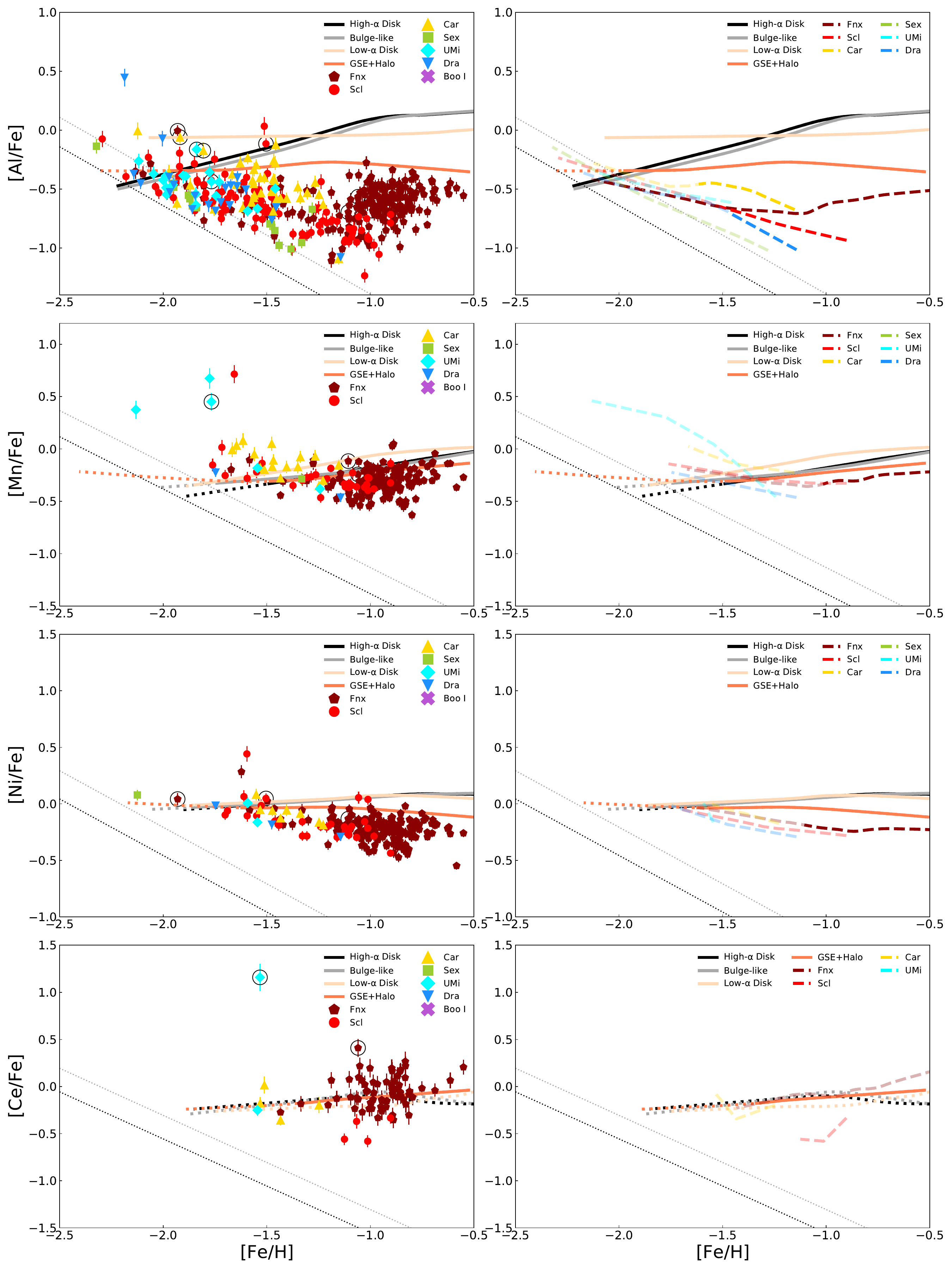}
    \caption{The APOGEE \xfe{X} abundance ratios for Al, Mn, Ni, and Ce versus metallicity for stars (left panels) with good abundance measurements in each dSph galaxy and the abundance trends (right panels) of each dSph (as shown in Figure \ref{fig:cn_grid}).
    }
    \label{fig:agbIa_grid}
\end{figure*}

\begin{figure*} 
    \centering
    \includegraphics[width=0.8\textwidth]{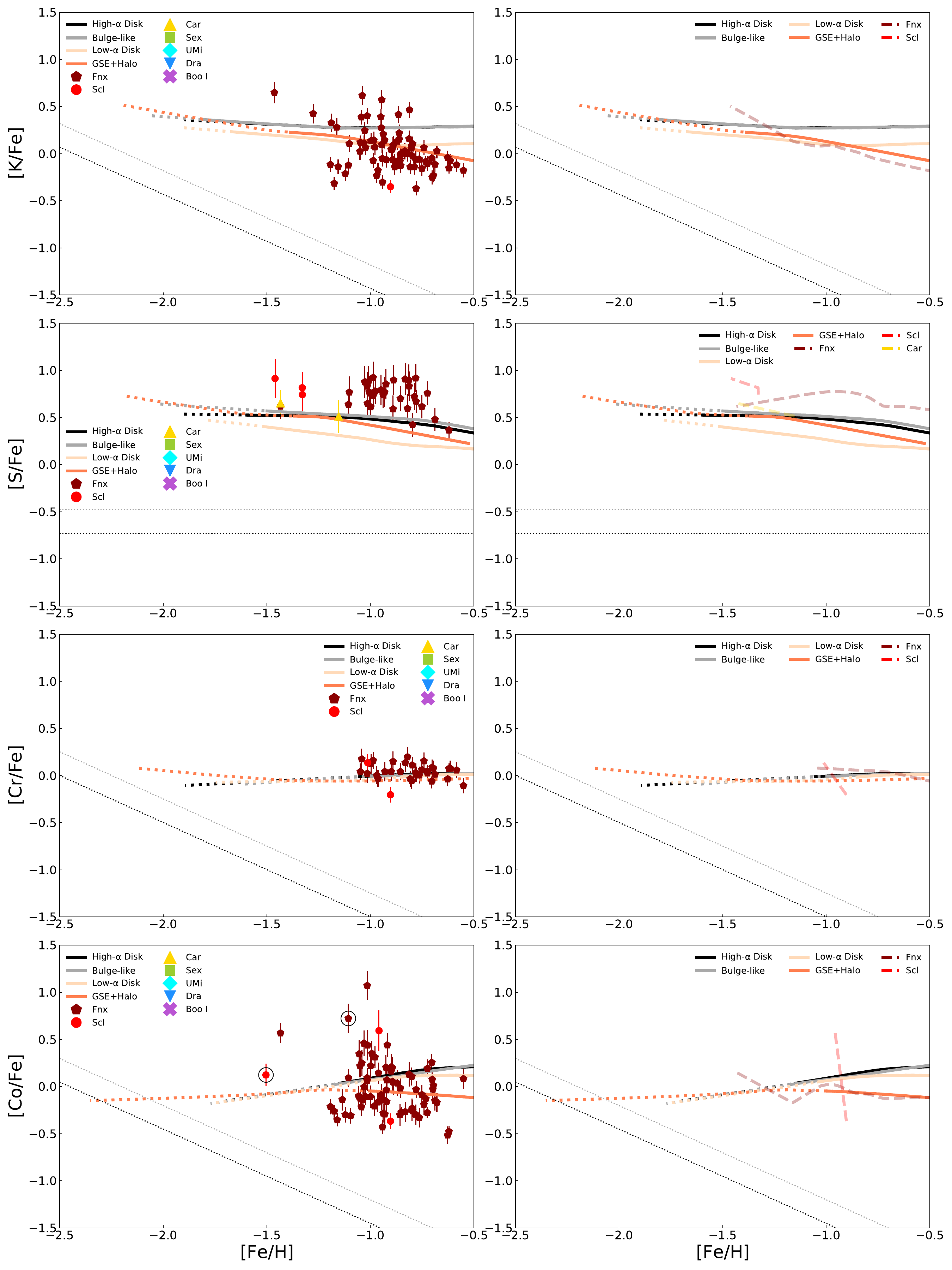}
    \caption{The APOGEE \xfe{X} abundance ratios for K, V, Cr, Co versus metallicity for stars (left panels) with good abundance measurements in each dSph galaxy and the abundance trends (right panels) of each dSph (as shown in Figure \ref{fig:cn_grid}).
    }
    \label{fig:misc_grid}
\end{figure*}

In the subsections to follow, chemical abundance patterns ([X/Fe] for element X) for reliable elements are shown as a function of [Fe/H].
    \autoref{fig:cn_grid} for C and N, 
    \autoref{fig:alpha_grid} for the $\alpha$ elements (O, Mg, Si, and Ca),
    \autoref{fig:agbIa_grid} for the AGB + Ia elements (Al, Mn, Ni, Ce), and 
    \autoref{fig:misc_grid} for other elements (K, S, Cr, Co). 
In each of these figures, the data for each chemical abundance are shown in two ways: 
    on the left, individual points are color-coded according to the dSph and on the right, trend lines are shown for each dSph using the same colors (dashed lines).
The symbols and color-coding are as follows:
    Fornax is magenta and/or pentagons, 
    Sculptor is red and/or circles, 
    Carina is orange and/or upward triangles, 
    Sextans is green and/or squares, 
    Ursa Minor is cyan and/or diamonds, 
    Draco is blue and/or downward triangles, and 
    Bootes I is purple and/or x'es. 
In the left panels, likely carbon stars are circled, where we define a ``carbon star'' as a star with [C/O] $>$ 0.27, which in linear abundances means that the star has more carbon in its atmosphere than oxygen.  These carbon stars may have problematic derived abundances for a number of reasons, including potential variable stars and the rapid changes in the ASPCAP synthetic spectrum grid over the C/O = 1 boundary. 
In both panels of Figures \ref{fig:cn_grid} - \ref{fig:misc_grid}, the MW APOGEE comparison samples are shown (see \autoref{app:comparison} for selection details) for the high-$\alpha$ disk (black), the inner galaxy/Bulge-like stars (grey), the low-$\alpha$ disk (pink), and the GSE$+$inner halo (dark pink). 
In both panels, the approximate grid edge is shown as the dotted black line and the approximate grid point before the edge shown as the grey dotted line; these are approximate because the calibration of the abundances is performed in such a way that individual sources can move across these boundaries \citep[see the calibration discussions in ][Holtzman et al., in prep.]{Jonsson_2020}.

The trend lines shown in the right panels of Figures \ref{fig:cn_grid} - \ref{fig:misc_grid} are made using a LOWESS (locally weighted scatterplot smoothing) algorithm.
The upper limit analysis (\autoref{eq:upper_limit} and \autoref{app:upper_limits}) can bias the determination of these trend lines if a large fraction of the sample is removed. 
We indicate this by reducing the opacity of the lines when more than 50\% of the sample below a given [Fe/H] is removed due to the upper limits; we plot these to give the reader a sense of the apparent trends, as we still have some abundance information at these metallicities, but also to convey that these apparent trends are less clear. 
In some cases, we may be able to recover constraints on these trends using the upper limits, but we do not attempt that here.

\subsection{C, N, C+N, C/N} \label{sec:c_n_cn} 

\autoref{fig:cn_grid} shows the APOGEE, \xfe{C}, \xfe{N}, \xfe{(C+N)}, and [C/N] (from top to bottom) as a function of metallicity \xh{Fe} for each galaxy. 
APOGEE provides carbon and nitrogen abundances determined from (1) the entire spectrum, which we refer to as the parameter-level abundances (because these are fit simultaneously with other stellar parameters) and (2) selected windows around C and N sensitive features. We use the parameter-level fits for our analysis, but find that the windowed [C/Fe] and [N/Fe] abundances do track the parameter-level abundances relatively well in the dSph sample (with a slight bias in the windowed abundances $+$0.022 above the parameter-level fits for C, and -0.003 for N), though the errors on the windowed [C/Fe] abundances are slightly larger.

\paragraph{\xfe{C}} 

The decline in [C/Fe] with decreasing [Fe/H] across all galaxies, including the MW, shown in the top row of \autoref{fig:cn_grid} is consistent with increased efficiency of extra mixing in red giant stars, as described in \citet{Gratton_2000}. 
At metallicities \xh{Fe} $> -1.5$ there are some tentative differences between the C abundances of several dSphs, as best seen in the trend lines of the upper right panel of \autoref{fig:cn_grid}. For example, there appear to be $\sim 0.5$ dex differences in the mean \xfe{C} abundance between Carina (yellow) and Draco (blue), with Sculptor (red) and Fornax (magenta)
between.
At \xh{Fe} $\sim -1$ there is a $\sim 0.2$ dex offset between the \xfe{C} abundances of Sculptor and Fornax. 
At any given metallicity, the dSph stars may have different ages, so these differences in C could be due to differences in mass-dependent internal mixing as stars ascend the red giant branch, or they could reflect underlying chemical evolution differences in the dSphs.  In addition, the Fornax and Sculptor samples near \xh{Fe} $\sim -1$ have different mean surface gravities, with the Fornax sample having log(g) $\sim 0.7$ dex while the Sculptor sample having log(g) $\sim 1.2$.   
This could point towards a difference between more massive AGB star abundances in Fornax and lower mass red giant abundances in Sculptor, although systemic biases in the analysis as a function of log(g) also cannot be ruled out.
The MW sample is weighted towards stars of higher surface gravity, which likely weights the C abundance towards higher carbon abundances, although mass/age differences also cannot be ruled out.

Among the Fornax population, 20 stars have \xfe{C} $>$ -0.1 and of these 16 are either above the RGB tip or 0.1 dex bluer in G$_{BP}$- G$_{RP}$ than the bulk of the stars at their metallicities.    These stars are likely AGB stars, and the anonymously high \xfe{C} probably reflects a stellar evolutionary increase and not a galactic chemical evolution.  However, not all AGB stars show enhanced \xfe{C}, e.g. oxygen-rich asymptotic giant branch (O-AGB) stars.   Below the TRGB some of the higher C abundance stars could be AGB stars since AGB stars can be metal-poor C-enhanced stars, metal-poor C-normal stars, and metal-rich stars C-normal stars, or they could be RGB stars with a mass transfer signature from a binary companion.    

\paragraph{\xfe{N}} 

In the H-band the only source of nitrogen abundances comes from the CN molecule.    For this reason, the carbon abundance must be known independently from the nitrogen abundance.    
To determine which stars have detectable nitrogen abundances, we require that the star have no carbon and nitrogen upper limits. 
The N abundance patterns in the dSphs look quite similar to each other, and to the GSE$+$halo abundance pattern.
The low-$\alpha$ ``thin'' disk appears to be elevated by $\sim$ 0.5 dex in [N/Fe] relative to the dSphs and MW halo, likely due to the fact that the low-$\alpha$ thin disk is composed of higher mass giant branch stars that dredged up more N when they ascended the red giant branch as compared to the lower mass, and therefore, older giants in the dSphs and MW halo. 

\paragraph{\xfe{(C+N)}} 
To attempt to separate out chemical evolution effects from stellar evolution effects on the abundances of C and N, we consider the sum total of these abundances, as first dredge up, to first order, enhances nitrogen and depletes carbon. 
In the Fornax and Sculptor datasets there is an indication for lower \xfe{(C+N)} than all of the Milky Way samples with increasing metallicity. This is likely because the dSphs formed stars with a higher Type Ia/Type II SNe ratio than the MW, resulting in enhanced Fe relative to C+N.  Indeed (and as seen in the past in these dSphs) when considering the $\alpha$-element abundance patterns, we can see that these patterns are also lower than the MW samples, suggesting the more significant influence of Type Ia SNe and their Fe-peak abundance contributions.
The most metal-rich Fornax giants show a rise in the \xfe{(C+N)}. As observed by \citet{Hasselquist2021_sats} this increase in \xfe{(C+N)} could either be due to a burst of star formation that appears to have altered the abundances of other elements (particularly the $\alpha$-elements), or because the most metal-rich Fornax stars are relatively young  (\citealp{Pont_2004, Letarte_2010, Kirby_2010, Lemasle_2014}) and this increase is instead due to contributions from AGB stars.

\paragraph{[C/N]} 
The [C/N] abundance of a red giant star can be indicative of its mass, at least at higher metallicity. It is not clear how reliable of a mass indicator the [C/N] abundance is at lower metallicity where extra mixing apparently plays a role. However, in the bottom row of Figure \ref{fig:cn_grid}, we find that Sculptor and Fornax both exhibit [C/N] abundances that fall below the MW trend.   For Sculptor, Carina, Draco, and Ursa Minor, $>$ 50\% of the sample is removed for upper limits over the entire metallicity range.   This means that interpretation of any offsets in the [C/N] between these galaxies should be treated with a high degree of skepticism.   For example, the Carina [C/N] points fall above the Sculptor points while the \xfe{(C+N)} distributions are very similar, which if interpreted as being due to mass/age would imply that the Carina sample is older than the Sculptor sample at the metallicity over which they overlap.   However, this is not what is expected, Carina, particularly the burst stars, should be younger \citep{Weisz2014}.  We attribute this inconsistency to ab initio higher C abundances and corresponding lower N abundances, or systemics introduced by the large number of upper limits skewing the results.  

However, for the metal-rich Fornax stars, the detections outnumber the upper limits. So, if we interpret [C/N] solely as a mass indicator, these abundances would suggest that the metal-rich Fornax stars are more massive than the MW disk stars. Because mass is correlated with age on the giant branch, we can conclude that Fornax has relatively recently formed stars, consistent with the starburst prediction of \citet{Hendricks2014} and \citet{Hasselquist2021_sats}, with the caveats of metallicity and luminosity-dependent extra mixing and evolutionary history.


\subsection{The Alpha Elements} 

\autoref{fig:alpha_grid} shows O, Mg, Si, and Ca following the style established in  \autoref{fig:cn_grid}. All together, the dSphs tend to have decreasing \xfe{$\alpha$} ratios with increasing metallicity, though this pattern flattens at solar or near-solar [$\alpha$/Fe] for a few galaxies like Carina and Fornax.  
We can also see that in the trends of each galaxy this decreasing abundance pattern does have a relative shift among the galaxies.  
At lower metallicities where all galaxies populate, e.g., [Fe/H]$\sim -1.8$, we typically see that Scl is the most $\alpha$-enhanced followed by Draco and Fornax, and then UMi, Car, and Sxt having the lowest $\alpha$-abundances, and Boo is unclear given how few stars have been observed in it.  
Although it is not clear whether any of these dwarfs reach an $\alpha$-abundance plateau and show evidence of an $\alpha$-knee, it does appear that some of them approach the plateau seen in the MW bulge/thick-disk.


\paragraph{\xfe{O}} 

O appears to have some of the tightest abundance patterns of the $\alpha$-elements, though we can see that carbon-enhanced stars appear to have low O abundances. It is unclear if that is because these stars have intrinsically low O abundances, or if because our upper limit methodology is developed for carbon-normal stars and not carbon-enhanced stars, that perhaps the O in these stars is not easily measurable because the CO lines saturate and the remaining C forms CN molecular features, which masks the H-band CO lines.

Although Boo is poorly populated it does appear to generally be more O-poor on average than other dSphs, which differs from some of the other elements where these Boo stars seem to have more similar $\alpha$-element abundances to the other systems.


\paragraph{\xfe{Mg}} 

Mg is the $\alpha$-element where we observe the largest differences between each of the galaxies. Examining the distribution of stellar abundances we can visually see that, at low metallicities, Scl has higher \xfe{Mg} ratios than UMi and Dra, which are again higher than Sxt.  Interestingly, we can also see some structure in the chemical abundance patterns of some of these galaxies, particularly Car and Fnx.  As shown in \citet{Hasselquist2021_sats}, in Fornax we see a decreasing \xfe{Mg} abundance pattern with increasing metallicity until [Fe/H] = -1.2, where the Fnx [Mg/Fe] abundances begin to increase with increasing [Fe/H].  Similarly in Car, we see a decrease in \xfe{Mg} with increasing [Fe/H] until [Fe/H] = -1.7, where we see a $\sim 0.4$ dex jump in \xfe{Mg} that is reminiscent of what has been seen in the literature for Car, though perhaps with higher precision here.  This feature is washed out in the trend lines because of the small number of stars and the rapid change.  As will be discussed in \autoref{sec:discussion} these are both consistent with what one would expect if these galaxies experienced a burst of star formation that momentarily increases the number of core collapse SNe and drives the \xfe{$\alpha$} ratios closer towards pure CCSNe yields.  At [Fe/H] $\sim$ -1.1, we find that the Scl [Mg/Fe] abundances are 0.2-0.3 dex lower than the Fornax [Mg/Fe] abundances, but this deficiency is not shared by the other $\alpha$ elements. This could be explained if Mg is the only ``pure'' Type II SNe $\alpha$ element, the larger ratio of Type Ia/Type II SNe in Sculptor at these metallicities could explain this difference in Mg, if there is an IMF difference between the two galaxies, or there are metallicity-dependent yields and some delays incorporating SN material into the next generation of stars in one of the two galaxies, e.g. \citet{Kirby_2019,de_los_Reyes_2022}.


\paragraph{\xfe{Si}} 
Si shows distributions very similar to those of O and Mg, with a potentially larger spread within each system (particularly UMi and Draco).  There is also a less pronounced increase in Fnx's \xfe{Si} ratios at high metallicities compared to Mg and O, perhaps because more Si is produced in larger relative quantities in Ia SNe.  At [Fe/H] $\sim -1.5$ to $-1$ Scl looks flatter than it does in Mg, suggesting an interesting difference in the [Mg/Si] ratios, which appear to differ from those seen in Fnx at the same metallicities.

\paragraph{\xfe{Ca}} 
Of all of the $\alpha$-elements investigated in this study, Ca is the most affected by upper limits.
Only the highest metallicity stars have measurable Ca, where we see that it appears similar to what is seen in some of the other $\alpha$-elements.  
For instance, like in O and Si, the highest metallicity stars in Scl have \xfe{Ca} ratios similar to those of Fnx (unlike in Mg), both of which sit around the solar ratio. The Fnx stars with [Fe/H] $>$ -1.0 have [Ca/Fe] abundance ratios similar to those of the MW low-$\alpha$ disk stars.

\subsection{Other Elements} 

\autoref{fig:agbIa_grid} shows Al, Mn, Ni, and Ce following the style established in \autoref{fig:cn_grid}.

\paragraph{\xfe{Al}} 

Strong lines of Al I appear in the APOGEE spectra, but unfortunately, the APSCAP methodology limits the Al measurements to [Al/H] $>$ -2.5 (the [M/H] grid edge). The significantly subsolar \xfe{Al} ratios seen in the dSphs mean that Al runs into the APOGEE grid edge more quickly than other elements. Therefore, while Al is not significantly affected by the upper limit cases because of the weakness of the lines, it is limited by the grid edge.  The LOWESS trend lines represent this limitation by cutting at the metal-poor end, when needed, to avoid the grid edge values from playing an overwhelming role in the resulting value, i.e. when more than 50\% of the values have \xh{Al} grid edge "BAD" values.  This should not be confused with the LOWESS trend lines representing when more than 50\% of the sample may be impacted by upper limits, represented by becoming transparent (in the case of the dSph curves) and dotted (in the case of the MW samples).   

The dwarf galaxies fall below all components of the MW except among the most metal-poor stars. Despite this, we can see some differences between the dSphs.  The Carina and Fornax dwarfs show a flattening or rising trend in \xfe{Al} with increasing [Fe/H] in their most metal-rich stars, similar to what was found for the $\alpha$ element Mg for these galaxies. This seems to continue to indicate that these two galaxies may have experienced bursts, and the added production of light elements through CCSNe is raising the Al ratios, mirroring the trends seen in \xfe{Mg} and to a lesser extent \xfe{O} and \xfe{Si}.  Traditionally, Al production would be associated with N production, but we do not measure N independently, but only through \xfe{(C+N)}.   There do not appear to be strong correlations between \xfe{(C+N)} and \xfe{Al}.

Draco, Ursa Minor, Sextans and Sculptor show decreasing [Al/Fe] values throughout their chemical evolution, with the [Al/Fe] abundances quickly dropping below the trend of GSE+MW Halo at [Fe/H] $>$ -2.0.

\paragraph{\xfe{Mn}} 
Mn, as an Fe-peak element, appears relatively flat across all of the dSphs across the metallicities, where it is not significantly affected by the upper limits.  There may be a slight hint that Car has an \xfe{Mn} enhancement over Scl at similar metalilcities, but this is based on only small-number statistics, so larger samples would be needed to verify this. Some stars have high \xfe{Mn} ratios ($\gtrsim 0.5$), these values lie close to the upper limit thresholds and may suggest that a higher threshold criterion is needed for Mn.  

\paragraph{\xfe{Ni}} 
The [Ni/Fe] abundances are moderately affected by upper limits such that most stars with [Fe/H] $<$ -1.5 do not have reliable [Ni/Fe] measurements. At this metallicity and above, we find that Carina, Scl, and Fnx all exhibit declining [Ni/Fe] abundances with increasing [Fe/H], which put the abundance trends of these galaxies below those of the MW, similar to Al. The [Ni/Fe]-[Fe/H] trend for Fnx apparently flattens out around [Fe/H] $\sim$ -1.0, the same metallicity at which we observe inflection points for the $\alpha$ elements and Al.

\paragraph{\xfe{Ce}} 
Although the sample of stars with Ce measurements is small, there are some interesting notes from Ce.  For instance, most of the Scl upper limits (not shown) and the few bona-fide measurements fall below the Fnx detections at the same metallicities.  This would be consistent with the idea that Fnx has had much more extended SFH \citep{Hasselquist2021_sats} than Scl \citep[e.g., ][]{Bettinelli2019}, and given that Ce is produced primarily in AGB stars through the slow neutron capture process, the longer SFH would allow Fnx to build up more Ce over time than Scl with its short evolution.

\paragraph{Remaining elements} 
Potassium, sulfur, chromium and cobalt are shown in \autoref{fig:misc_grid}. 
K, S, Cr, and Co are measured only in a few stars, primarily in Fnx, because it is the most metal-rich.  Although the formula we derive for the upper limits rejects 86\% of the sample, the remaining 14\% exhibit a large spread in values, and/or the upper limits we compute for the dSph stars fill the same part of parameter space as the detections; thus, it is difficult to separate out real detections from likely upper limits. 
Because of the upper limit thresholds for these elements, we cannot draw any significant conclusions about their abundance patterns in the dwarfs. However, for Fnx, we can see that it is consistent with having solar ratios of \xfe{K} and \xfe{Cr} at its highest metallicities.  It may also have solar \xfe{Co}, though the large scatter prohibits one to see this clearly.  We believe that the \xfe{S} detections are not real and consistent with non-detections as indicated in \autoref{tab:final_chem_sample}.

\section{Comparison With Literature} \label{sec:litcomp}
Many of these galaxies have been studied previously through high-resolution optical observations, and therefore there exist chemical abundance measurements of individual stars that can be compared to the APOGEE ones presented here. Previous investigations have indicated that the abundances measured in the NIR, like APOGEE, can show systematic offsets from measurements in the literature; the origins are partially the use of different wavelength ranges that probe different layers of the stellar photosphere where, in the NIR, the NLTE effects are lower \citep[see e.g.,][]{Korotin2020,osorio_2020}. Because the APOGEE DR17 abundance methodology attempts to correct for NLTE effects on the NIR Mg lines (see \citealt{osorio_2020}), and many of the optical samples we compare to assume LTE, we speculate that this could be the source of some of the offsets we observe.

Additionally, systematic offsets between the NIR and optical can also arise from the methodology used to measure the abundances, including the synthesis code used, model atmospheres adopted, and assumptions about/fixing of stellar parameters. As an additional caveat, the APOGEE dataset is calibrated to stars near the Sun, so zero-point offsets have been applied to the initial ASPCAP outputs, which might result in systematic differences with optical studies, especially at lower metallicities.

\begin{figure*}
    \centering
    \includegraphics[width=1.0\textwidth]{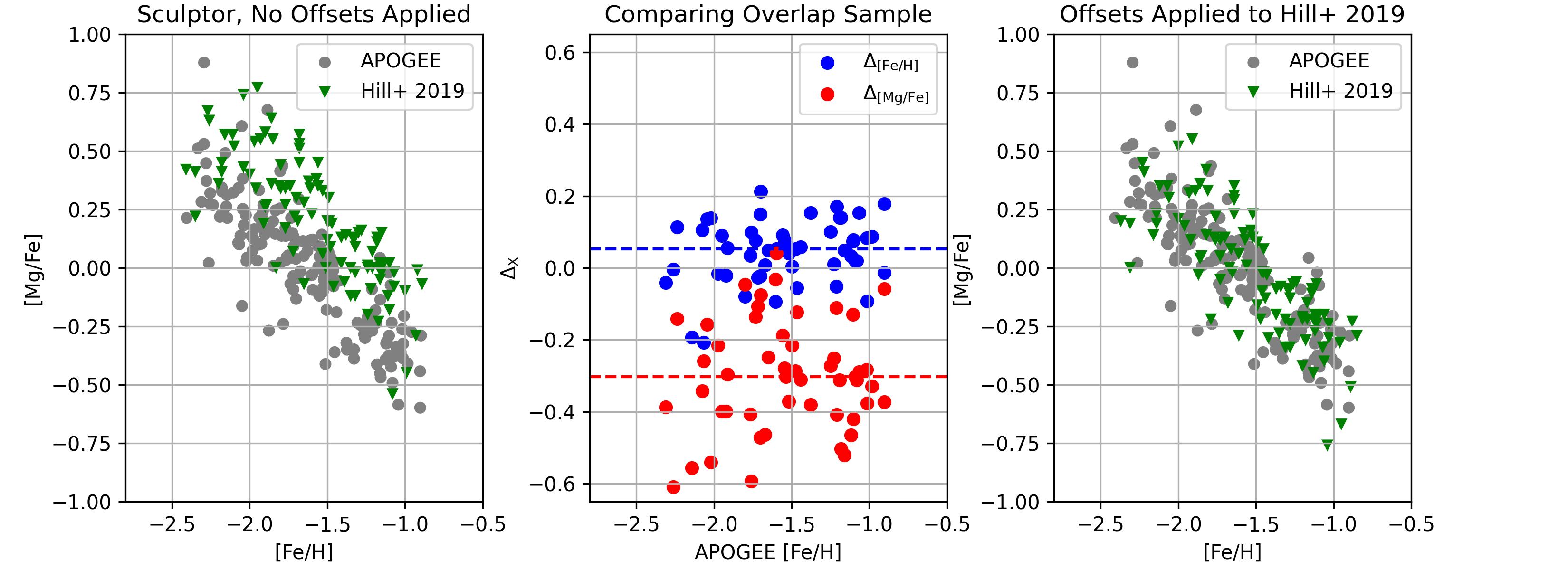}   
    \caption{
        Left: the [Mg/Fe]-[Fe/H] abundance patterns of Sculptor for the APOGEE sample in this work (grey circles) and the \citet{Hill_2019} work (green triangles). While the overall shape is similar between studies, there are clear abundance offsets. Middle: offsets in [Fe/H] (blue) and [Mg/Fe] (red) between APOGEE and \citet{Hill_2019}  as a function of APOGEE [Fe/H] for a subset of stars in common between the two studies. Dashed lines mark the median offsets for each abundance. Right: similar to the left panel, but with the offsets applied to the \citet{Hill_2019} demonstrating that, at least for this literature study, the abundances are merely offset from each other and both studies agree on the dependence of [Mg/Fe] on metallicity
    }
    \label{fig:literature_demo}
\end{figure*}

Thus, before comparing the APOGEE results with those in the literature,  we first use the literature studies with substantial overlap with APOGEE to derive offsets to align those works to the APOGEE abundance scale; we do this in a simple fashion, deriving a global difference that may neglect abundance-dependent offsets. An example of how this is done for Sculptor is shown in \autoref{fig:literature_demo}. The left panel of \autoref{fig:literature_demo} shows how the \citet{Hill_2019} [Mg/Fe]-[Fe/H] abundance pattern compares with what we measure for APOGEE. Although the shape of the sequence is similar between the two studies, there is an apparent offset between them. The middle panel of \autoref{fig:literature_demo} shows the difference in [Fe/H] and [Mg/Fe] measured between the two studies for a subset of stars observed in both studies. The right panel shows how the two studies compare when global offsets are applied to [Fe/H] and [Mg/Fe] abundances of the \citet{Hill_2019} data. 

\begin{figure*}
    \centering
    \includegraphics[width=0.4\textwidth]{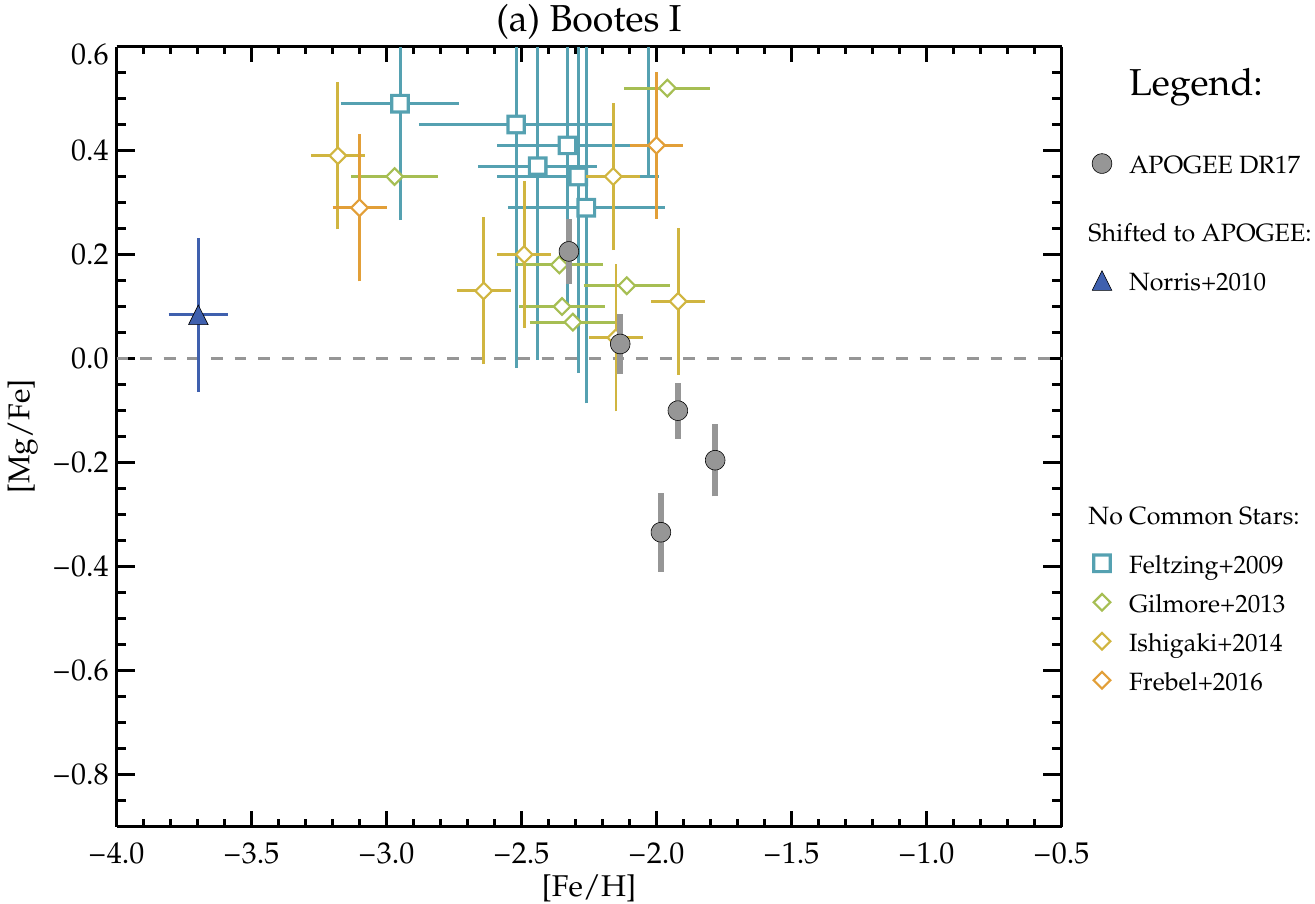}
    \includegraphics[width=0.4\textwidth]{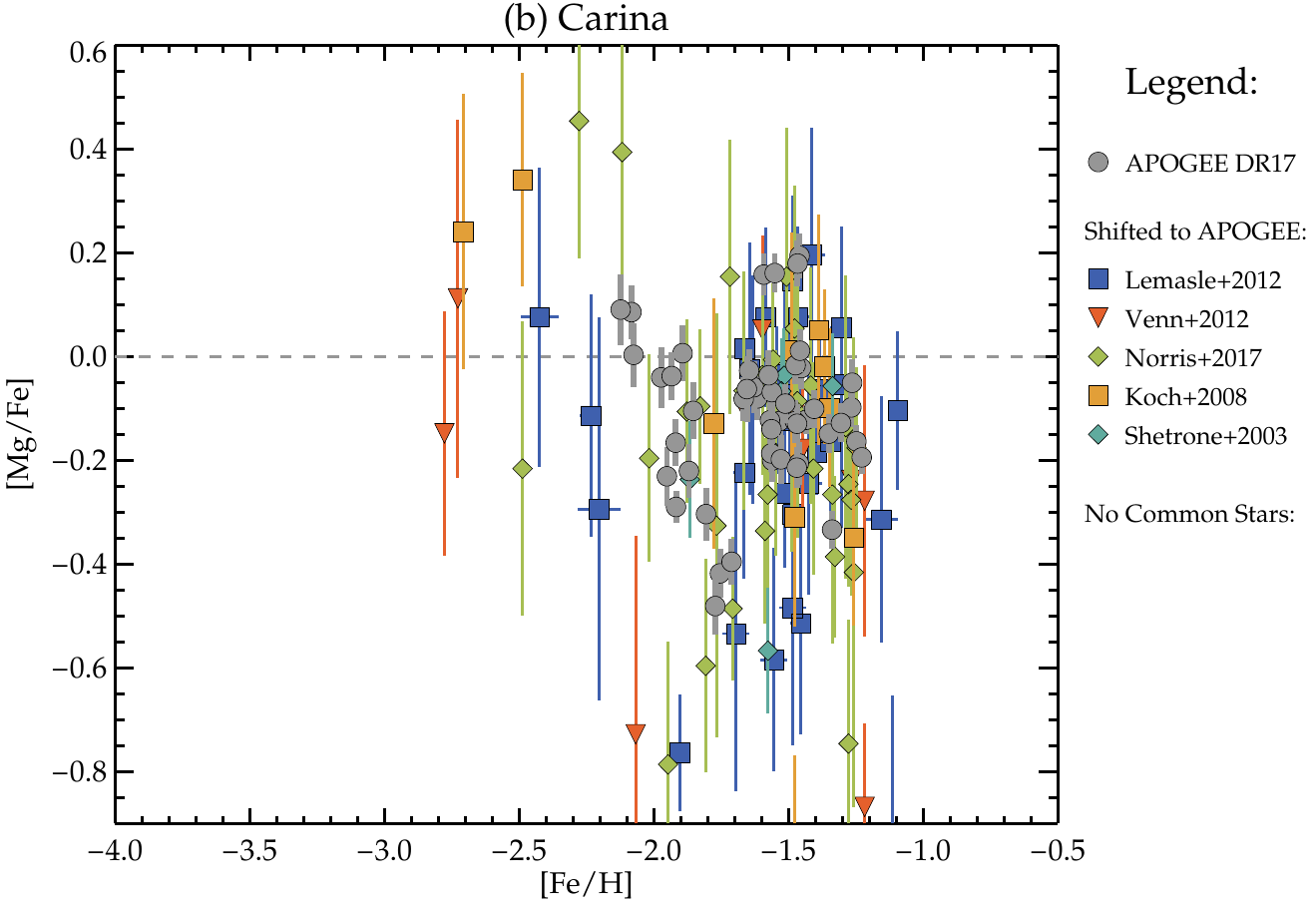}
    \includegraphics[width=0.4\textwidth]{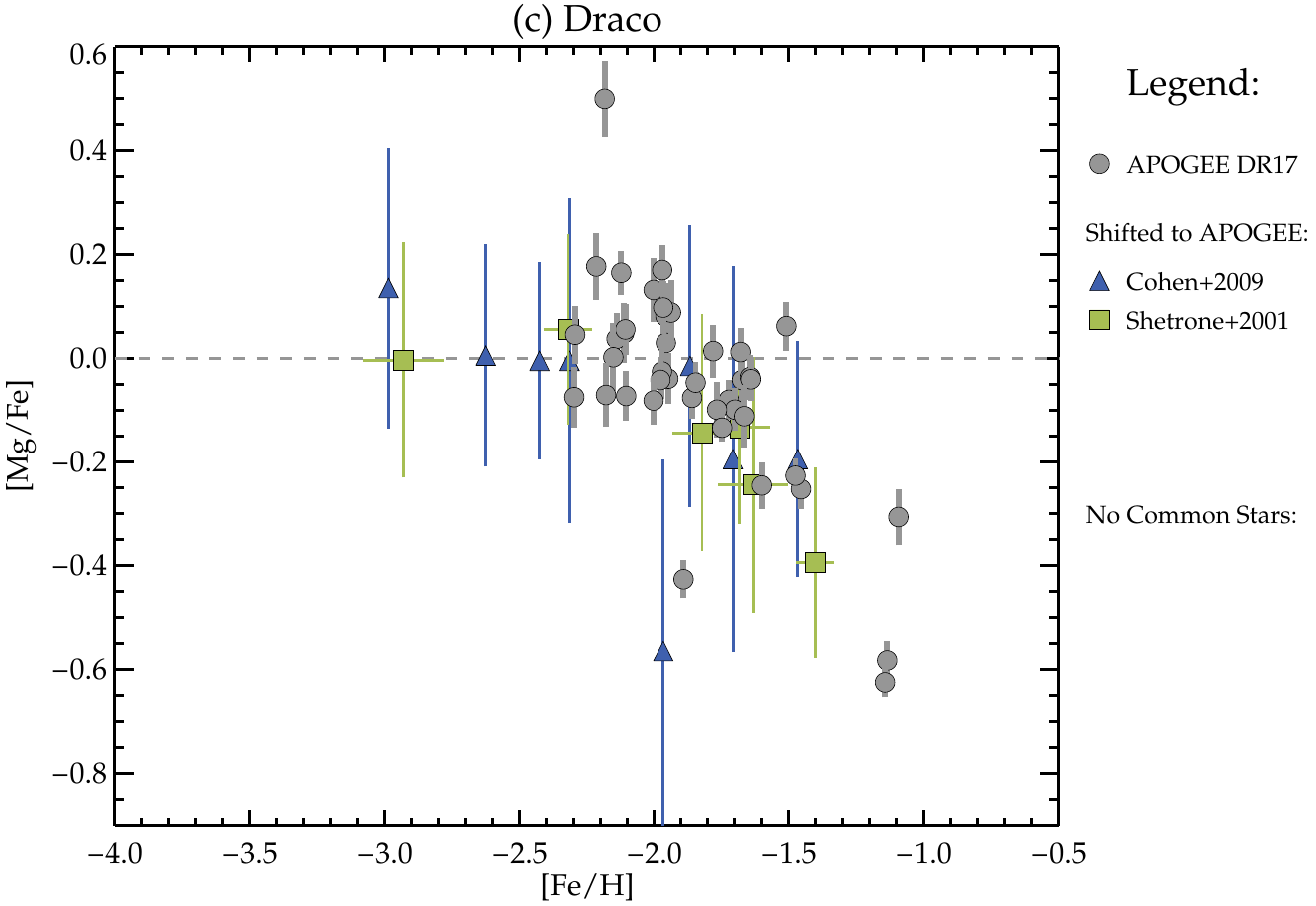}
    \includegraphics[width=0.4\textwidth]{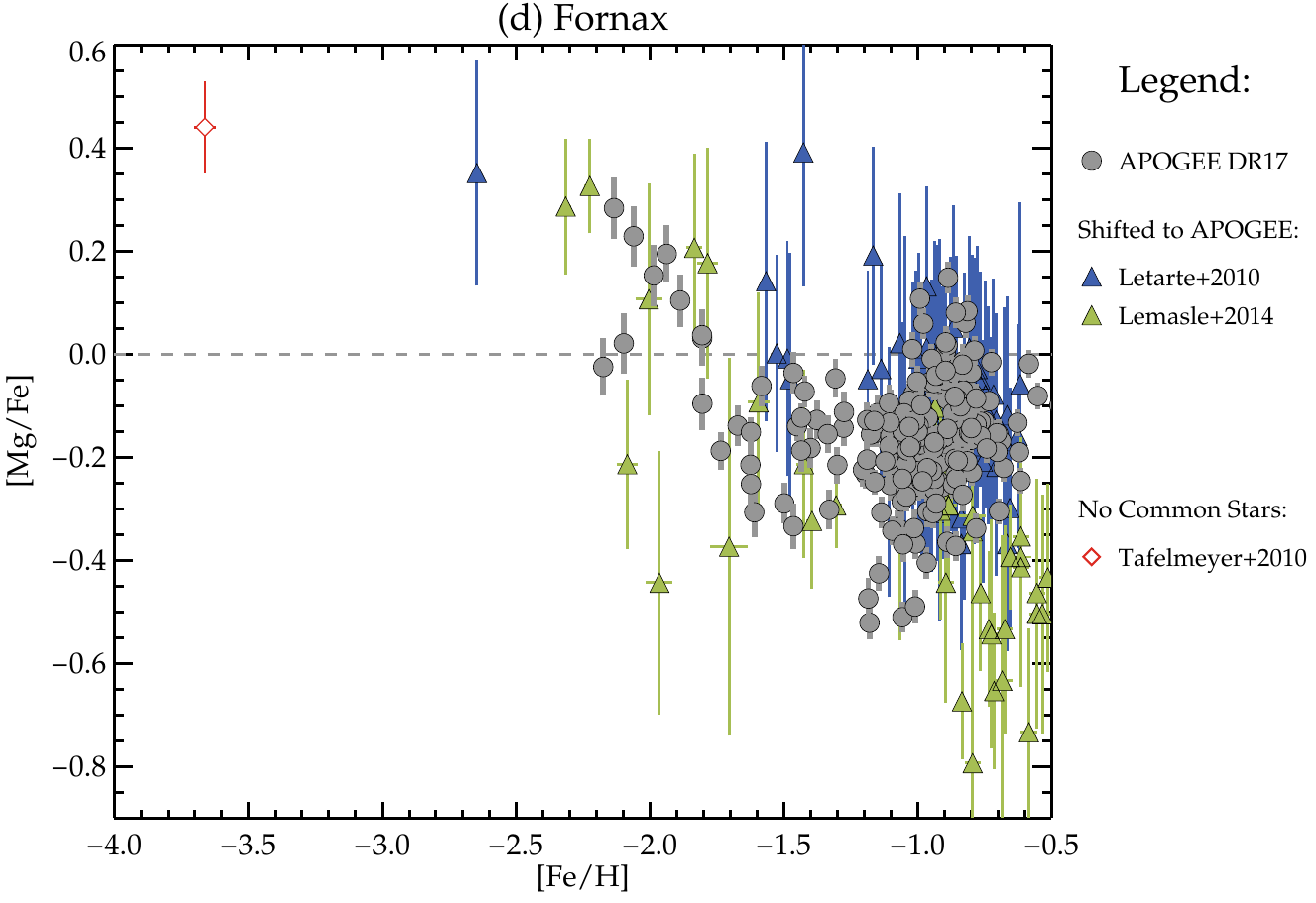}
    \includegraphics[width=0.4\textwidth]{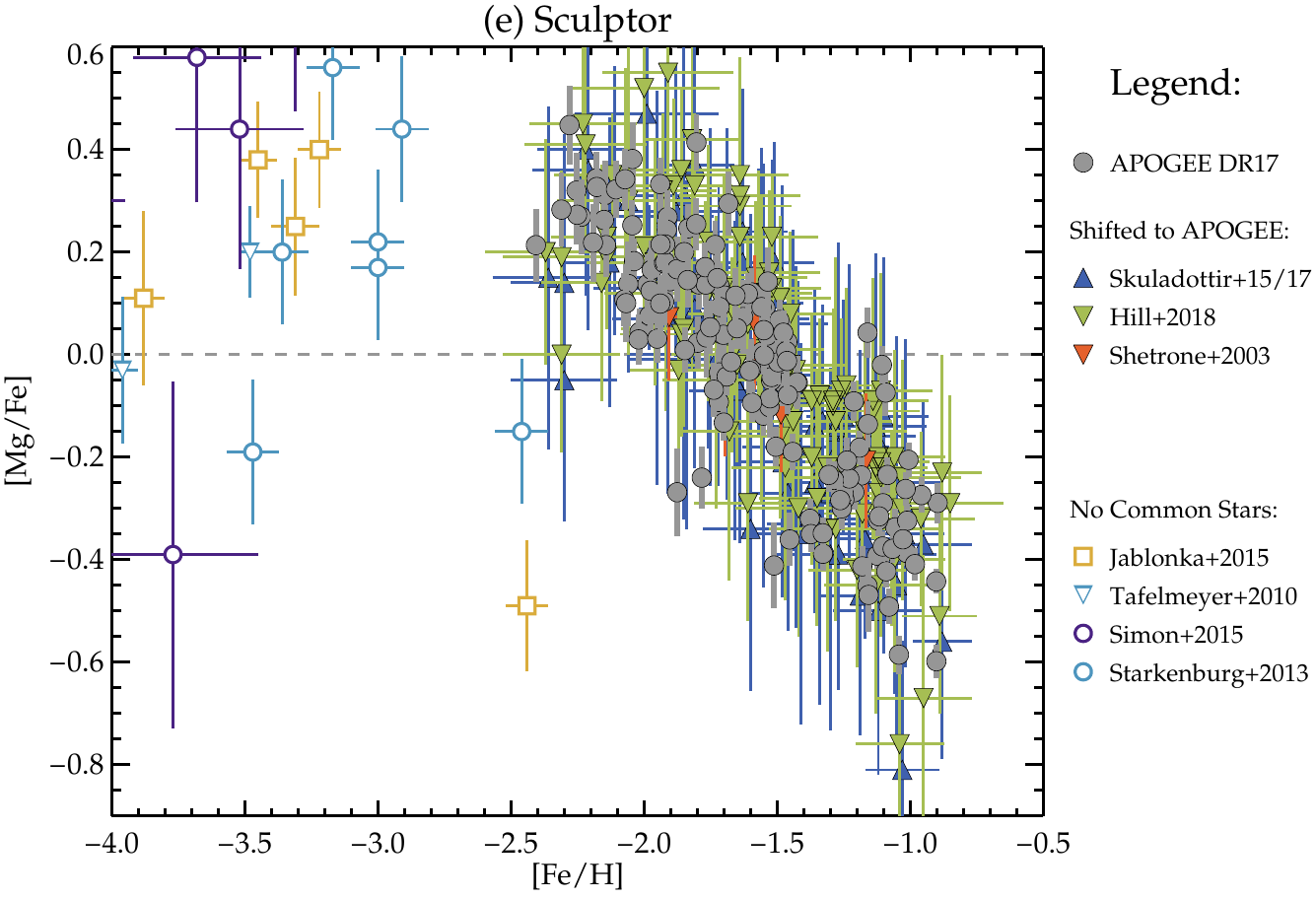}
    \includegraphics[width=0.4\textwidth]{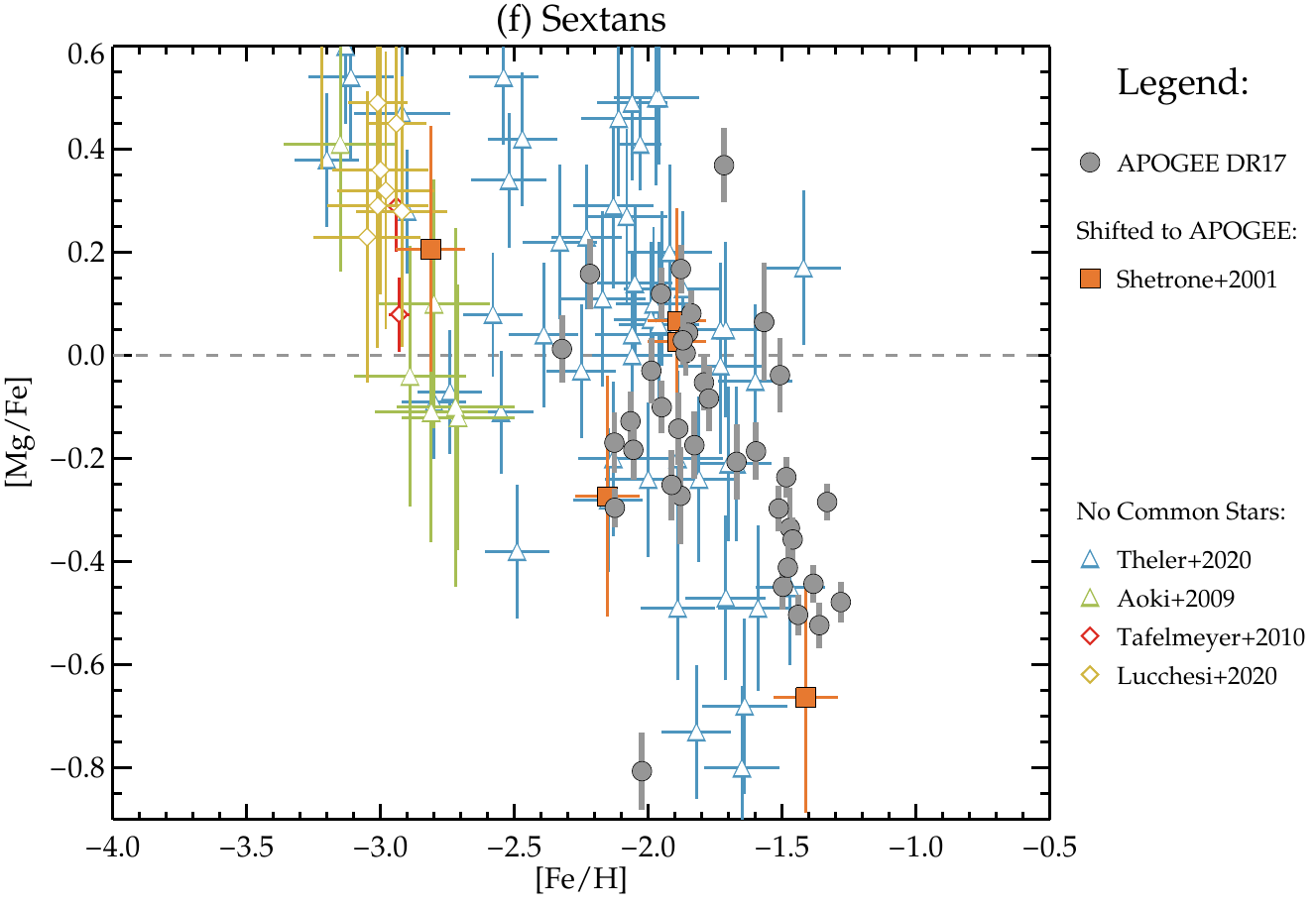}
    \includegraphics[width=0.4\textwidth]{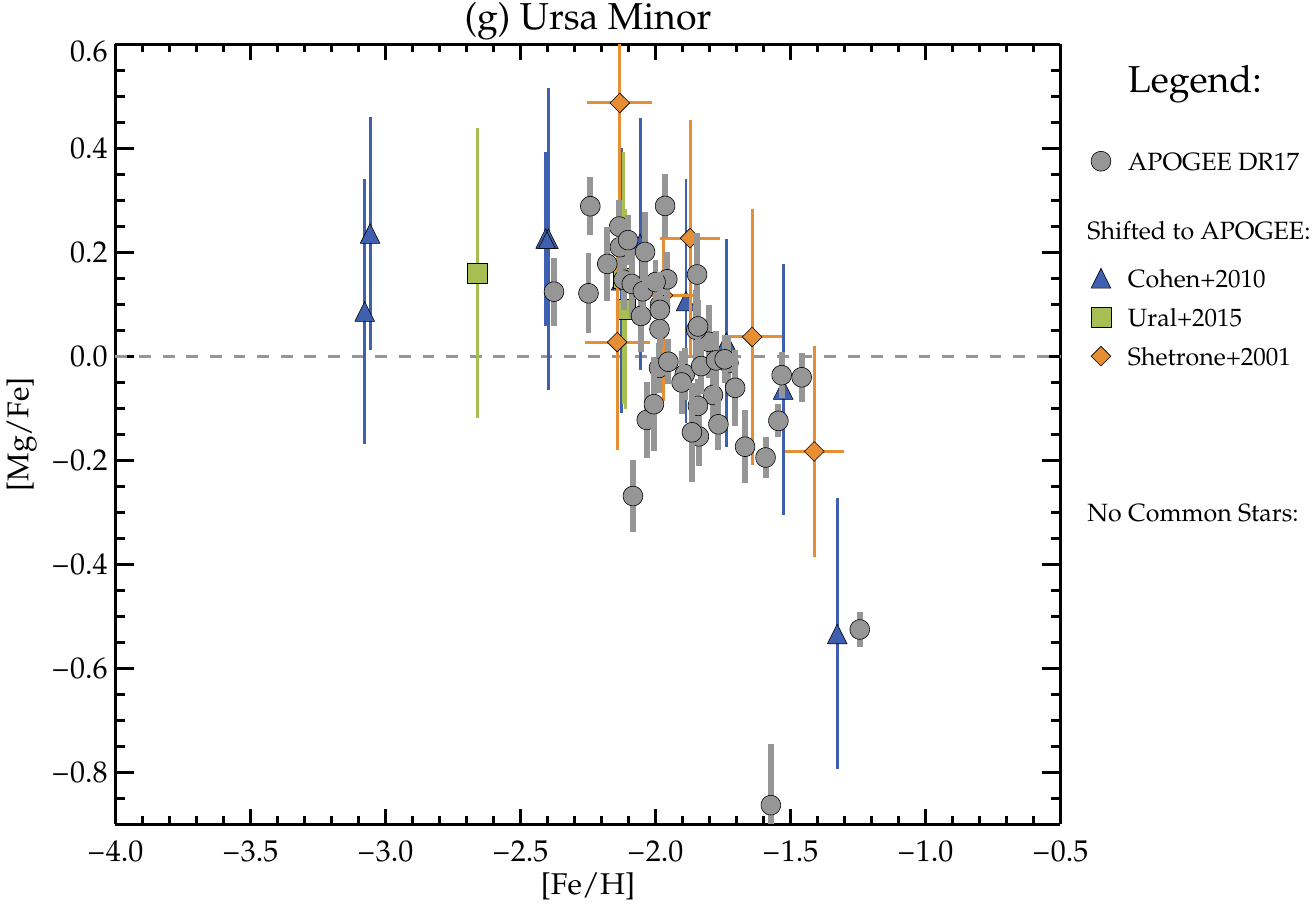}    
    \caption{
        APOGEE sample (grey circles) alongside the literature high-resolution studies; legends in each panel indicate the studies being shown and references are provided in \autoref{sec:litcomp}d. 
        Open symbols indicate studies presented in the published values where as filled symbols indicate studies that have zeropoint shifts applied (see \autoref{tab:offsets}).
    }
    \label{fig:literature}
\end{figure*}

We follow a similar approach to each literature sample for each galaxy. The offsets for each study can be found in \autoref{tab:offsets}, and the results are plotted in \autoref{fig:literature}. The purpose of the following literature comparison is to understand the extent to which any conclusions drawn from APOGEE data can also be drawn from the existing optical data and to determine whether the high-resolution optical data, when added to APOGEE, results in a more complete idea of the star formation history of these galaxies. The data in the literature we used for each galaxy are briefly described below. We tried to be as comprehensive as possible, but for several literature sources, because there were no stars with overlap with APOGEE, we were unable to determine an abundance offset.

\paragraph{Bootes:} 
    There is no or minimal overlap with APOGEE, so there are no offsets for \citet{Feltzing_2009}, \citet{Gilmore_2013}, \citet{Ishigaki_2014} and \citet{Frebel_2016}.   
    Applied offsets from \citet{Norris_2017} to \citet{Norris_2010c}. 

\paragraph{Carina:} 
    All of the following studies overlap with APOGEE and offsets were derived: \citet{Lemasle_2012}, \citet{Venn_2012}, \citet{Norris_2017}, \citet{Koch_2008}, and \citet{Shetrone_2003}, 

\paragraph{Draco:} 
    Both \citet{Cohen_2009} and \citet{Shetrone_2001} have enough overlap with APOGEE to determine offsets.

\paragraph{Fornax:} 
    Both \citet{Letarte_2010} and \citet{Lemasle_2014} have a significant overlap with APOGEE, and the offsets are determined. 
    However, APOGEE did not observe the metal-poor star of \citet{Tafelmeyer_2010} and it is shown as reported in that study. There was no overlap with the \citet{Hendricks2014} study. 

\paragraph{Sculptor:} 
    The two studies of \citet{Skul_2015,Skul_2017} and \citet{Hill_2019} both have a significant overlap with the APOGEE sample and are offset to match the APOGEE distribution.
    We also use offsets for \citet{Shetrone_2003}.
    However, the studies of \citet{Jablonka_2015}, \citet{Tafelmeyer_2010}, \citet{Starkenburg_2013}, and \citet{Simon_2015} do not have an overlap sample and are shown as reported.   

\paragraph{Sextans:} 
    We offset \citet{Shetrone_2001} to the APOGEE scale. 
    However, the studies of \citet{Tafelmeyer_2010}, \citet{Aoki_2009}, \citet{theler_2020}, and \citet{Lucchesi_2020} have no overlap observations with APOGEE and we show them as is.

\paragraph{Ursa Minor:} 
    We determined offsets for the studies of \citet{Cohen_2010}, \citet{Ural_2015}, and \citet{Shetrone_2001} to place them on the APOGEE scale.\\

From the literature comparisons in Mg in Figure \ref{fig:literature}, once zero points have been applied, we generally see a good agreement between the abundance patterns observed in the literature and those seen here in APOGEE, although depending on the system, the APOGEE measurements are sometimes higher precision, such as in Carina, Sextans, Draco, or Ursa Minor.  In other systems like Sculptor we see that the similar precision literature measurements are (with a zero point offset) in excellent agreement with APOGEE measurements and find a very similar chemical abundance pattern.  For systems like Sculptor, Draco, and Ursa Minor, the combined literature and the APOGEE data sets continue to show relatively simple chemical abundance patterns, where an $\alpha$-knee might even be visible.  It is interesting that the large scatter and unusual chemistry seen in individual literature samples for Carina and Sextans do continue to have an odd chemical abundance distribution with a complex pattern when combining the literature results with zero points and the higher-precision APOGEE results.  In addition, the large scatter seen in Sextans continues to be seen even in the higher-precision APOGEE data, suggesting that this is real, and not entirely due to observational uncertainties.  

The one case where we do not see a great agreement between APOGEE and the literature is in Bootes I, however, this is in part due to no overlap between APOGEE and the dominant literature samples, so there may still be zero point offsets between the two \citep[as there are known to be in Mg, e.g.,][Holtzman et al. in prep.]{Bergemann_2017}.  This, in addition to the small sample size, makes Bootes I difficult to interpret. 

The similarities between APOGEE and the literature, at least in this key APOGEE $\alpha$-element, give confidence in comparisons between the galaxies across the metallicity range.  We discuss some of the basic interpretations of these abundance patterns below.  One of the key differences among the dSph abundance patterns is that some show large scatter or complex behavior that is not typically associated with a simple chemical evolution over a short span of metallicity.  Examples of systems with increases, jumps, or large scatter at a given metallicity in \xfe{$\alpha$} are Carina, Fornax and Sextans, whereas Sculptor, Draco, Ursa Minor, and possibly Bootes I, show monotonically decreasing \xfe{$\alpha$} with increasing metallicity, with hints of a plateau at low metallicities.  The former we interpret as the chemical evolution caused by bursty or periodic star formation that quickly changes the ratio of CC SNe to SN Ia which can change the $\alpha$-element and Fe-peak element ratios.  The latter have a chemical abundance pattern more indicative of a more constant or quiescent star formation history, potentially only experiencing SF at early times in the universe.

\begin{table*}
    \caption{Offsets to Place Literature Studies on the APOGEE Scale}
    \begin{tabular}{c c c cc cc cc}
     \hline \hline
    Study  & Galaxies & $N$ & $\Delta_{[Fe/H]}$ & $\sigma_{[Fe/H]}$ & $\Delta_{[Mg/H]}$ & $\sigma_{[Mg/H]}$ & $\Delta_{[Mg/Fe]}$ & $\sigma_{[Mg/Fe]}$\\
     \hline 
    \citet{Norris_2010c}  & Boo & * & -0.013 & 0.076 & 0.353 & 0.163 & 0.356 & 0.165 \\ 
     \hline
    \citet{Shetrone_2003} & Car &  5 & -0.074 & 0.071 & 0.203 & 0.050 & 0.277 & 0.104 \\
    \citet{Koch_2008}     & Car &  6 & -0.043 & 0.130 & 0.006 & 0.230 & 0.043 & 0.129 \\
    \citet{Lemasle_2012}  & Car & 17 & -0.086 & 0.099 & 0.228 & 0.108 & 0.314 & 0.109 \\ 
    \citet{Venn_2012}     & Car &  4 & -0.083 & 0.089 & 0.285 & 0.053 & 0.353 & 0.102 \\
    \citet{Norris_2017}   & Car & 13 & -0.013 & 0.076 & 0.353 & 0.163 & 0.356 & 0.165 \\
     \hline 
    \citet{Shetrone_2001} & Dra &  4 & -0.038 & 0.058 & 0.145 & 0.115 & 0.183 & 0.150 \\
    \citet{Cohen_2009}    & Dra & 10 & -0.073 & 0.076 & 0.111 & 0.136 & 0.184 & 0.086 \\
     \hline
    \citet{Letarte_2010}  & For & 28 & 0.046  & 0.068 & 0.075 & 0.135 & 0.029 & 0.124 \\
    \citet{Lemasle_2014}  & For & 17 & -0.086 & 0.099 & 0.228 & 0.108 & 0.314 & 0.109 \\
     \hline 
    \citet{Shetrone_2003} & Scl &  5 & -0.074 & 0.071 & 0.203 & 0.050 & 0.277 & 0.104 \\
    \citet{Skul_2015}     & Scl & 42 & -0.050 & 0.084 & 0.210 & 0.156 & 0.271 & 0.143 \\ 
      \citet{Hill_2019}     & Scl & 50 & -0.040 & 0.090 & 0.260 & 0.170 & 0.220 &       \\
     \hline
    \citet{Shetrone_2001} & Sxt &  4 & -0.038 & 0.058 & 0.145 & 0.115 & 0.183 & 0.150 \\ 
     \hline
    \citet{Shetrone_2001} & UMi &  4 & -0.038 & 0.058 & 0.145 & 0.115 & 0.183 & 0.150 \\
    \citet{Cohen_2010}    & UMi & 10 & -0.073 & 0.076 & 0.111 & 0.136 & 0.184 & 0.086 \\
    \citet{Ural_2015}     & UMi &  2 & -0.010 &       & 0.160 &       &       &       \\	
     \hline \hline 
    \end{tabular}
    \label{tab:offsets}
\end{table*}

\section{Discussion} \label{sec:discussion}

In principle, the detailed chemical abundances presented here can be used to infer details of the star formation history of each dSph. In practice, this is difficult, primarily due to uncertain yield calculations for some elements that result in incorrect chemical evolution models combined with limitations of the spectroscopic abundance themselves, including NLTE. However, in this discussion, we use knowledge of where the elements are generally formed to comment on the SFHs of these galaxies, relating them to general properties of each galaxy such as mass and environment. We begin with the $\alpha$ elements, as they are most easily interpreted, and then we discuss the other elements.

\subsection{Star Formation Histories from the $\alpha$-elements} \label{sec:alpha}

The $\alpha$ elements can be used to infer early star formation efficiency in each galaxy, as they are created primarily in the hydrostatic burning phases of massive stars and released to the ISM via Type II SNe. Because these stars are short-lived, the early enrichment in any galaxy should be dominated by enrichment from these sources. This is why nearly all dSphs have elevated [$\alpha$/Fe] abundances at the lowest metallicities. At some point in [Fe/H], depending on how vigorous this initial burst of star formation was, Type Ia SNe begin to explode, enriching the ISM in heavier elements largely to the exclusion of $\alpha$ elements, decreasing the [$\alpha$/Fe] abundance ratio while still increasing overall metallicity.

For some galaxies, the end of chemical evolution is reached shortly after Type Ia SNe begin to contribute in large quantities, as the gas is used up without replenishment. However, \citet{Hendricks2014} and \citet{Hasselquist2021_sats} have shown that later starbursts can result in flat or even increasing [$\alpha$/Fe]-[Fe/H] abundance patterns. We first discuss the low-metallicity [$\alpha$/Fe]-[Fe/H] values and trends and then explore which galaxies show extended star formation histories and why this might be. 

\subsubsection{The $\alpha$-element knee} \label{sec:knees}

One commonly sought feature of the chemical abundance patterns of dwarf galaxies is the ``knee'' in the \xfe{$\alpha$} abundance patterns as a function of metallicity, which represents the point at which Type Ia SNe begin to contribute to the chemical evolution of the galaxy.  The metallicity at which the knee occurs is a product of the early star formation efficiency of the galaxy, with the slope of the knee being related to outflows and the overall [$\alpha$/Fe] location of the knee related to the IMF. Dwarf galaxies enrich slowly, so the $\alpha$-knee can occur at very low metallicities, where stochastic enrichment, mixing, and larger abundance measurement uncertainties and upper limits can lead to a large scatter in the observed abundance ratios of stars in these systems and make it difficult to identify the $\alpha$-knee. 

In the literature, no universal method has been adopted to measure these knees. Here we simply identify the knee as the metallicity at which the abundance trends for each galaxy cross the plateau seen in our high-$\alpha$ disk sample, which except for the very most metal-poor stars is higher than the halo$+$GSE samble.  This self-consistent methodology avoids questions about modeling, such as species-to-species metallicity-dependent NLTE affects, and overionization.  It assumes that the plateau formed by the high-$\alpha$ disk sample is a largely pure CCSN product with a universal IMF.   We report these knee metallicities in Table \ref{tab:knees} for the three $\alpha$-elements that are well measured to low-metallicities in our sample, O, Mg and Si. For any galaxy which lies below the MW high-$\alpha$ disk sample plateau across the entire metallicity range of that galaxy as sampled here, we report the lowest metallicity point in our sample as the upper limit on the knee position.  For galaxies with measurements of knee position, we average the positions found in each element to produce our adopted $\alpha$-knee measurement. For Fornax, Carina, Draco, and Sextans, we only derive upper limits for the location of the knee (between $\sim -2.3$ and $-2.1$) with our assumed methodology.  However, for Sculptor, Ursa Minor and Bootes I we do measure an approximate knee location at \feh{} $\sim -2.2$, $-2.3$, and $-2.4$, respectively.

These measurements have a few caveats.  The first caveat is that we have assumed that the dwarf galaxies and the MW samples had a similar ``plateau'' at some metallicity.  However, variation in the IMF can change the location of the plateau, and the element/yield/chemical evolution model choice can even make the plateau weakly sloped.  A shift in the position of the plateau could change the way we would evaluate some of these chemical abundance patterns.  For instance, we do see some flattening in the \xfe{$\alpha$} distribution for Draco (though this corresponds to a large spread in abundances at these metallicities) that could perhaps be explained as evidence for a higher metallicity knee combined with top-light IMF. 

The second caveat is that, while we do report a measurement of the knee location for Sculptor, Ursa Minor and Bootes I, some of these (marked with a ``:'' in Table \ref{tab:knees}) lie at metallicities where there are a large fraction of upper limit or non-detections in the abundance measurements.  While we have chosen a method for defining the knee position that isn't dependent on finding a flattening in the \xfe{$\alpha$} slope, which can be hampered by upper limits, the trends that we evaluate may still be following the stars with the highest \xfe{$\alpha$} ratios at these metallicities making the trends seem to rise more quickly by masking the stars that scatter to low abundance ratios.  We recommend that the values we report are taken with caution; nevertheless, we can compare these knee positions with those reported in recent literature studies of these systems.

\begin{table*}
    \centering
    \caption{Chemical Evolution Knee Locations}
    \begin{tabular}{c ccccccc}
    \hline \hline
    Species & Fornax & Sculptor & Carina & Ursa Minor & Draco & Sextans & Bootes \\
    \hline  
     \xfe{O} & $<$-2.2 & -2.11: & $<$-2.12 & -2.21: & $<$-2.30 & $<$-2.32: & $<$-2.32  \\
    \xfe{Mg} & $<$-2.2 & -2.07 & $<$-2.12 & -2.24: & $<$-2.30 & $<$-2.32 & -2.32:  \\
    \xfe{Si} & $<$-2.2 & -2.41: & $<$-2.12 & -2.33: & $<$-2.30 & $<$-2.32 & $<$-2.32  \\
     \hline 
Adopted & $<$-2.2 & -2.17 & $<$-2.12 & -2.26 & $<$-2.30 &  $<$-2.32 & -2.4:  \\
    \hline \hline 
    \end{tabular}
    \label{tab:knees}
\end{table*}

The largest sample of homogeneously analyzed dSph \xfe{$\alpha$} abundance patterns was done by \citet{Kirby2011_iv} using medium resolution Keck DEIMOS spectra.  \citet{Kirby2011_iv} remark on searching for knees in the \xfe{$\alpha$} abundance patterns of eight galaxies they observe, including Fornax, Sculptor, Sextans, Draco, and Ursa Minor, which are also analyzed here.  They concluded that a knee cannot be clearly detected in their data for most elements in these dwarf galaxies, though they do see a knee in the \xfe{Ca} of Sculptor at a metallicity of \feh{} $= -1.8$ and the possible detection of a knee in Ursa Minor at a slightly lower metallicity \feh{} $\sim -2$.  These are slightly more metal-rich than what we find here, and if the zero point difference in [Fe/H] between the stars in common between the Kirby analysis and the APOGEE DR17 data set is considered (-0.14 dex) then the knee would be pushed even more metal-rich in the APOGEE metallicity system.   This is a fairly large discrepancy, but as pointed out in \citet{Kirby2011_iv} this is only seen in \xfe{Ca} and not in \xfe{Mg}, \xfe{Si}, or \xfe{Ti}, all of which show no hint of a knee above \feh{} $\sim -2.5$ (or \feh{} $\sim -2.36$ in the APOGEE metallicity system), somewhat more consistent with our results.   Like \citet{Kirby2011_iv} we do not report a clear knee for Fornax, Sextans, and Draco.  

We point out that the chemical evolution models shown in Figures 2-12
of \citet{Kirby2011_iv} show knees below \feh{} $\sim -2.5$ except Fornax, which has a model with a knee at \feh{} $\sim -1.7$.   A more detailed "knee" comparison with the Kirby data set or other literature samples should be done with caution because differences in NLTE corrections from species to species, from spectral line to spectral line and even metallicity dependent NLTE corrections of a single spectral feature make comparisons laborious.   Other effects such as differences in fitting techniques, confusion between upper limits and detections for the most metal-poor and low S/N spectra, and abundance scale zero points further complicate comparisons.   Such a detailed analysis is beyond the scope of this work.   



\begin{figure*}
    \centering
    \includegraphics[width=1.0\textwidth]{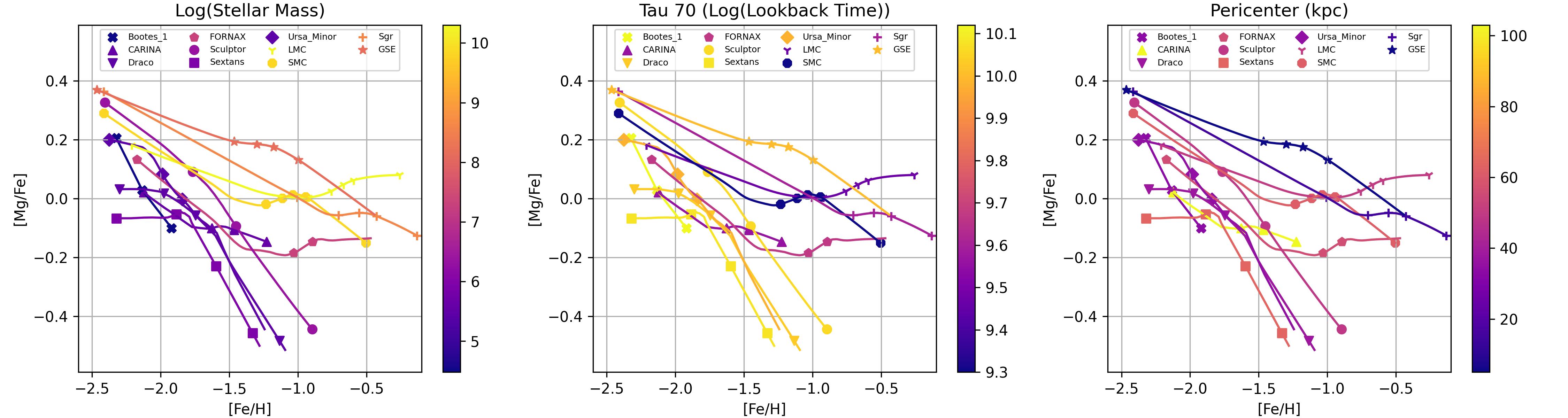}   
    \caption{
        LOWESS curves in the [Mg/Fe]-[Fe/H] abundance plane for all galaxies studied in this work along with the APOGEE LMC, SMC, Sgr, and GSE samples studied in \citet{Hasselquist2021_sats}. Lines and points are colored by stellar mass in the left panel, $\tau_{70}$ in the middle panel (lookback time at which the galaxy formed 70 \% of its stars), and pericenter of the orbit in the right panel.
    }
    \label{fig:summary}
\end{figure*}

The low metallicities of all knees suggest that the early star-formation histories of these galaxies were very inefficient compared to those of the MW. However, Scupltor enriched to slightly higher metallicities before Type Ia SNe begin contributing than the other galaxies. If we look at the mass of these galaxies, Sculptor is the second most massive galaxy, behind Fornax. This suggests a correlation with stellar mass and early star formation, either because one necessarily leads to the other, or because the more massive dark-matter halo leads to more vigorous early star formation. However, as highlighted in the left panel of \autoref{fig:summary}, Sculptor, perhaps unexpectedly, has a more metal-rich knee than Fornax, suggesting that Fornax must have had some sort of extended SFH, which we discuss in detail below. 

\subsubsection{Continuous versus Episodic Star Formation in dSph Galaxies} \label{sec:starformation}

Although all of the dSph have metal-poor knees associated with low early star formation efficiency, the chemical evolution of these galaxies diverge at higher metallicities. Scl, UM, and Draco all have [$\alpha$/Fe] abundances that continue to decline with [Fe/H] across all metallicities, suggesting that these galaxies only experienced one major epoch of star formation some time in the early Universe. This is supported in various literature photometric studies of these galaxies (e.g., \citealt{Bettinelli2019} for Scl and, e.g., \citealt{Carrera_2002,Dolphin2002,Weisz2014,Ural_2015} for Ursa Minor and Draco), as well as chemical evolution models fit to spectroscopic data (e.g., \citealt{Hill_2019,delosReyes2022}). However, Carina and Fornax exhibit a flattening and even slight reversal of this trend at [Fe/H] $>$ -1.7 for Carina and [Fe/H] $>$ -1.1 for Fornax. This suggests that these two galaxies experienced more recent bursts of star formation, with Type II SNe contributing anew to the chemical evolution of these galaxies, raising the relative amount of Type II SNe. A second epoch of star formation that turns on relatively slowly can produce a more gradual change in chemical abundance ratios, whereas a burst can produce a jump in metallicity and abundance ratios. Al may be dependent on the available N abundance, which could be produced through slower processes such as production in AGB stars or directly by Type II SNe.    

Support for later periods of star formation in Carina and Fornax can be found in the literature, both through photometric studies (e.g., \citealt{Dolphin2002,Weisz2014,Weisz_2015}) and spectroscopic studies invoking chemical evolution models (e.g., \citealt{Hendricks2014,Hasselquist2021_sats}). The large spread in [Fe/H] combined with the relatively narrow red giant branch for Carina suggests that Carina experienced distinct episodic bursts compared to the more coherent chemical abundance pattern of Fornax, which \citet{Hasselquist2021_sats} explained with two episodes of star formation separated by a lull. 

The [$\alpha$/Fe] - [Fe/H] abundance patterns that are indicative of an extended star formation history have been observed in more massive local group dwarf galaxies, including Sagittarius, the LMC, and the SMC (e.g., \citealt{Bekki2012,Nidever_2020,Hasselquist2021_sats}).  The left panel of \autoref{fig:summary} shows the same [Mg/Fe]-[Fe/H] LOWESS lines for all galaxies studied in this work, along with the massive MW satellites from \citet{Hasselquist2021_sats}, colored by the mass of the galaxy.   
The most massive dwarf galaxies do not show a single downward slope in their later chemical evolution.  In addition, these more massive galaxies are offset to the upper right portion of the figure, which we usually associate with a more rapid chemical evolution.    The mass is clearly a very strong driver of chemical evolution; however, Carina, which shows slow but complex chemical evolution, has a lower mass than that of Sculptor and Sextans.    There must be more drivers than just the galaxy mass.

The middle panel of \autoref{fig:summary} displays the lookback time at which each galaxy formed 70\% of its stars ($\tau_{70}$), as inferred from the photometric SFHs of these galaxies $\tau_{70}$. As alluded to earlier, the photometric SFHs agree with our interpretations of the [Mg/Fe]-[Fe/H] abundance patterns. Specifically, Bootes I, Sculptor, Draco, and Ursa Minor all have $\tau_{70}$ $>$ 10 Gyr. Carina and Fornax have much more recent $\tau_{70}$ values.  With this color coding, the galaxies with complex chemical evolution clearly stand out; thus, one can predict the lookback time with the shape of the [Mg/Fe]-[Fe/H] abundance trend or predict the shape of the abundance trend if the lookback time is known.    However, this does not give much insight into why Carina, with its low mass, has late chemical evolution with a bursty abundance pattern.

In the right panel of \autoref{fig:summary} we color the same trend lines by orbital pericenters of the galaxies. We adopt the pericenters from \citet{Battaglia2022} (Table B.5) for the less massive galaxies. For the LMC and SMC, we adopt their present day distance from the MW as their pericenters. For Sgr, we use 15 kpc and for GSE we use 5kpc, although this number is much less meaningful for GSE as it has been merged with the MW. With the exception of Sextans (and the galaxies that have already merged with the MW, e.g., GSE and Sgr), the galaxies with simple, truncated SFHs are much closer to the MW (i.e., have much smaller orbital pericenters) than those with extended SFH. Carina happens to be the galaxy with the farthest pericenter. So, while mass likely plays a role in determining the early SF efficiency of a galaxy, the galactic environment, specifically proximity to the MW, in this case, plays a role in determining the extent to which a galaxy exhibits future episodes of SF. These episodes likely come in the form of galaxy-galaxy interactions, but interactions that have much more similar mass ratios, i.e. not with MW-sized galaxies.   GSE and Sgr stand out as peculiar in this trend of pericenter and late stage burstyness.  In this case, the pericenter may be misleading; for example, Ruiz-Lara et al. 2020 report that Sgr may have had its first infall with the MW 5 Gyr ago, giving Sgr a long period to evolve and undergo small mergers creating bursts before a single first orbital encounter.   The chemical evolution of a galaxy is driven both by its inherent properties, such as mass, but also by its environmental properties, such as proximity to massive galaxies (stripping and quenching) and smaller galaxies (bursts and infall).

Sextans appears to be somewhat of an exception to this rule. It has a relatively low mass, but more massive than its apparent knee might suggest. Moreover, it has much less recent $\tau_{70}$ values than we might infer from its larger pericenter distance. However, unlike the dSphs that show simple SFH, Sextans apparently never reaches solar [Mg/Fe] at the lowest metallicities studied by APOGEE.  If we consider the literature and the APOGEE observations of Sextans (panel f) of \autoref{fig:literature}) together, we infer that Sextans actually may have a very complicated [Mg/Fe]-[Fe/H] abundance pattern, one that is similar to that of Carina but shifted to much lower metallicity.  
Sextans has less literature indicating multiple episodes of star formation, with studies finding that Sextans formed most of its stars early in its SFH, although a small number of stars may have been formed up to 7 Gyr ago \citep{Lee_2009,Bettinelli_2018}.  However, other studies have found evidence for kinematically distinct components of Sextans that could indicate that it has had previous mergers that could have driven a burst of star formation or prolonged star formation in Sextans \citep{Kleyna2004,Battaglia_2011,Okamoto2017,Cicuendez2018}. Future spectroscopic surveys that target Sextans stars across the full metallicity range will shed light on whether Sextans is an apparent exception to the mass-environment rule discussed above.

\subsection{Metallicity-dependent Yields} \label{sec:other}

\begin{figure*}
    \centering
    \includegraphics[width=0.8\textwidth]{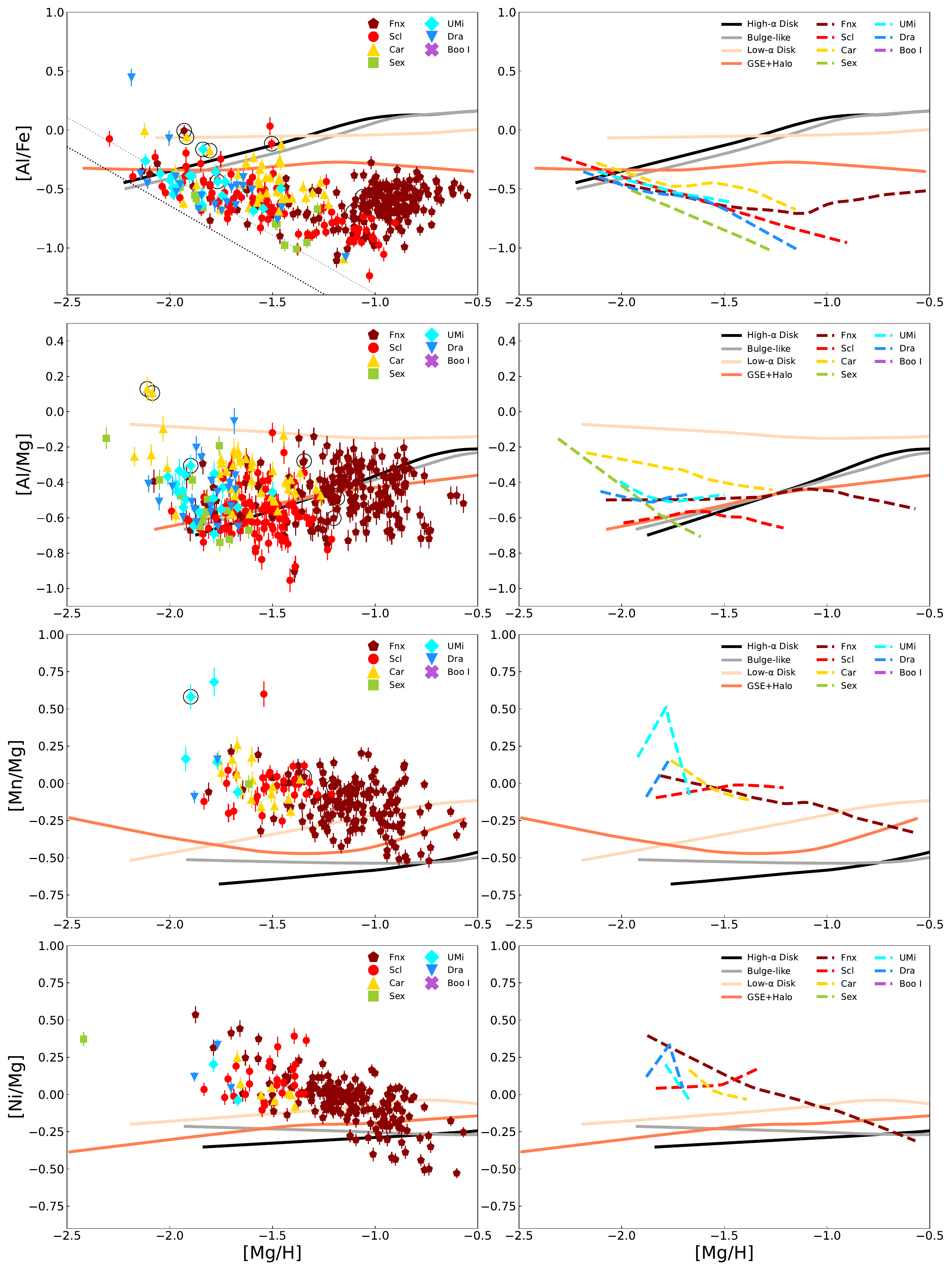}   
    \caption{
        Similar to Figure \ref{fig:agbIa_grid} but using [Mg/H] as the metallicity source. 
    }
    \label{fig:xmg}
\end{figure*}

Most of the elements for which we have been able to derive abundances have origins in CCSN and/or Type Ia SN (and to a lesser extent AGB stars).   If we use Mg as a proxy for pure CCSN then the elements with declining [X/Mg] abundance ratios with increasing metallicity either have contributions from Type Ia SN or have metallicity-dependent CCSN yields, see Figure \ref{fig:xmg}. Additionally, elements with trends of [X/Mg] that are different from those of the CCSN dominated MW samples (bulge/halo/high-alpha disk)  have significant Type Ia SN contributions or have metallicity-dependent CCSN yields.   

However, for Carina and Fornax, which show evidence of a burst in their star formation, we have some leverage to understand the relative contributions.   The elements that show discontinuity at the burst Fe metallicity in Carina are O, Mg, Si, and possibly Al, Mn, and C. The first three are well understood to be alpha elements.   For the last three the number of true detections below the burst is very small to be definitive but the median value at the burst is higher than that of the other dSph, suggesting that there may be an extra contribution above the scaled solar amount produced in CCSN.  It is the pre-burst stars in Carina that exhibit anomalous [Al/Mg] ratios; the more metal-rich burst stars exhibit [Al/Mg] ratios more consistent with the other dwarf galaxies and the CCSN dominated MW samples.   
 We also note that Sextans exhibit a rising Al abundance trend as a function of declining metallicity; similar to Carina and to a lesser extent the low-alpha disk.

The elements that show discontinuity at the burst Fe metallicity in Fornax are C, N, O, Mg, Si, Ca, Al, and Mn.   In addition to O and Mg, the elements which show the strongest rise include C, and Al; which may have contributions from many possible sources, e.g. whatever mechanism enriches second generation globular cluster stars.  The relatively smaller increases in Ca and Mn in the burst populations hint at metallicity-dependent yields.

\section{Summary}

\begin{itemize}
    \item We performed a detailed membership analysis of the 1315 targets in the vicinity of seven dSphs using Gaia EDR3 proper motions and APOGEE DR17 radial velocities. We consider 731 targets as members of their respective dSph. Of these, 528 or 72\% have reliable ASPCAP stellar parameters.
    \item We developed a method to distinguish between true measurements and upper limits in ASPCAP results using synthetic spectra of metal-poor giants (\feh{} $= -1$, \logg $= 1.0$) over the following ranges: T$_{\rm eff} = 4200 - 4800$ K and \xh{X} $= -2.5 - -0.5$.  This was completed for 17 elements. The number of stars in each dSph with reliable chemical abundance detections is given in \autoref{tab:detectionsample}. We provide these data to the community.  High-quality abundances with few upper limits are determined for oxygen, magnesium, aluminum, silicon, and iron.   For elements carbon, nitrogen, calcium, manganese, and nickel, we determined that we are dominated by upper limits for the more metal-poor and lower S/N spectra; thus, the trends at the metal-poor end for these elements are not constrained.   We note that for Sodium, and Vanadium we find no detections in our sample and for potassium, sulfur (marginally), chromium, cobalt and cerium there are only detections for the most metal-rich and highest S/N observations.   We warn the community to be extremely careful in using DR17 abundances for these elements with low S/N ratio spectra of low metallicity or warm stars.  
    \item \autoref{fig:sample} has the color magnitude diagram for the final dSph sample. The APOGEE DR17 sampling spans $\sim$1.0 to $\sim$1.5 mag on the upper giant branch for all but Fornax.  In addition to the RGB, we have stars selected from the AGB and above the RGB.  
    \item We produced visualizations of the abundance distributions for 16 independent abundance planes (including \xfe{(C+N)} and [C/N]) and compared them to comparison samples in four Milky Way structures. Where more than 50\% of the sample has measured abundances (versus upper limits), we produce median trend lines. 
    \item We combined APOGEE observations with high-resolution studies in the literature, many of which extend to significantly lower [Fe/H]. 
    We demonstrate how chemical abundance patterns provide insight into the star formation history and constrain the level of SFR in course values. We separated into continuous and episodic SFH based on the chemical abundance patterns.
\end{itemize}
 
  
\begin{acknowledgements} 
This material is based on work supported by the National Science Foundation under Grant Number 2206514. SH acknowledges support from the STScI DDRF.
 
Funding for the Sloan Digital Sky Survey IV has been provided by the Alfred P. Sloan Foundation, the U.S. Department of Energy Office of Science, and the Participating Institutions. SDSS-IV acknowledges support and resources from the Center for High-Performance Computing at the University of Utah. The SDSS web site is \url{www.sdss.org}.

SDSS-IV is managed by the Astrophysical Research Consortium for the Participating Institutions of the SDSS Collaboration including the Brazilian Participation Group, the Carnegie Institution for Science, Carnegie Mellon University, the Chilean Participation Group, the French Participation Group, Harvard-Smithsonian Center for Astrophysics, Instituto de Astrof\'isica de Canarias, The Johns Hopkins University, Kavli Institute for the Physics and Mathematics of the Universe (IPMU) / University of Tokyo, Lawrence Berkeley National Laboratory, Leibniz Institut f\"ur Astrophysik Potsdam (AIP), Max-Planck-Institut f\"ur Astronomie (MPIA Heidelberg), Max-Planck-Institut f\"ur Astrophysik (MPA Garching), Max-Planck-Institut f\"ur Extraterrestrische Physik (MPE), National Astronomical Observatory of China, New Mexico State University, New York University, University of Notre Dame, Observat\'ario Nacional / MCTI, The Ohio State University, Pennsylvania State University, Shanghai Astronomical Observatory, United Kingdom Participation Group, Universidad Nacional Aut\'onoma de M\'exico, University of Arizona, University of Colorado Boulder, University of Oxford, University of Portsmouth, University of Utah, University of Virginia, University of Washington, University of Wisconsin, Vanderbilt University, and Yale University.

This publication makes use of data products from the Two Micron All Sky Survey, which is a joint project of the University of Massachusetts and the Infrared Processing and Analysis Center/California Institute of Technology, funded by the National Aeronautics and Space Administration and the National Science Foundation.

This work is based, in part, on observations made with the \emph{Spitzer} Space Telescope, which is operated by the Jet Propulsion Laboratory, California Institute of Technology under a contract with NASA.

This publication makes use of data products from the Wide-field Infrared Survey Explorer \citep{WISE}, which is a joint project of the University of California, Los Angeles, and the Jet Propulsion Laboratory/California Institute of Technology, funded by the National Aeronautics and Space Administration, and NEOWISE, which is a project of the Jet Propulsion Laboratory/California Institute of Technology. 
WISE and NEOWISE are funded by the National Aeronautics and Space Administration 

This work has made use of data from the European Space Agency (ESA) mission \emph{Gaia} (\url{https://www.cosmos.esa.int/gaia}), processed by the \emph{Gaia} Data Processing and Analysis Consortium (DPAC, \url{https://www.cosmos.esa.int/web/gaia/dpac/consortium}). Funding for the DPAC has been provided by national institutions, in particular the institutions participating in the \emph{Gaia} Multilateral Agreement.

This research has made use of NASA’s Astrophysics Data System.

This research has made use of the SIMBAD database, operated at CDS, Strasbourg, France. The original description of the SIMBAD service was published in \citet{simbad_2000}.

This research has made use of the VizieR catalogue access tool, CDS, Strasbourg, France (DOI: 10.26093/cds/vizier). The original description of the VizieR service was published \citet{vizier2000}.

\end{acknowledgements}
    \facilities{ Du Pont (APOGEE), 
                Sloan (APOGEE), 
                \emph{Spitzer}, 
                WISE, 
                2MASS, 
                \emph{Gaia} } 
    \software{Astropy \citep{Astropy_2013,Astropy_2018}, 
        Doppler \citep{2021zndo...4906681N},
        NumPy \citep{vanderWalt_2011,Harris_2020numpy},
        TopCat \citep{topcat},
        Turbospectrum \citep{turbospectrum1,turbospectrum2}} 
\bibliography{BIB_lit_comps,BIB_EverythingElse,BIB_APOGEE}

@MISC{turbospectrum1,
       author = {{Plez}, B.},
        title = "{Turbospectrum: Code for spectral synthesis}",
 howpublished = {Astrophysics Source Code Library, record ascl:1205.004},
         year = 2012,
        month = may,
          eid = {ascl:1205.004},
        pages = {ascl:1205.004},
archivePrefix = {ascl},
       eprint = {1205.004},
       adsurl = {https://ui.adsabs.harvard.edu/abs/2012ascl.soft05004P}
}

@ARTICLE{marcs,
       author = {{Gustafsson}, B. and {Edvardsson}, B. and {Eriksson}, K. and {J{\o}rgensen}, U.~G. and {Nordlund}, {\r{A}}. and {Plez}, B.},
        title = "{A grid of MARCS model atmospheres for late-type stars. I. Methods and general properties}",
      journal = {\aap},
         year = 2008,
        month = aug,
       volume = {486},
       number = {3},
        pages = {951-970},
          doi = {10.1051/0004-6361:200809724},
archivePrefix = {arXiv},
       eprint = {0805.0554},
 primaryClass = {astro-ph},
       adsurl = {https://ui.adsabs.harvard.edu/abs/2008A&A...486..951G}
}

@ARTICLE{turbospectrum2,
       author = {{Alvarez}, R. and {Plez}, B.},
        title = "{Near-infrared narrow-band photometry of M-giant and Mira stars: models meet observations}",
      journal = {\aap},
         year = 1998,
        month = feb,
       volume = {330},
        pages = {1109-1119},
          doi = {10.48550/arXiv.astro-ph/9710157},
archivePrefix = {arXiv},
       eprint = {astro-ph/9710157},
 primaryClass = {astro-ph},
       adsurl = {https://ui.adsabs.harvard.edu/abs/1998A&A...330.1109A}
}

@ARTICLE{Hayes_2020,
       author = {{Hayes}, Christian R. and {Majewski}, Steven R. and {Hasselquist}, Sten and {Anguiano}, Borja and {Shetrone}, Matthew and {Law}, David R. and {Schiavon}, Ricardo P. and {Cunha}, Katia and {Smith}, Verne V. and {Beaton}, Rachael L. and {Price-Whelan}, Adrian M. and {Allende Prieto}, Carlos and {Battaglia}, Giuseppina and {Bizyaev}, Dmitry and {Brownstein}, Joel R. and {Cohen}, Roger E. and {Frinchaboy}, Peter M. and {Garc{\'\i}a-Hern{\'a}ndez}, D.~A. and {Lacerna}, Ivan and {Lane}, Richard R. and {M{\'e}sz{\'a}ros}, Szabolcs and {Bidin}, Christian Moni and {M{\~{u}}noz}, Ricardo R. and {Nidever}, David L. and {Oravetz}, Audrey and {Oravetz}, Daniel and {Pan}, Kaike and {Roman-Lopes}, Alexandre and {Sobeck}, Jennifer and {Stringfellow}, Guy},
        title = "{Metallicity and {\ensuremath{\alpha}}-Element Abundance Gradients along the Sagittarius Stream as Seen by APOGEE}",
      journal = {\apj},
         year = 2020,
        month = jan,
       volume = {889},
       number = {1},
          eid = {63},
        pages = {63},
          doi = {10.3847/1538-4357/ab62ad},
archivePrefix = {arXiv},
       eprint = {1912.06707},
 primaryClass = {astro-ph.GA},
       adsurl = {https://ui.adsabs.harvard.edu/abs/2020ApJ...889...63H}
}

@ARTICLE{osorio_2020,
       author = {{Osorio}, Y. and {Allende Prieto}, C. and {Hubeny}, I. and {M{\'e}sz{\'a}ros}, Sz. and {Shetrone}, M.},
        title = "{NLTE for APOGEE: simultaneous multi-element NLTE radiative transfer}",
      journal = {\aap},
         year = 2020,
        month = may,
       volume = {637},
          eid = {A80},
        pages = {A80},
          doi = {10.1051/0004-6361/201937054},
archivePrefix = {arXiv},
       eprint = {2003.13353},
 primaryClass = {astro-ph.SR},
       adsurl = {https://ui.adsabs.harvard.edu/abs/2020A&A...637A..80O}
}

@ARTICLE{Hayes_2022,
       author = {{Hayes}, Christian R. and {Masseron}, Thomas and {Sobeck}, Jennifer and {Garc{\'\i}a-Hern{\'a}ndez}, D.~A. and {Allende Prieto}, Carlos and {Beaton}, Rachael L. and {Cunha}, Katia and {Hasselquist}, Sten and {Holtzman}, Jon A. and {J{\"o}nsson}, Henrik and {Majewski}, Steven R. and {Shetrone}, Matthew and {Smith}, Verne V. and {Almeida}, Andr{\'e}s},
        title = "{BACCHUS Analysis of Weak Lines in APOGEE Spectra (BAWLAS)}",
      journal = {\apjs},
         year = 2022,
        month = sep,
       volume = {262},
       number = {1},
          eid = {34},
        pages = {34},
          doi = {10.3847/1538-4365/ac839f},
archivePrefix = {arXiv},
       eprint = {2208.00071},
 primaryClass = {astro-ph.GA},
       adsurl = {https://ui.adsabs.harvard.edu/abs/2022ApJS..262...34H}
}

@ARTICLE{SDSS_DR17,
       author = {{Abdurro'uf} and {Accetta}, Katherine and {Aerts}, Conny and {Silva Aguirre}, Victor and {Ahumada}, Romina and {Ajgaonkar}, Nikhil and {Filiz Ak}, N. and {Alam}, Shadab and {Allende Prieto}, Carlos and {Almeida}, Andres and {Anders}, Friedrich and {Anderson}, Scott F. and {Andrews}, Brett H. and {Anguiano}, Borja and {Aquino-Ortiz}, Erik and {Aragon-Salamanca}, Alfonso and {Argudo-Fernandez}, Maria and {Ata}, Metin and {Aubert}, Marie and {Avila-Reese}, Vladimir and {Badenes}, Carles and {Barba}, Rodolfo H. and {Barger}, Kat and {Barrera-Ballesteros}, Jorge K. and {Beaton}, Rachael L. and {Beers}, Timothy C. and {Belfiore}, Francesco and {Bender}, Chad F. and {Bernardi}, Mariangela and {Bershady}, Matthew A. and {Beutler}, Florian and {Moni Bidin}, Christian and {Bird}, Jonathan C. and {Bizyaev}, Dmitry and {Blanc}, Guillermo A. and {Blanton}, Michael R. and {Boardman}, Nicholas Fraser and {Bolton}, Adam S. and {Boquien}, Mederic and {Borissova}, Jura and {Bovy}, Jo and {Brandt}, W.~N. and {Brown}, Jordan and {Brownstein}, Joel R. and {Brusa}, Marcella and {Buchner}, Johannes and {Bundy}, Kevin and {Burchett}, Joseph N. and {Bureau}, Martin and {Burgasser}, Adam and {Cabang}, Tuesday K. and {Campbell}, Stephanie and {Cappellari}, Michele and {Carlberg}, Joleen K. and {Carneiro Wanderley}, Fabio and {Carrera}, Ricardo and {Cash}, Jennifer and {Chen}, Yan-Ping and {Chen}, Wei-Huai and {Cherinka}, Brian and {Chiappini}, Cristina and {Choi}, Peter Doohyun and {Chojnowski}, S. Drew and {Chung}, Haeun and {Clerc}, Nicolas and {Cohen}, Roger E. and {Comerford}, Julia M. and {Comparat}, Johan and {da Costa}, Luiz and {Covey}, Kevin and {Crane}, Jeffrey D. and {Cruz-Gonzalez}, Irene and {Culhane}, Connor and {Cunha}, Katia and {Dai}, Y. Sophia and {Damke}, Guillermo and {Darling}, Jeremy and {Davidson}, James W., Jr. and {Davies}, Roger and {Dawson}, Kyle and {De Lee}, Nathan and {Diamond-Stanic}, Aleksandar M. and {Cano-Diaz}, Mariana and {Dominguez Sanchez}, Helena and {Donor}, John and {Duckworth}, Chris and {Dwelly}, Tom and {Eisenstein}, Daniel J. and {Elsworth}, Yvonne P. and {Emsellem}, Eric and {Eracleous}, Mike and {Escoffier}, Stephanie and {Fan}, Xiaohui and {Farr}, Emily and {Feng}, Shuai and {Fernandez-Trincado}, Jose G. and {Feuillet}, Diane and {Filipp}, Andreas and {Fillingham}, Sean P and {Frinchaboy}, Peter M. and {Fromenteau}, Sebastien and {Galbany}, Lluis and {Garcia}, Rafael A. and {Garcia-Hernandez}, D.~A. and {Ge}, Junqiang and {Geisler}, Doug and {Gelfand}, Joseph and {Geron}, Tobias and {Gibson}, Benjamin J. and {Goddy}, Julian and {Godoy-Rivera}, Diego and {Grabowski}, Kathleen and {Green}, Paul J. and {Greener}, Michael and {Grier}, Catherine J. and {Griffith}, Emily and {Guo}, Hong and {Guy}, Julien and {Hadjara}, Massinissa and {Harding}, Paul and {Hasselquist}, Sten and {Hayes}, Christian R. and {Hearty}, Fred and {Hernndez}, Jess and {Hill}, Lewis and {Hogg}, David W. and {Holtzman}, Jon A. and {Horta}, Danny and {Hsieh}, Bau-Ching and {Hsu}, Chin-Hao and {Hsu}, Yun-Hsin and {Huber}, Daniel and {Huertas-Company}, Marc and {Hutchinson}, Brian and {Hwang}, Ho Seong and {Ibarra-Medel}, Hector J. and {Ider Chitham}, Jacob and {Ilha}, Gabriele S. and {Imig}, Julie and {Jaekle}, Will and {Jayasinghe}, Tharindu and {Ji}, Xihan and {Johnson}, Jennifer A. and {Jones}, Amy and {Jonsson}, Henrik and {Katkov}, Ivan and {Khalatyan}, Arman, Dr. and {Kinemuchi}, Karen and {Kisku}, Shobhit and {Knapen}, Johan H. and {Kneib}, Jean-Paul and {Kollmeier}, Juna A. and {Kong}, Miranda and {Kounkel}, Marina and {Kreckel}, Kathryn and {Krishnarao}, Dhanesh and {Lacerna}, Ivan and {Lane}, Richard R. and {Langgin}, Rachel and {Lavender}, Ramon and {Law}, David R. and {Lazarz}, Daniel and {Leung}, Henry W. and {Leung}, Ho-Hin and {Lewis}, Hannah M. and {Li}, Cheng and {Li}, Ran and {Lian}, Jianhui and {Liang}, Fu-Heng and {Lin}, Lihwai and {Lin}, Yen-Ting and {Lin}, Sicheng and {Lintott}, Chris and {Long}, Dan and {Longa-Pena}, Penelope and {Lopez-Coba}, Carlos and {Lu}, Shengdong and {Lundgren}, Britt F. and {Luo}, Yuanze and {Mackereth}, J. Ted and {de la Macorra}, Axel and {Mahadevan}, Suvrath and {Majewski}, Steven R. and {Manchado}, Arturo and {Mandeville}, Travis and {Maraston}, Claudia and {Margalef-Bentabol}, Berta and {Masseron}, Thomas and {Masters}, Karen L. and {Mathur}, Savita and {McDermid}, Richard M. and {Mckay}, Myles and {Merloni}, Andrea and {Merrifield}, Michael and {Meszaros}, Szabolcs and {Miglio}, Andrea and {Di Mille}, Francesco and {Minniti}, Dante and {Minsley}, Rebecca and {Monachesi}, Antonela and {Moon}, Jeongin and {Mosser}, Benoit and {Mulchaey}, John and {Muna}, Demitri and {Munoz}, Ricardo R. and {Myers}, Adam D. and {Myers}, Natalie and {Nadathur}, Seshadri and {Nair}, Preethi and {Nandra}, Kirpal and {Neumann}, Justus and {Newman}, Jeffrey A. and {Nidever}, David L. and {Nikakhtar}, Farnik and {Nitschelm}, Christian and {O'Connell}, Julia E. and {Garma-Oehmichen}, Luis and {Luan Souza de Oliveira}, Gabriel and {Olney}, Richard and {Oravetz}, Daniel and {Ortigoza-Urdaneta}, Mario and {Osorio}, Yeisson and {Otter}, Justin and {Pace}, Zachary J. and {Padilla}, Nelson and {Pan}, Kaike and {Pan}, Hsi-An and {Parikh}, Taniya and {Parker}, James and {Peirani}, Sebastien and {Pena Ramirez}, Karla and {Penny}, Samantha and {Percival}, Will J. and {Perez-Fournon}, Ismael and {Pinsonneault}, Marc and {Poidevin}, Frederick and {Poovelil}, Vijith Jacob and {Price-Whelan}, Adrian M. and {Queiroz}, Anna Barbara de Andrade and {Raddick}, M. Jordan and {Ray}, Amy and {Barboza Rembold}, Sandro and {Riddle}, Nicole and {Riffel}, Rogemar A. and {Riffel}, Rogerio and {Rix}, Hans-Walter and {Robin}, Annie C. and {Rodriguez-Puebla}, Aldo and {Santana Rojas}, Felipe Antonio and {Roman-Lopes}, Alexandre and {Roman-Zuniga}, Carlos and {Rose}, Benjamin and {Ross}, Ashley J. and {Rossi}, Graziano and {Rubin}, Kate H.~R. and {Salvato}, Mara and {Sanchez-Gallego}, Jose R. and {Sanderson}, Robyn and {Sarceno}, Edgar and {Sarmiento}, Regina and {Sayres}, Conor and {Sazonova}, Elizaveta and {Schaefer}, Adam L. and {Schlegel}, David J and {Schneider}, Donald P. and {Schultheis}, Mathias and {Schwope}, Axel and {Serenelli}, Aldo and {Serna}, Javier and {Shao}, Zhengyi and {Shapiro}, Griffin and {Sharma}, Anubhav and {Shen}, Yue and {Shetrone}, Matthew and {Shu}, Yiping and {Simon}, Joshua D. and {Skrutskie}, M.~F. and {Smethurst}, Rebecca and {Smith}, Verne and {Sobeck}, Jennifer and {Spoo}, Taylor and {Sprague}, Dani and {Stark}, David V. and {Stassun}, Keivan G. and {Steinmetz}, Matthias and {Stello}, Dennis and {Stone-Martinez}, Alexander and {Storchi-Bergmann}, Thaisa and {Stringfellow}, Guy S. and {Stutz}, Amelia and {Su}, Yung-Chau and {Taghizadeh-Popp}, Manuchehr and {Talbot}, Michael S. and {Tayar}, Jamie and {Telles}, Eduardo and {Teske}, Johanna and {Thakar}, Ani and {Theissen}, Christopher and {Thomas}, Daniel and {Tkachenko}, Andrew and {Tojeiro}, Rita and {Hernandez Toledo}, Hector and {Troup}, Nicholas W. and {Trump}, Jonathan R. and {Trussler}, James and {Turner}, Jacqueline and {Tuttle}, Sarah and {Unda-Sanzana}, Eduardo and {Vazquez-Mata}, Jose Antonio and {Valentini}, Marica and {Valenzuela}, Octavio and {Vargas-Gonzalez}, Jaime and {Vargas-Magana}, Mariana and {Alfaro}, Pablo Vera and {Villanova}, Sandro and {Vincenzo}, Fiorenzo and {Wake}, David and {Warfield}, Jack T. and {Washington}, Jessica Diane and {Weaver}, Benjamin Alan and {Weijmans}, Anne-Marie and {Weinberg}, David H. and {Weiss}, Achim and {Westfall}, Kyle B. and {Wild}, Vivienne and {Wilde}, Matthew C. and {Wilson}, John C. and {Wilson}, Robert F. and {Wilson}, Mikayla and {Wolf}, Julien and {Wood-Vasey}, W.~M. and {Yan}, Renbin and {Zamora}, Olga and {Zasowski}, Gail and {Zhang}, Kai and {Zhao}, Cheng and {Zheng}, Zheng and {Zheng}, Zheng and {Zhu}, Kai},
        title = "{The Seventeenth Data Release of the Sloan Digital Sky Surveys: Complete Release of MaNGA, MaStar and APOGEE-2 Data}",
      journal = {arXiv e-prints},
         year = 2021,
        month = dec,
          eid = {arXiv:2112.02026},
        pages = {arXiv:2112.02026},
archivePrefix = {arXiv},
       eprint = {2112.02026},
 primaryClass = {astro-ph.GA},
       adsurl = {https://ui.adsabs.harvard.edu/abs/2021arXiv211202026A}
}

@ARTICLE{Hasselquist2021_sats,
       author = {{Hasselquist}, Sten and {Hayes}, Christian R. and {Lian}, Jianhui and {Weinberg}, David H. and {Zasowski}, Gail and {Horta}, Danny and {Beaton}, Rachael and {Feuillet}, Diane K. and {Garro}, Elisa R. and {Gallart}, Carme and {Smith}, Verne V. and {Holtzman}, Jon A. and {Minniti}, Dante and {Lacerna}, Ivan and {Shetrone}, Matthew and {J{\"o}nsson}, Henrik and {Cioni}, Maria-Rosa L. and {Fillingham}, Sean P. and {Cunha}, Katia and {O{\'C}onnell}, Robert and {Fern{\'a}ndez-Trincado}, Jos{\'e} G. and {Mu{\~n}oz}, Ricardo R. and {Schiavon}, Ricardo and {Almeida}, Andres and {Anguiano}, Borja and {Beers}, Timothy C. and {Bizyaev}, Dmitry and {Brownstein}, Joel R. and {Cohen}, Roger E. and {Frinchaboy}, Peter and {Garc{\'\i}a-Hern{\'a}ndez}, D.~A. and {Geisler}, Doug and {Lane}, Richard R. and {Majewski}, Steven R. and {Nidever}, David L. and {Nitschelm}, Christian and {Povick}, Joshua and {Price-Whelan}, Adrian and {Roman-Lopes}, Alexandre and {Rosado}, Margarita and {Sobeck}, Jennifer and {Stringfellow}, Guy and {Valenzuela}, Octavio and {Villanova}, Sandro and {Vincenzo}, Fiorenzo},
        title = "{APOGEE Chemical Abundance Patterns of the Massive Milky Way Satellites}",
      journal = {arXiv e-prints},
         year = 2021,
        month = sep,
          eid = {arXiv:2109.05130},
        pages = {arXiv:2109.05130},
archivePrefix = {arXiv},
       eprint = {2109.05130},
 primaryClass = {astro-ph.GA},
       adsurl = {https://ui.adsabs.harvard.edu/abs/2021arXiv210905130H}
}

@ARTICLE{Santana_2021,
       author = {{Santana}, Felipe A. and {Beaton}, Rachael L. and {Covey}, Kevin R. and {O'Connell}, Julia E. and {Longa-Pe{\~n}a}, Pen{\'e}lope and {Cohen}, Roger and {Fern{\'a}ndez-Trincado}, Jos{\'e} G. and {Hayes}, Christian R. and {Zasowski}, Gail and {Sobeck}, Jennifer S. and {Majewski}, Steven R. and {Chojnowski}, S.~D. and {De Lee}, Nathan and {Oelkers}, Ryan J. and {Stringfellow}, Guy S. and {Almeida}, Andr{\'e}s and {Anguiano}, Borja and {Donor}, John and {Frinchaboy}, Peter M. and {Hasselquist}, Sten and {Johnson}, Jennifer A. and {Kollmeier}, Juna A. and {Nidever}, David L. and {Price-Whelan}, Adrian. M. and {Rojas-Arriagada}, Alvaro and {Schultheis}, Mathias and {Shetrone}, Matthew and {Simon}, Joshua D. and {Aerts}, Conny and {Borissova}, Jura and {Drout}, Maria R. and {Geisler}, Doug and {Law}, C.~Y. and {Medina}, Nicol{\'a}s and {Minniti}, Dante and {Monachesi}, Antonela and {Mu{\~n}oz}, Ricardo R. and {Poleski}, Rados{\l}aw and {Roman-Lopes}, Alexandre and {Schlaufman}, Kevin C. and {Stutz}, Amelia M. and {Teske}, Johanna and {Tkachenko}, Andrew and {Van Saders}, Jennifer L. and {Weinberger}, Alycia and {Zoccali}, Manuela},
        title = "{Final Targeting Strategy for the SDSS-IV APOGEE-2S Survey}",
      journal = {arXiv e-prints},
         year = 2021,
        month = aug,
          eid = {arXiv:2108.11908},
        pages = {arXiv:2108.11908},
archivePrefix = {arXiv},
       eprint = {2108.11908},
 primaryClass = {astro-ph.GA},
       adsurl = {https://ui.adsabs.harvard.edu/abs/2021arXiv210811908S}
}

@ARTICLE{Beaton_2021,
       author = {{Beaton}, Rachael L. and {Oelkers}, Ryan J. and {Hayes}, Christian R. and {Covey}, Kevin R. and {Chojnowski}, S.~D. and {De Lee}, Nathan and {Sobeck}, Jennifer S. and {Majewski}, Steven R. and {Cohen}, Roger E. and {Fern{\'a}ndez-Trincado}, Jos{\'e} and {Longa-Pe{\~n}a}, Pen{\'e}lope and {O'Connell}, Julia E. and {Santana}, Felipe A. and {Stringfellow}, Guy S. and {Zasowski}, Gail and {Aerts}, Conny and {Anguiano}, Borja and {Bender}, Chad and {Ca{\~n}as}, Caleb I. and {Cunha}, Katia and {Donor}, John and {Fleming}, Scott W. and {Frinchaboy}, Peter M. and {Feuillet}, Diane and {Harding}, Paul and {Hasselquist}, Sten and {Holtzman}, Jon A. and {Johnson}, Jennifer A. and {Kollmeier}, Juna A. and {Kounkel}, Marina and {Mahadevan}, Suvrath and {Price-Whelan}, Adrian. M. and {Rojas-Arriagada}, Alvaro and {Rom{\'a}n-Z{\'u}{\~n}iga}, Carlos and {Schlafly}, Edward F. and {Schultheis}, Mathias and {Shetrone}, Matthew and {Simon}, Joshua D. and {Stassun}, Keivan G. and {Stutz}, Amelia M. and {Tayar}, Jamie and {Teske}, Johanna and {Tkachenko}, Andrew and {Troup}, Nicholas and {Albareti}, Franco D. and {Bizyaev}, Dmitry and {Bovy}, Jo and {Burgasser}, Adam J. and {Comparat}, Johan and {Downes}, Juan Jos{\'e} and {Geisler}, Doug and {Inno}, Laura and {Manchado}, Arturo and {Ness}, Melissa K. and {Pinsonneault}, Marc H. and {Prada}, Francisco and {Roman-Lopes}, Alexandre and {Simonian}, Gregory V.~A. and {Smith}, Verne V. and {Yan}, Renbin and {Zamora}, Olga},
        title = "{Final Targeting Strategy for the Sloan Digital Sky Survey IV Apache Point Observatory Galactic Evolution Experiment 2 North Survey}",
      journal = {\aj},
         year = 2021,
        month = dec,
       volume = {162},
       number = {6},
          eid = {302},
        pages = {302},
          doi = {10.3847/1538-3881/ac260c},
       adsurl = {https://ui.adsabs.harvard.edu/abs/2021AJ....162..302B}
}

@ARTICLE{Hasselquist_2017,
       author = {{Hasselquist}, Sten and {Shetrone}, Matthew and {Smith}, Verne and {Holtzman}, Jon and {McWilliam}, Andrew and {Fern{\'a}ndez-Trincado}, J.~G. and {Beers}, Timothy C. and {Majewski}, Steven R. and {Nidever}, David L. and {Tang}, Baitian and {Tissera}, Patricia B. and {Fern{\'a}ndez Alvar}, Emma and {Allende Prieto}, Carlos and {Almeida}, Andres and {Anguiano}, Borja and {Battaglia}, Giuseppina and {Carigi}, Leticia and {Delgado Inglada}, Gloria and {Frinchaboy}, Peter and {Garc{\'\i}a-Hern{\'a}ndez}, D.~A. and {Geisler}, Doug and {Minniti}, Dante and {Placco}, Vinicius M. and {Schultheis}, Mathias and {Sobeck}, Jennifer and {Villanova}, Sandro},
        title = "{APOGEE Chemical Abundances of the Sagittarius Dwarf Galaxy}",
      journal = {\apj},
         year = 2017,
        month = aug,
       volume = {845},
       number = {2},
          eid = {162},
        pages = {162},
          doi = {10.3847/1538-4357/aa7ddc},
archivePrefix = {arXiv},
       eprint = {1707.03456},
 primaryClass = {astro-ph.GA},
       adsurl = {https://ui.adsabs.harvard.edu/abs/2017ApJ...845..162H}
}

@ARTICLE{gunn_2006,
   author = {{Gunn}, J.~E. and {Siegmund}, W.~A. and {Mannery}, E.~J. and 
	{Owen}, R.~E. and {Hull}, C.~L. and {Leger}, R.~F. and {Carey}, L.~N. and 
	{Knapp}, G.~R. and {York}, D.~G. and {Boroski}, W.~N. and {Kent}, S.~M. and 
	{Lupton}, R.~H. and {Rockosi}, C.~M. and {Evans}, M.~L. and 
	{Waddell}, P. and {Anderson}, J.~E. and {Annis}, J. and {Barentine}, J.~C. and 
	{Bartoszek}, L.~M. and {Bastian}, S. and {Bracker}, S.~B. and 
	{Brewington}, H.~J. and {Briegel}, C.~I. and {Brinkmann}, J. and 
	{Brown}, Y.~J. and {Carr}, M.~A. and {Czarapata}, P.~C. and 
	{Drennan}, C.~C. and {Dombeck}, T. and {Federwitz}, G.~R. and 
	{Gillespie}, B.~A. and {Gonzales}, C. and {Hansen}, S.~U. and 
	{Harvanek}, M. and {Hayes}, J. and {Jordan}, W. and {Kinney}, E. and 
	{Klaene}, M. and {Kleinman}, S.~J. and {Kron}, R.~G. and {Kresinski}, J. and 
	{Lee}, G. and {Limmongkol}, S. and {Lindenmeyer}, C.~W. and 
	{Long}, D.~C. and {Loomis}, C.~L. and {McGehee}, P.~M. and {Mantsch}, P.~M. and 
	{Neilsen}, Jr., E.~H. and {Neswold}, R.~M. and {Newman}, P.~R. and 
	{Nitta}, A. and {Peoples}, Jr., J. and {Pier}, J.~R. and {Prieto}, P.~S. and 
	{Prosapio}, A. and {Rivetta}, C. and {Schneider}, D.~P. and 
	{Snedden}, S. and {Wang}, S.-i.},
    title = "{The 2.5 m Telescope of the Sloan Digital Sky Survey}",
  journal = {\aj},
   eprint = {astro-ph/0602326},
     year = 2006,
    month = apr,
   volume = 131,
    pages = {2332-2359},
      doi = {10.1086/500975},
   adsurl = {http://adsabs.harvard.edu/abs/2006AJ....131.2332G}
}

@article{Bowen_1973,
author = {I. S. Bowen and A. H. Vaughan},
journal = {Appl. Opt.},
number = {7},
pages = {1430--1435},
publisher = {OSA},
title = {The Optical Design of the 40-in. Telescope and of the Irenee DuPont Telescope at Las Campanas Observatory, Chile},
volume = {12},
month = {Jul},
year = {1973},
url = {http://ao.osa.org/abstract.cfm?URI=ao-12-7-1430},
doi = {10.1364/AO.12.001430},
}

@ARTICLE{wilson_2019,
   author = {{Wilson}, J.~C. and {Hearty}, F.~R. and {Skrutskie}, M.~F. and 
	{Majewski}, S.~R. and {Holtzman}, J.~A. and {Eisenstein}, D. and 
	{Gunn}, J. and {Blank}, B. and {Henderson}, C. and {Smee}, S. and 
	{Nelson}, M. and {Nidever}, D. and {Arns}, J. and {Barkhouser}, R. and 
	{Barr}, J. and {Beland}, S. and {Bershady}, M.~A. and {Blanton}, M.~R. and 
	{Brunner}, S. and {Burton}, A. and {Carey}, L. and {Carr}, M. and 
	{Colque}, J.~P. and {Crane}, J. and {Damke}, G.~J. and {Davidson}, Jr., J.~W. and 
	{Dean}, J. and {Di Mille}, F. and {Don}, K.~W. and {Ebelke}, G. and 
	{Evans}, M. and {Fitzgerald}, G. and {Gillespie}, B. and {Hall}, M. and 
	{Harding}, A. and {Harding}, P. and {Hammond}, R. and {Hancock}, D. and 
	{Harrison}, C. and {Hope}, S. and {Horne}, T. and {Karakla}, J. and 
	{Lam}, C. and {Leger}, F. and {MacDonald}, N. and {Maseman}, P. and 
	{Matsunari}, J. and {Melton}, S. and {Mitcheltree}, T. and {O'Brien}, T. and 
	{O'Connell}, R.~W. and {Patten}, A. and {Richardson}, W. and 
	{Rieke}, G. and {Rieke}, M. and {Roman-Lopes}, A. and {Schiavon}, R.~P. and 
	{Sobeck}, J.~S. and {Stolberg}, T. and {Stoll}, R. and {Tembe}, M. and 
	{Trujillo}, J.~D. and {Uomoto}, A. and {Vernieri}, M. and {Walker}, E. and 
	{Weinberg}, D.~H. and {Young}, E. and {Anthony-Brumfield}, B. and 
	{Bizyaev}, D. and {Breslauer}, B. and {De Lee}, N. and {Downey}, J. and 
	{Halverson}, S. and {Huehnerhoff}, J. and {Klaene}, M. and {Leon}, E. and 
	{Long}, D. and {Mahadevan}, S. and {Malanushenko}, E. and {Nguyen}, D.~C. and 
	{Owen}, R. and {S{\'a}nchez-Gallego}, J.~R. and {Sayres}, C. and 
	{Shane}, N. and {Shectman}, S.~A. and {Shetrone}, M. and {Skinner}, D. and 
	{Stauffer}, F. and {Zhao}, B.},
    title = "{The Apache Point Observatory Galactic Evolution Experiment (APOGEE) Spectrographs}",
  journal = {\pasp},
archivePrefix = "arXiv",
   eprint = {1902.00928},
 primaryClass = "astro-ph.IM",
     year = 2019,
    month = may,
   volume = 131,
   number = 5,
    pages = {055001},
      doi = {10.1088/1538-3873/ab0075},
   adsurl = {http://adsabs.harvard.edu/abs/2019PASP..131e5001W}
}

@ARTICLE{Zasowski_2013,
   author = {{Zasowski}, G. and {Johnson}, J.~A. and {Frinchaboy}, P.~M. and 
	{Majewski}, S.~R. and {Nidever}, D.~L. and {Rocha Pinto}, H.~J. and 
	{Girardi}, L. and {Andrews}, B. and {Chojnowski}, S.~D. and 
	{Cudworth}, K.~M. and {Jackson}, K. and {Munn}, J. and {Skrutskie}, M.~F. and 
	{Beaton}, R.~L. and {Blake}, C.~H. and {Covey}, K. and {Deshpande}, R. and 
	{Epstein}, C. and {Fabbian}, D. and {Fleming}, S.~W. and {Garcia Hernandez}, D.~A. and 
	{Herrero}, A. and {Mahadevan}, S. and {M{\'e}sz{\'a}ros}, S. and 
	{Schultheis}, M. and {Sellgren}, K. and {Terrien}, R. and {van Saders}, J. and 
	{Allende Prieto}, C. and {Bizyaev}, D. and {Burton}, A. and 
	{Cunha}, K. and {da Costa}, L.~N. and {Hasselquist}, S. and 
	{Hearty}, F. and {Holtzman}, J. and {Garc{\'{\i}}a P{\'e}rez}, A.~E. and 
	{Maia}, M.~A.~G. and {O'Connell}, R.~W. and {O'Donnell}, C. and 
	{Pinsonneault}, M. and {Santiago}, B.~X. and {Schiavon}, R.~P. and 
	{Shetrone}, M. and {Smith}, V. and {Wilson}, J.~C.},
    title = "{Target Selection for the Apache Point Observatory Galactic Evolution Experiment (APOGEE)}",
  journal = {\aj},
archivePrefix = "arXiv",
   eprint = {1308.0351},
     year = 2013,
    month = oct,
   volume = 146,
      eid = {81},
    pages = {81},
      doi = {10.1088/0004-6256/146/4/81},
   adsurl = {http://adsabs.harvard.edu/abs/2013AJ....146...81Z}
}

@ARTICLE{Nidever_2020,
       author = {{Nidever}, David L. and {Hasselquist}, Sten and {Hayes}, Christian R. and {Hawkins}, Keith and {Povick}, Joshua and {Majewski}, Steven R. and {Smith}, Verne V. and {Anguiano}, Borja and {Stringfellow}, Guy S. and {Sobeck}, Jennifer S. and {Cunha}, Katia and {Beers}, Timothy C. and {Bestenlehner}, Joachim M. and {Cohen}, Roger E. and {Garcia-Hernandez}, D.~A. and {J{\"o}nsson}, Henrik and {Nitschelm}, Christian and {Shetrone}, Matthew and {Lacerna}, Ivan and {Allende Prieto}, Carlos and {Beaton}, Rachael L. and {Dell'Agli}, Flavia and {Fern{\'a}ndez-Trincado}, Jos{\'e} G. and {Feuillet}, Diane and {Gallart}, Carme and {Hearty}, Fred R. and {Holtzman}, Jon and {Manchado}, Arturo and {Mu{\~n}oz}, Ricardo R. and {O'Connell}, Robert and {Rosado}, Margarita},
        title = "{The Lazy Giants: APOGEE Abundances Reveal Low Star Formation Efficiencies in the Magellanic Clouds}",
      journal = {\apj},
         year = 2020,
        month = jun,
       volume = {895},
       number = {2},
          eid = {88},
        pages = {88},
          doi = {10.3847/1538-4357/ab7305},
archivePrefix = {arXiv},
       eprint = {1901.03448},
 primaryClass = {astro-ph.GA},
       adsurl = {https://ui.adsabs.harvard.edu/abs/2020ApJ...895...88N}
}

@ARTICLE{nidever_2015,
   author = {{Nidever}, D.~L. and {Holtzman}, J.~A. and {Allende Prieto}, C. and 
	{Beland}, S. and {Bender}, C. and {Bizyaev}, D. and {Burton}, A. and 
	{Desphande}, R. and {Fleming}, S.~W. and {Garc{\'{\i}}a P{\'e}rez}, A.~E. and 
	{Hearty}, F.~R. and {Majewski}, S.~R. and {M{\'e}sz{\'a}ros}, S. and 
	{Muna}, D. and {Nguyen}, D. and {Schiavon}, R.~P. and {Shetrone}, M. and 
	{Skrutskie}, M.~F. and {Sobeck}, J.~S. and {Wilson}, J.~C.},
    title = "{The Data Reduction Pipeline for the Apache Point Observatory Galactic Evolution Experiment}",
  journal = {\aj},
archivePrefix = "arXiv",
   eprint = {1501.03742},
 primaryClass = "astro-ph.IM",
     year = 2015,
    month = dec,
   volume = 150,
      eid = {173},
    pages = {173},
      doi = {10.1088/0004-6256/150/6/173},
   adsurl = {http://adsabs.harvard.edu/abs/2015AJ....150..173N}
}

@ARTICLE{garciaperez_2016,
   author = {{Garc{\'{\i}}a P{\'e}rez}, A.~E. and {Allende Prieto}, C. and 
	{Holtzman}, J.~A. and {Shetrone}, M. and {M{\'e}sz{\'a}ros}, S. and 
	{Bizyaev}, D. and {Carrera}, R. and {Cunha}, K. and {Garc{\'{\i}}a-Hern{\'a}ndez}, D.~A. and 
	{Johnson}, J.~A. and {Majewski}, S.~R. and {Nidever}, D.~L. and 
	{Schiavon}, R.~P. and {Shane}, N. and {Smith}, V.~V. and {Sobeck}, J. and 
	{Troup}, N. and {Zamora}, O. and {Weinberg}, D.~H. and {Bovy}, J. and 
	{Eisenstein}, D.~J. and {Feuillet}, D. and {Frinchaboy}, P.~M. and 
	{Hayden}, M.~R. and {Hearty}, F.~R. and {Nguyen}, D.~C. and 
	{O'Connell}, R.~W. and {Pinsonneault}, M.~H. and {Wilson}, J.~C. and 
	{Zasowski}, G.},
    title = "{ASPCAP: The APOGEE Stellar Parameter and Chemical Abundances Pipeline}",
  journal = {\aj},
archivePrefix = "arXiv",
   eprint = {1510.07635},
 primaryClass = "astro-ph.SR",
     year = 2016,
    month = jun,
   volume = 151,
      eid = {144},
    pages = {144},
      doi = {10.3847/0004-6256/151/6/144},
   adsurl = {http://adsabs.harvard.edu/abs/2016AJ....151..144G}
}

@ARTICLE{Jonsson_2020,
       author = {{J{\"o}nsson}, Henrik and {Holtzman}, Jon A. and
         {Prieto}, Carlos Allende and {Cunha}, Katia and
         {Garc{\'\i}a-Hern{\'a}ndez}, D.~A. and {Hasselquist}, Sten and
         {Masseron}, Thomas and {Osorio}, Yeisson and {Shetrone}, Matthew and
         {Smith}, Verne and {Stringfellow}, Guy S. and {Bizyaev}, Dmitry and
         {Edvardsson}, Bengt and {Majewski}, Steven R. and
         {M{\'e}sz{\'a}ros}, Szabolcs and {Souto}, Diogo and {Zamora}, Olga and
         {Beaton}, Rachael L. and {Bovy}, Jo and {Donor}, John and
         {Pinsonneault}, Marc H. and {Poovelil}, Vijith Jacob and
         {Sobeck}, Jennifer},
        title = "{APOGEE Data and Spectral Analysis from SDSS Data Release 16: Seven Years of Observations Including First Results from APOGEE-South}",
      journal = {\aj},
         year = 2020,
        month = sep,
       volume = {160},
       number = {3},
          eid = {120},
        pages = {120},
          doi = {10.3847/1538-3881/aba592},
archivePrefix = {arXiv},
       eprint = {2007.05537},
 primaryClass = {astro-ph.GA},
       adsurl = {https://ui.adsabs.harvard.edu/abs/2020AJ....160..120J}
}

@ARTICLE{Smith_2021,
       author = {{Smith}, Verne V. and {Bizyaev}, Dmitry and {Cunha}, Katia and {Shetrone}, Matthew D. and {Souto}, Diogo and {Allende Prieto}, Carlos and {Masseron}, Thomas and {M{\'e}sz{\'a}ros}, Szabolcs and {J{\"o}nsson}, Henrik and {Hasselquist}, Sten and {Osorio}, Yeisson and {Garc{\'\i}a-Hern{\'a}ndez}, D.~A. and {Plez}, Bertrand and {Beaton}, Rachael L. and {Holtzman}, Jon and {Majewski}, Steven R. and {Stringfellow}, Guy S. and {Sobeck}, Jennifer},
        title = "{The APOGEE Data Release 16 Spectral Line List}",
      journal = {\aj},
         year = 2021,
        month = jun,
       volume = {161},
       number = {6},
          eid = {254},
        pages = {254},
          doi = {10.3847/1538-3881/abefdc},
archivePrefix = {arXiv},
       eprint = {2103.10112},
 primaryClass = {astro-ph.SR},
       adsurl = {https://ui.adsabs.harvard.edu/abs/2021AJ....161..254S}
}

@ARTICLE{Jonsson_2018,
       author = {{J{\"o}nsson}, Henrik and {Allende Prieto}, Carlos and {Holtzman}, Jon A. and {Feuillet}, Diane K. and {Hawkins}, Keith and {Cunha}, Katia and {M{\'e}sz{\'a}ros}, Szabolcs and {Hasselquist}, Sten and {Fern{\'a}ndez-Trincado}, J.~G. and {Garc{\'\i}a-Hern{\'a}ndez}, D.~A. and {Bizyaev}, Dmitry and {Carrera}, Ricardo and {Majewski}, Steven R. and {Pinsonneault}, Marc H. and {Shetrone}, Matthew and {Smith}, Verne and {Sobeck}, Jennifer and {Souto}, Diogo and {Stringfellow}, Guy S. and {Teske}, Johanna and {Zamora}, Olga},
        title = "{APOGEE Data Releases 13 and 14: Stellar Parameter and Abundance Comparisons with Independent Analyses}",
      journal = {\aj},
         year = 2018,
        month = sep,
       volume = {156},
       number = {3},
          eid = {126},
        pages = {126},
          doi = {10.3847/1538-3881/aad4f5},
archivePrefix = {arXiv},
       eprint = {1807.09784},
 primaryClass = {astro-ph.SR},
       adsurl = {https://ui.adsabs.harvard.edu/abs/2018AJ....156..126J}
}

@ARTICLE{Majewski_2000_carina,
       author = {{Majewski}, Steven R. and {Ostheimer}, James C. and {Patterson}, Richard J. and {Kunkel}, William E. and {Johnston}, Kathryn V. and {Geisler}, Doug},
        title = "{Exploring Halo Substructure with Giant Stars. II. Mapping the Extended Structure of the Carina Dwarf Spheroidal Galaxy}",
      journal = {\aj},
         year = 2000,
        month = feb,
       volume = {119},
       number = {2},
        pages = {760-776},
          doi = {10.1086/301228},
archivePrefix = {arXiv},
       eprint = {astro-ph/9911191},
 primaryClass = {astro-ph},
       adsurl = {https://ui.adsabs.harvard.edu/abs/2000AJ....119..760M}
}

@ARTICLE{Korotin2020,
       author = {{Korotin}, S.~A. and {Andrievsky}, S.~M. and {Caffau}, E. and {Bonifacio}, P. and {Oliva}, E.},
        title = "{Study of the departures from LTE in the unevolved stars infrared spectra}",
      journal = {\mnras},
         year = 2020,
        month = aug,
       volume = {496},
       number = {2},
        pages = {2462-2473},
          doi = {10.1093/mnras/staa1707},
archivePrefix = {arXiv},
       eprint = {2006.10998},
 primaryClass = {astro-ph.SR},
       adsurl = {https://ui.adsabs.harvard.edu/abs/2020MNRAS.496.2462K}
}

@ARTICLE{Casey2016,
       author = {{Casey}, Andrew R. and {Hogg}, David W. and {Ness}, Melissa and {Rix}, Hans-Walter and {Ho}, Anna Q Y and {Gilmore}, Gerry},
        title = "{The Cannon 2: A data-driven model of stellar spectra for detailed chemical abundance analyses}",
      journal = {arXiv e-prints},
         year = 2016,
        month = mar,
          eid = {arXiv:1603.03040},
        pages = {arXiv:1603.03040},
          doi = {10.48550/arXiv.1603.03040},
archivePrefix = {arXiv},
       eprint = {1603.03040},
 primaryClass = {astro-ph.SR},
       adsurl = {https://ui.adsabs.harvard.edu/abs/2016arXiv160303040C}
}

@ARTICLE{Yang2024,
       author = {{Yang}, Yanbin and {Caffau}, Elisabetta and {Bonifacio}, Piercarlo and {Hammer}, Fran{\c{c}}ois and {Wang}, Jianling and {Mamon}, Gary A.},
        title = "{The accretion history of the Milky Way: IV. Hints of recent star formation in Milky Way dwarf spheroidal galaxies}",
      journal = {\aap},
         year = 2024,
        month = nov,
       volume = {691},
          eid = {A363},
        pages = {A363},
          doi = {10.1051/0004-6361/202451233},
archivePrefix = {arXiv},
       eprint = {2409.15414},
 primaryClass = {astro-ph.GA},
       adsurl = {https://ui.adsabs.harvard.edu/abs/2024A&A...691A.363Y}
}

@ARTICLE{Pace2020,
       author = {{Pace}, Andrew B. and {Kaplinghat}, Manoj and {Kirby}, Evan and {Simon}, Joshua D. and {Tollerud}, Erik and {Mu{\~n}oz}, Ricardo R. and {C{\^o}t{\'e}}, Patrick and {Djorgovski}, S.~G. and {Geha}, Marla},
        title = "{Multiple chemodynamic stellar populations of the Ursa Minor dwarf spheroidal galaxy}",
      journal = {\mnras},
         year = 2020,
        month = jul,
       volume = {495},
       number = {3},
        pages = {3022-3040},
          doi = {10.1093/mnras/staa1419},
archivePrefix = {arXiv},
       eprint = {2002.09503},
 primaryClass = {astro-ph.GA},
       adsurl = {https://ui.adsabs.harvard.edu/abs/2020MNRAS.495.3022P}
}

@software{synspec,
       author = {{Hubeny}, Ivan and {Lanz}, Thierry},
        title = "{Synspec: General Spectrum Synthesis Program}",
 howpublished = {Astrophysics Source Code Library, record ascl:1109.022},
         year = 2011,
        month = sep,
          eid = {ascl:1109.022},
       adsurl = {https://ui.adsabs.harvard.edu/abs/2011ascl.soft09022H}
}

@ARTICLE{Jenkins2021,
       author = {{Jenkins}, Sydney A. and {Li}, Ting S. and {Pace}, Andrew B. and {Ji}, Alexander P. and {Koposov}, Sergey E. and {Mutlu-Pakdil}, Bur{\c{c}}in},
        title = "{Very Large Telescope Spectroscopy of Ultra-faint Dwarf Galaxies. I. Bo{\"o}tes I, Leo IV, and Leo V}",
      journal = {\apj},
         year = 2021,
        month = oct,
       volume = {920},
       number = {2},
          eid = {92},
        pages = {92},
          doi = {10.3847/1538-4357/ac1353},
archivePrefix = {arXiv},
       eprint = {2101.00013},
 primaryClass = {astro-ph.GA},
       adsurl = {https://ui.adsabs.harvard.edu/abs/2021ApJ...920...92J}
}

@ARTICLE{Battaglia2022,
       author = {{Battaglia}, G. and {Taibi}, S. and {Thomas}, G.~F. and {Fritz}, T.~K.},
        title = "{Gaia early DR3 systemic motions of Local Group dwarf galaxies and orbital properties with a massive Large Magellanic Cloud}",
      journal = {\aap},
         year = 2022,
        month = jan,
       volume = {657},
          eid = {A54},
        pages = {A54},
          doi = {10.1051/0004-6361/202141528},
archivePrefix = {arXiv},
       eprint = {2106.08819},
 primaryClass = {astro-ph.GA},
       adsurl = {https://ui.adsabs.harvard.edu/abs/2022A&A...657A..54B}
}

@ARTICLE{Weisz_2014,
       author = {{Weisz}, Daniel R. and {Dolphin}, Andrew E. and {Skillman}, Evan D. and {Holtzman}, Jon and {Gilbert}, Karoline M. and {Dalcanton}, Julianne J. and {Williams}, Benjamin F.},
        title = "{The Star Formation Histories of Local Group Dwarf Galaxies. II. Searching For Signatures of Reionization}",
      journal = {\apj},
         year = 2014,
        month = jul,
       volume = {789},
       number = {2},
          eid = {148},
        pages = {148},
          doi = {10.1088/0004-637X/789/2/148},
archivePrefix = {arXiv},
       eprint = {1405.3281},
 primaryClass = {astro-ph.GA},
       adsurl = {https://ui.adsabs.harvard.edu/abs/2014ApJ...789..148W}
}

@ARTICLE{Skuladottir_2021,
       author = {{Sk{\'u}lad{\'o}ttir}, {\'A}sa and {Salvadori}, Stefania and {Amarsi}, Anish M. and {Tolstoy}, Eline and {Irwin}, Michael J. and {Hill}, Vanessa and {Jablonka}, Pascale and {Battaglia}, Giuseppina and {Starkenburg}, Else and {Massari}, Davide and {Helmi}, Amina and {Posti}, Lorenzo},
        title = "{Zero-metallicity Hypernova Uncovered by an Ultra-metal-poor Star in the Sculptor Dwarf Spheroidal Galaxy}",
      journal = {\apjl},
         year = 2021,
        month = jul,
       volume = {915},
       number = {2},
          eid = {L30},
        pages = {L30},
          doi = {10.3847/2041-8213/ac0dc2},
archivePrefix = {arXiv},
       eprint = {2106.11592},
 primaryClass = {astro-ph.GA},
       adsurl = {https://ui.adsabs.harvard.edu/abs/2021ApJ...915L..30S}
}

@article{AllendePrieto2006,
	Author = {{Allende Prieto}, Carlos and Beers, Timothy C. and Wilhelm, Ronald and Newberg, Heidi Jo and Rockosi, Constance M. and Yanny, Brian and Lee, Young Sun},
	Doi = {10.1086/498131},
	Issn = {0004-637X},
	Journal = {AJ},
	Month = jan,
	Number = {2},
	Pages = {804--820},
	Title = {{A Spectroscopic Study of the Ancient Milky Way: F??? and G???Type Stars in the Third Data Release of the Sloan Digital Sky Survey}},
	Url = {http://adsabs.harvard.edu/abs/2006ApJ...636..804A},
	Volume = {636},
	Year = {2006}}

@ARTICLE{Weisz2014,
       author = {{Weisz}, Daniel R. and {Dolphin}, Andrew E. and {Skillman}, Evan D. and {Holtzman}, Jon and {Gilbert}, Karoline M. and {Dalcanton}, Julianne J. and {Williams}, Benjamin F.},
        title = "{The Star Formation Histories of Local Group Dwarf Galaxies. I. Hubble Space Telescope/Wide Field Planetary Camera 2 Observations}",
      journal = {\apj},
         year = 2014,
        month = jul,
       volume = {789},
       number = {2},
          eid = {147},
        pages = {147},
          doi = {10.1088/0004-637X/789/2/147},
archivePrefix = {arXiv},
       eprint = {1404.7144},
 primaryClass = {astro-ph.GA},
       adsurl = {https://ui.adsabs.harvard.edu/abs/2014ApJ...789..147W}
}

@ARTICLE{Hendricks2014,
       author = {{Hendricks}, Benjamin and {Koch}, Andreas and {Walker}, Matthew and {Johnson}, Christian I. and {Pe{\~n}arrubia}, Jorge and {Gilmore}, Gerard},
        title = "{Insights from the outskirts: Chemical and dynamical properties in the outer parts of the Fornax dwarf spheroidal galaxy}",
      journal = {\aap},
         year = 2014,
        month = dec,
       volume = {572},
          eid = {A82},
        pages = {A82},
          doi = {10.1051/0004-6361/201424645},
archivePrefix = {arXiv},
       eprint = {1408.5597},
 primaryClass = {astro-ph.GA},
       adsurl = {https://ui.adsabs.harvard.edu/abs/2014A&A...572A..82H}
}

@ARTICLE{Ji2023,
       author = {{Ji}, Alexander P. and {Simon}, Joshua D. and {Roederer}, Ian U. and {Magg}, Ekaterina and {Frebel}, Anna and {Johnson}, Christian I. and {Klessen}, Ralf S. and {Magg}, Mattis and {Cescutti}, Gabriele and {Mateo}, Mario and {Bergemann}, Maria and {Bailey}, John I.},
        title = "{Metal Mixing in the r-process Enhanced Ultrafaint Dwarf Galaxy Reticulum II}",
      journal = {\aj},
         year = 2023,
        month = mar,
       volume = {165},
       number = {3},
          eid = {100},
        pages = {100},
          doi = {10.3847/1538-3881/acad84},
archivePrefix = {arXiv},
       eprint = {2207.03499},
 primaryClass = {astro-ph.GA},
       adsurl = {https://ui.adsabs.harvard.edu/abs/2023AJ....165..100J}
}

@MISC{2021zndo...4906681N,
       author = {{Nidever}, David},
        title = "{dnidever/doppler: Cannon and Payne models}",
 howpublished = {Zenodo},
         year = 2021,
        month = jun,
          eid = {10.5281/zenodo.4906681},
          doi = {10.5281/zenodo.4906681},
      version = {v1.1.0},
    publisher = {Zenodo},
       adsurl = {https://ui.adsabs.harvard.edu/abs/2021zndo...4906681N}
}

@ARTICLE{Dolphin2002,
       author = {{Dolphin}, A.~E.},
        title = "{Numerical methods of star formation history measurement and applications to seven dwarf spheroidals}",
      journal = {\mnras},
         year = 2002,
        month = may,
       volume = {332},
       number = {1},
        pages = {91-108},
          doi = {10.1046/j.1365-8711.2002.05271.x},
archivePrefix = {arXiv},
       eprint = {astro-ph/0112331},
 primaryClass = {astro-ph},
       adsurl = {https://ui.adsabs.harvard.edu/abs/2002MNRAS.332...91D}
}

@ARTICLE{delosReyes2022,
       author = {{de los Reyes}, Mithi A.~C. and {Kirby}, Evan N. and {Ji}, Alexander P. and {Nu{\~n}ez}, Evan H.},
        title = "{Simultaneous Constraints on the Star Formation History and Nucleosynthesis of Sculptor dSph}",
      journal = {\apj},
         year = 2022,
        month = jan,
       volume = {925},
       number = {1},
          eid = {66},
        pages = {66},
          doi = {10.3847/1538-4357/ac332b},
archivePrefix = {arXiv},
       eprint = {2110.01690},
 primaryClass = {astro-ph.GA},
       adsurl = {https://ui.adsabs.harvard.edu/abs/2022ApJ...925...66D}
}

@ARTICLE{Bettinelli2019,
       author = {{Bettinelli}, M. and {Hidalgo}, S.~L. and {Cassisi}, S. and {Aparicio}, A. and {Piotto}, G. and {Valdes}, F. and {Walker}, A.~R.},
        title = "{The star formation history of the Sculptor dwarf spheroidal galaxy}",
      journal = {\mnras},
         year = 2019,
        month = aug,
       volume = {487},
       number = {4},
        pages = {5862-5873},
          doi = {10.1093/mnras/stz1679},
archivePrefix = {arXiv},
       eprint = {1906.07042},
 primaryClass = {astro-ph.GA},
       adsurl = {https://ui.adsabs.harvard.edu/abs/2019MNRAS.487.5862B}
}

@ARTICLE{Bekki2012,
       author = {{Bekki}, Kenji and {Tsujimoto}, Takuji},
        title = "{Chemical Evolution of the Large Magellanic Cloud}",
      journal = {\apj},
         year = 2012,
        month = dec,
       volume = {761},
       number = {2},
          eid = {180},
        pages = {180},
          doi = {10.1088/0004-637X/761/2/180},
archivePrefix = {arXiv},
       eprint = {1210.3968},
 primaryClass = {astro-ph.CO},
       adsurl = {https://ui.adsabs.harvard.edu/abs/2012ApJ...761..180B}
}

@ARTICLE{Kirby2011_iv,
       author = {{Kirby}, Evan N. and {Cohen}, Judith. G. and {Smith}, Graeme H. and {Majewski}, Steven R. and {Sohn}, Sangmo Tony and {Guhathakurta}, Puragra},
        title = "{Multi-element Abundance Measurements from Medium-resolution Spectra. IV. Alpha Element Distributions in Milky Way Satellite Galaxies}",
      journal = {\apj},
         year = 2011,
        month = feb,
       volume = {727},
       number = {2},
          eid = {79},
        pages = {79},
          doi = {10.1088/0004-637X/727/2/79},
archivePrefix = {arXiv},
       eprint = {1011.5221},
 primaryClass = {astro-ph.GA},
       adsurl = {https://ui.adsabs.harvard.edu/abs/2011ApJ...727...79K}
}

@ARTICLE{Kirby2011_iii,
       author = {{Kirby}, Evan N. and {Lanfranchi}, Gustavo A. and {Simon}, Joshua D. and {Cohen}, Judith G. and {Guhathakurta}, Puragra},
        title = "{Multi-element Abundance Measurements from Medium-resolution Spectra. III. Metallicity Distributions of Milky Way Dwarf Satellite Galaxies}",
      journal = {\apj},
         year = 2011,
        month = feb,
       volume = {727},
       number = {2},
          eid = {78},
        pages = {78},
          doi = {10.1088/0004-637X/727/2/78},
archivePrefix = {arXiv},
       eprint = {1011.4937},
 primaryClass = {astro-ph.GA},
       adsurl = {https://ui.adsabs.harvard.edu/abs/2011ApJ...727...78K}
}

@ARTICLE{Bergemann_2017,
       author = {{Bergemann}, Maria and {Collet}, Remo and {Amarsi}, Anish M. and {Kovalev}, Mikhail and {Ruchti}, Greg and {Magic}, Zazralt},
        title = "{Non-local Thermodynamic Equilibrium Stellar Spectroscopy with 1D and <3D> Models. I. Methods and Application to Magnesium Abundances in Standard Stars}",
      journal = {\apj},
         year = 2017,
        month = sep,
       volume = {847},
       number = {1},
          eid = {15},
        pages = {15},
          doi = {10.3847/1538-4357/aa88cb},
archivePrefix = {arXiv},
       eprint = {1612.07355},
 primaryClass = {astro-ph.SR},
       adsurl = {https://ui.adsabs.harvard.edu/abs/2017ApJ...847...15B}
}

@ARTICLE{gaia_dr2,
       author = {{Gaia Collaboration} and {Brown}, A.~G.~A. and {Vallenari}, A. and {Prusti}, T. and {de Bruijne}, J.~H.~J. and {Babusiaux}, C. and {Bailer-Jones}, C.~A.~L. and {Biermann}, M. and {Evans}, D.~W. and {Eyer}, L. and {Jansen}, F. and {Jordi}, C. and {Klioner}, S.~A. and {Lammers}, U. and {Lindegren}, L. and {Luri}, X. and {Mignard}, F. and {Panem}, C. and {Pourbaix}, D. and {Randich}, S. and {Sartoretti}, P. and {Siddiqui}, H.~I. and {Soubiran}, C. and {van Leeuwen}, F. and {Walton}, N.~A. and {Arenou}, F. and {Bastian}, U. and {Cropper}, M. and {Drimmel}, R. and {Katz}, D. and {Lattanzi}, M.~G. and {Bakker}, J. and {Cacciari}, C. and {Casta{\~n}eda}, J. and {Chaoul}, L. and {Cheek}, N. and {De Angeli}, F. and {Fabricius}, C. and {Guerra}, R. and {Holl}, B. and {Masana}, E. and {Messineo}, R. and {Mowlavi}, N. and {Nienartowicz}, K. and {Panuzzo}, P. and {Portell}, J. and {Riello}, M. and {Seabroke}, G.~M. and {Tanga}, P. and {Th{\'e}venin}, F. and {Gracia-Abril}, G. and {Comoretto}, G. and {Garcia-Reinaldos}, M. and {Teyssier}, D. and {Altmann}, M. and {Andrae}, R. and {Audard}, M. and {Bellas-Velidis}, I. and {Benson}, K. and {Berthier}, J. and {Blomme}, R. and {Burgess}, P. and {Busso}, G. and {Carry}, B. and {Cellino}, A. and {Clementini}, G. and {Clotet}, M. and {Creevey}, O. and {Davidson}, M. and {De Ridder}, J. and {Delchambre}, L. and {Dell'Oro}, A. and {Ducourant}, C. and {Fern{\'a}ndez-Hern{\'a}ndez}, J. and {Fouesneau}, M. and {Fr{\'e}mat}, Y. and {Galluccio}, L. and {Garc{\'\i}a-Torres}, M. and {Gonz{\'a}lez-N{\'u}{\~n}ez}, J. and {Gonz{\'a}lez-Vidal}, J.~J. and {Gosset}, E. and {Guy}, L.~P. and {Halbwachs}, J. -L. and {Hambly}, N.~C. and {Harrison}, D.~L. and {Hern{\'a}ndez}, J. and {Hestroffer}, D. and {Hodgkin}, S.~T. and {Hutton}, A. and {Jasniewicz}, G. and {Jean-Antoine-Piccolo}, A. and {Jordan}, S. and {Korn}, A.~J. and {Krone-Martins}, A. and {Lanzafame}, A.~C. and {Lebzelter}, T. and {L{\"o}ffler}, W. and {Manteiga}, M. and {Marrese}, P.~M. and {Mart{\'\i}n-Fleitas}, J.~M. and {Moitinho}, A. and {Mora}, A. and {Muinonen}, K. and {Osinde}, J. and {Pancino}, E. and {Pauwels}, T. and {Petit}, J. -M. and {Recio-Blanco}, A. and {Richards}, P.~J. and {Rimoldini}, L. and {Robin}, A.~C. and {Sarro}, L.~M. and {Siopis}, C. and {Smith}, M. and {Sozzetti}, A. and {S{\"u}veges}, M. and {Torra}, J. and {van Reeven}, W. and {Abbas}, U. and {Abreu Aramburu}, A. and {Accart}, S. and {Aerts}, C. and {Altavilla}, G. and {{\'A}lvarez}, M.~A. and {Alvarez}, R. and {Alves}, J. and {Anderson}, R.~I. and {Andrei}, A.~H. and {Anglada Varela}, E. and {Antiche}, E. and {Antoja}, T. and {Arcay}, B. and {Astraatmadja}, T.~L. and {Bach}, N. and {Baker}, S.~G. and {Balaguer-N{\'u}{\~n}ez}, L. and {Balm}, P. and {Barache}, C. and {Barata}, C. and {Barbato}, D. and {Barblan}, F. and {Barklem}, P.~S. and {Barrado}, D. and {Barros}, M. and {Barstow}, M.~A. and {Bartholom{\'e} Mu{\~n}oz}, S. and {Bassilana}, J. -L. and {Becciani}, U. and {Bellazzini}, M. and {Berihuete}, A. and {Bertone}, S. and {Bianchi}, L. and {Bienaym{\'e}}, O. and {Blanco-Cuaresma}, S. and {Boch}, T. and {Boeche}, C. and {Bombrun}, A. and {Borrachero}, R. and {Bossini}, D. and {Bouquillon}, S. and {Bourda}, G. and {Bragaglia}, A. and {Bramante}, L. and {Breddels}, M.~A. and {Bressan}, A. and {Brouillet}, N. and {Br{\"u}semeister}, T. and {Brugaletta}, E. and {Bucciarelli}, B. and {Burlacu}, A. and {Busonero}, D. and {Butkevich}, A.~G. and {Buzzi}, R. and {Caffau}, E. and {Cancelliere}, R. and {Cannizzaro}, G. and {Cantat-Gaudin}, T. and {Carballo}, R. and {Carlucci}, T. and {Carrasco}, J.~M. and {Casamiquela}, L. and {Castellani}, M. and {Castro-Ginard}, A. and {Charlot}, P. and {Chemin}, L. and {Chiavassa}, A. and {Cocozza}, G. and {Costigan}, G. and {Cowell}, S. and {Crifo}, F. and {Crosta}, M. and {Crowley}, C. and {Cuypers}, J. and {Dafonte}, C. and {Damerdji}, Y. and {Dapergolas}, A. and {David}, P. and {David}, M. and {de Laverny}, P. and {De Luise}, F. and {De March}, R. and {de Martino}, D. and {de Souza}, R. and {de Torres}, A. and {Debosscher}, J. and {del Pozo}, E. and {Delbo}, M. and {Delgado}, A. and {Delgado}, H.~E. and {Di Matteo}, P. and {Diakite}, S. and {Diener}, C. and {Distefano}, E. and {Dolding}, C. and {Drazinos}, P. and {Dur{\'a}n}, J. and {Edvardsson}, B. and {Enke}, H. and {Eriksson}, K. and {Esquej}, P. and {Eynard Bontemps}, G. and {Fabre}, C. and {Fabrizio}, M. and {Faigler}, S. and {Falc{\~a}o}, A.~J. and {Farr{\`a}s Casas}, M. and {Federici}, L. and {Fedorets}, G. and {Fernique}, P. and {Figueras}, F. and {Filippi}, F. and {Findeisen}, K. and {Fonti}, A. and {Fraile}, E. and {Fraser}, M. and {Fr{\'e}zouls}, B. and {Gai}, M. and {Galleti}, S. and {Garabato}, D. and {Garc{\'\i}a-Sedano}, F. and {Garofalo}, A. and {Garralda}, N. and {Gavel}, A. and {Gavras}, P. and {Gerssen}, J. and {Geyer}, R. and {Giacobbe}, P. and {Gilmore}, G. and {Girona}, S. and {Giuffrida}, G. and {Glass}, F. and {Gomes}, M. and {Granvik}, M. and {Gueguen}, A. and {Guerrier}, A. and {Guiraud}, J. and {Guti{\'e}rrez-S{\'a}nchez}, R. and {Haigron}, R. and {Hatzidimitriou}, D. and {Hauser}, M. and {Haywood}, M. and {Heiter}, U. and {Helmi}, A. and {Heu}, J. and {Hilger}, T. and {Hobbs}, D. and {Hofmann}, W. and {Holland}, G. and {Huckle}, H.~E. and {Hypki}, A. and {Icardi}, V. and {Jan{\ss}en}, K. and {Jevardat de Fombelle}, G. and {Jonker}, P.~G. and {Juh{\'a}sz}, {\'A}. L. and {Julbe}, F. and {Karampelas}, A. and {Kewley}, A. and {Klar}, J. and {Kochoska}, A. and {Kohley}, R. and {Kolenberg}, K. and {Kontizas}, M. and {Kontizas}, E. and {Koposov}, S.~E. and {Kordopatis}, G. and {Kostrzewa-Rutkowska}, Z. and {Koubsky}, P. and {Lambert}, S. and {Lanza}, A.~F. and {Lasne}, Y. and {Lavigne}, J. -B. and {Le Fustec}, Y. and {Le Poncin-Lafitte}, C. and {Lebreton}, Y. and {Leccia}, S. and {Leclerc}, N. and {Lecoeur-Taibi}, I. and {Lenhardt}, H. and {Leroux}, F. and {Liao}, S. and {Licata}, E. and {Lindstr{\o}m}, H.~E.~P. and {Lister}, T.~A. and {Livanou}, E. and {Lobel}, A. and {L{\'o}pez}, M. and {Managau}, S. and {Mann}, R.~G. and {Mantelet}, G. and {Marchal}, O. and {Marchant}, J.~M. and {Marconi}, M. and {Marinoni}, S. and {Marschalk{\'o}}, G. and {Marshall}, D.~J. and {Martino}, M. and {Marton}, G. and {Mary}, N. and {Massari}, D. and {Matijevi{\v{c}}}, G. and {Mazeh}, T. and {McMillan}, P.~J. and {Messina}, S. and {Michalik}, D. and {Millar}, N.~R. and {Molina}, D. and {Molinaro}, R. and {Moln{\'a}r}, L. and {Montegriffo}, P. and {Mor}, R. and {Morbidelli}, R. and {Morel}, T. and {Morris}, D. and {Mulone}, A.~F. and {Muraveva}, T. and {Musella}, I. and {Nelemans}, G. and {Nicastro}, L. and {Noval}, L. and {O'Mullane}, W. and {Ord{\'e}novic}, C. and {Ord{\'o}{\~n}ez-Blanco}, D. and {Osborne}, P. and {Pagani}, C. and {Pagano}, I. and {Pailler}, F. and {Palacin}, H. and {Palaversa}, L. and {Panahi}, A. and {Pawlak}, M. and {Piersimoni}, A.~M. and {Pineau}, F. -X. and {Plachy}, E. and {Plum}, G. and {Poggio}, E. and {Poujoulet}, E. and {Pr{\v{s}}a}, A. and {Pulone}, L. and {Racero}, E. and {Ragaini}, S. and {Rambaux}, N. and {Ramos-Lerate}, M. and {Regibo}, S. and {Reyl{\'e}}, C. and {Riclet}, F. and {Ripepi}, V. and {Riva}, A. and {Rivard}, A. and {Rixon}, G. and {Roegiers}, T. and {Roelens}, M. and {Romero-G{\'o}mez}, M. and {Rowell}, N. and {Royer}, F. and {Ruiz-Dern}, L. and {Sadowski}, G. and {Sagrist{\`a} Sell{\'e}s}, T. and {Sahlmann}, J. and {Salgado}, J. and {Salguero}, E. and {Sanna}, N. and {Santana-Ros}, T. and {Sarasso}, M. and {Savietto}, H. and {Schultheis}, M. and {Sciacca}, E. and {Segol}, M. and {Segovia}, J.~C. and {S{\'e}gransan}, D. and {Shih}, I. -C. and {Siltala}, L. and {Silva}, A.~F. and {Smart}, R.~L. and {Smith}, K.~W. and {Solano}, E. and {Solitro}, F. and {Sordo}, R. and {Soria Nieto}, S. and {Souchay}, J. and {Spagna}, A. and {Spoto}, F. and {Stampa}, U. and {Steele}, I.~A. and {Steidelm{\"u}ller}, H. and {Stephenson}, C.~A. and {Stoev}, H. and {Suess}, F.~F. and {Surdej}, J. and {Szabados}, L. and {Szegedi-Elek}, E. and {Tapiador}, D. and {Taris}, F. and {Tauran}, G. and {Taylor}, M.~B. and {Teixeira}, R. and {Terrett}, D. and {Teyssandier}, P. and {Thuillot}, W. and {Titarenko}, A. and {Torra Clotet}, F. and {Turon}, C. and {Ulla}, A. and {Utrilla}, E. and {Uzzi}, S. and {Vaillant}, M. and {Valentini}, G. and {Valette}, V. and {van Elteren}, A. and {Van Hemelryck}, E. and {van Leeuwen}, M. and {Vaschetto}, M. and {Vecchiato}, A. and {Veljanoski}, J. and {Viala}, Y. and {Vicente}, D. and {Vogt}, S. and {von Essen}, C. and {Voss}, H. and {Votruba}, V. and {Voutsinas}, S. and {Walmsley}, G. and {Weiler}, M. and {Wertz}, O. and {Wevers}, T. and {Wyrzykowski}, {\L}. and {Yoldas}, A. and {{\v{Z}}erjal}, M. and {Ziaeepour}, H. and {Zorec}, J. and {Zschocke}, S. and {Zucker}, S. and {Zurbach}, C. and {Zwitter}, T.},
        title = "{Gaia Data Release 2. Summary of the contents and survey properties}",
      journal = {\aap},
         year = 2018,
        month = aug,
       volume = {616},
          eid = {A1},
        pages = {A1},
          doi = {10.1051/0004-6361/201833051},
archivePrefix = {arXiv},
       eprint = {1804.09365},
 primaryClass = {astro-ph.GA},
       adsurl = {https://ui.adsabs.harvard.edu/abs/2018A&A...616A...1G}
}

@INPROCEEDINGS{topcat,
       author = {{Taylor}, M.~B.},
        title = "{TOPCAT \& STIL: Starlink Table/VOTable Processing Software}",
    booktitle = {Astronomical Data Analysis Software and Systems XIV},
         year = 2005,
       editor = {{Shopbell}, P. and {Britton}, M. and {Ebert}, R.},
       series = {Astronomical Society of the Pacific Conference Series},
       volume = {347},
        month = dec,
        pages = {29},
       adsurl = {https://ui.adsabs.harvard.edu/abs/2005ASPC..347...29T}
}

@ARTICLE{Harris_2020numpy,
  author  = {Harris, Charles R. and Millman, K. Jarrod and
            van der Walt, Stéfan J and Gommers, Ralf and
            Virtanen, Pauli and Cournapeau, David and
            Wieser, Eric and Taylor, Julian and Berg, Sebastian and
            Smith, Nathaniel J. and Kern, Robert and Picus, Matti and
            Hoyer, Stephan and van Kerkwijk, Marten H. and
            Brett, Matthew and Haldane, Allan and
            Fernández del Río, Jaime and Wiebe, Mark and
            Peterson, Pearu and Gérard-Marchant, Pierre and
            Sheppard, Kevin and Reddy, Tyler and Weckesser, Warren and
            Abbasi, Hameer and Gohlke, Christoph and
            Oliphant, Travis E.},
  title   = {Array programming with {NumPy}},
  journal = {Nature},
  year    = {2020},
  volume  = {585},
  pages   = {357–362},
  doi     = {10.1038/s41586-020-2649-2}
}

@ARTICLE{vanderWalt_2011,
       author = {{van der Walt}, St{\'e}fan and {Colbert}, S. Chris and {Varoquaux}, Ga{\"e}l},
        title = "{The NumPy Array: A Structure for Efficient Numerical Computation}",
      journal = {Computing in Science and Engineering},
         year = 2011,
        month = mar,
       volume = {13},
       number = {2},
        pages = {22-30},
          doi = {10.1109/MCSE.2011.37},
archivePrefix = {arXiv},
       eprint = {1102.1523},
 primaryClass = {cs.MS},
       adsurl = {https://ui.adsabs.harvard.edu/abs/2011CSE....13b..22V}
}

@ARTICLE{Astropy_2013,
       author = {{Astropy Collaboration} and {Robitaille}, Thomas P. and {Tollerud}, Erik J. and {Greenfield}, Perry and {Droettboom}, Michael and {Bray}, Erik and {Aldcroft}, Tom and {Davis}, Matt and {Ginsburg}, Adam and {Price-Whelan}, Adrian M. and {Kerzendorf}, Wolfgang E. and {Conley}, Alexander and {Crighton}, Neil and {Barbary}, Kyle and {Muna}, Demitri and {Ferguson}, Henry and {Grollier}, Fr{\'e}d{\'e}ric and {Parikh}, Madhura M. and {Nair}, Prasanth H. and {Unther}, Hans M. and {Deil}, Christoph and {Woillez}, Julien and {Conseil}, Simon and {Kramer}, Roban and {Turner}, James E.~H. and {Singer}, Leo and {Fox}, Ryan and {Weaver}, Benjamin A. and {Zabalza}, Victor and {Edwards}, Zachary I. and {Azalee Bostroem}, K. and {Burke}, D.~J. and {Casey}, Andrew R. and {Crawford}, Steven M. and {Dencheva}, Nadia and {Ely}, Justin and {Jenness}, Tim and {Labrie}, Kathleen and {Lim}, Pey Lian and {Pierfederici}, Francesco and {Pontzen}, Andrew and {Ptak}, Andy and {Refsdal}, Brian and {Servillat}, Mathieu and {Streicher}, Ole},
        title = "{Astropy: A community Python package for astronomy}",
      journal = {\aap},
         year = 2013,
        month = oct,
       volume = {558},
          eid = {A33},
        pages = {A33},
          doi = {10.1051/0004-6361/201322068},
archivePrefix = {arXiv},
       eprint = {1307.6212},
 primaryClass = {astro-ph.IM},
       adsurl = {https://ui.adsabs.harvard.edu/abs/2013A&A...558A..33A}
}

@ARTICLE{Astropy_2018,
       author = {{Astropy Collaboration} and {Price-Whelan}, A.~M. and {Sip{\H{o}}cz}, B.~M. and {G{\"u}nther}, H.~M. and {Lim}, P.~L. and {Crawford}, S.~M. and {Conseil}, S. and {Shupe}, D.~L. and {Craig}, M.~W. and {Dencheva}, N. and {Ginsburg}, A. and {VanderPlas}, J.~T. and {Bradley}, L.~D. and {P{\'e}rez-Su{\'a}rez}, D. and {de Val-Borro}, M. and {Aldcroft}, T.~L. and {Cruz}, K.~L. and {Robitaille}, T.~P. and {Tollerud}, E.~J. and {Ardelean}, C. and {Babej}, T. and {Bach}, Y.~P. and {Bachetti}, M. and {Bakanov}, A.~V. and {Bamford}, S.~P. and {Barentsen}, G. and {Barmby}, P. and {Baumbach}, A. and {Berry}, K.~L. and {Biscani}, F. and {Boquien}, M. and {Bostroem}, K.~A. and {Bouma}, L.~G. and {Brammer}, G.~B. and {Bray}, E.~M. and {Breytenbach}, H. and {Buddelmeijer}, H. and {Burke}, D.~J. and {Calderone}, G. and {Cano Rodr{\'\i}guez}, J.~L. and {Cara}, M. and {Cardoso}, J.~V.~M. and {Cheedella}, S. and {Copin}, Y. and {Corrales}, L. and {Crichton}, D. and {D'Avella}, D. and {Deil}, C. and {Depagne}, {\'E}. and {Dietrich}, J.~P. and {Donath}, A. and {Droettboom}, M. and {Earl}, N. and {Erben}, T. and {Fabbro}, S. and {Ferreira}, L.~A. and {Finethy}, T. and {Fox}, R.~T. and {Garrison}, L.~H. and {Gibbons}, S.~L.~J. and {Goldstein}, D.~A. and {Gommers}, R. and {Greco}, J.~P. and {Greenfield}, P. and {Groener}, A.~M. and {Grollier}, F. and {Hagen}, A. and {Hirst}, P. and {Homeier}, D. and {Horton}, A.~J. and {Hosseinzadeh}, G. and {Hu}, L. and {Hunkeler}, J.~S. and {Ivezi{\'c}}, {\v{Z}}. and {Jain}, A. and {Jenness}, T. and {Kanarek}, G. and {Kendrew}, S. and {Kern}, N.~S. and {Kerzendorf}, W.~E. and {Khvalko}, A. and {King}, J. and {Kirkby}, D. and {Kulkarni}, A.~M. and {Kumar}, A. and {Lee}, A. and {Lenz}, D. and {Littlefair}, S.~P. and {Ma}, Z. and {Macleod}, D.~M. and {Mastropietro}, M. and {McCully}, C. and {Montagnac}, S. and {Morris}, B.~M. and {Mueller}, M. and {Mumford}, S.~J. and {Muna}, D. and {Murphy}, N.~A. and {Nelson}, S. and {Nguyen}, G.~H. and {Ninan}, J.~P. and {N{\"o}the}, M. and {Ogaz}, S. and {Oh}, S. and {Parejko}, J.~K. and {Parley}, N. and {Pascual}, S. and {Patil}, R. and {Patil}, A.~A. and {Plunkett}, A.~L. and {Prochaska}, J.~X. and {Rastogi}, T. and {Reddy Janga}, V. and {Sabater}, J. and {Sakurikar}, P. and {Seifert}, M. and {Sherbert}, L.~E. and {Sherwood-Taylor}, H. and {Shih}, A.~Y. and {Sick}, J. and {Silbiger}, M.~T. and {Singanamalla}, S. and {Singer}, L.~P. and {Sladen}, P.~H. and {Sooley}, K.~A. and {Sornarajah}, S. and {Streicher}, O. and {Teuben}, P. and {Thomas}, S.~W. and {Tremblay}, G.~R. and {Turner}, J.~E.~H. and {Terr{\'o}n}, V. and {van Kerkwijk}, M.~H. and {de la Vega}, A. and {Watkins}, L.~L. and {Weaver}, B.~A. and {Whitmore}, J.~B. and {Woillez}, J. and {Zabalza}, V. and {Astropy Contributors}},
        title = "{The Astropy Project: Building an Open-science Project and Status of the v2.0 Core Package}",
      journal = {\aj},
         year = 2018,
        month = sep,
       volume = {156},
       number = {3},
          eid = {123},
        pages = {123},
          doi = {10.3847/1538-3881/aabc4f},
archivePrefix = {arXiv},
       eprint = {1801.02634},
 primaryClass = {astro-ph.IM},
       adsurl = {https://ui.adsabs.harvard.edu/abs/2018AJ....156..123A}
}

@ARTICLE{Wang_2019,
       author = {{Wang}, Shu and {Chen}, Xiaodian},
        title = "{The Optical to Mid-infrared Extinction Law Based on the APOGEE, Gaia DR2, Pan-STARRS1, SDSS, APASS, 2MASS, and WISE Surveys}",
      journal = {\apj},
         year = 2019,
        month = jun,
       volume = {877},
       number = {2},
          eid = {116},
        pages = {116},
          doi = {10.3847/1538-4357/ab1c61},
archivePrefix = {arXiv},
       eprint = {1904.04575},
 primaryClass = {astro-ph.GA},
       adsurl = {https://ui.adsabs.harvard.edu/abs/2019ApJ...877..116W}
}

@ARTICLE{Schlegel_1998,
       author = {{Schlegel}, David J. and {Finkbeiner}, Douglas P. and {Davis}, Marc},
        title = "{Maps of Dust Infrared Emission for Use in Estimation of Reddening and Cosmic Microwave Background Radiation Foregrounds}",
      journal = {\apj},
         year = 1998,
        month = jun,
       volume = {500},
       number = {2},
        pages = {525-553},
          doi = {10.1086/305772},
archivePrefix = {arXiv},
       eprint = {astro-ph/9710327},
 primaryClass = {astro-ph},
       adsurl = {https://ui.adsabs.harvard.edu/abs/1998ApJ...500..525S}
}

@ARTICLE{Wilkinson_2004,
       author = {{Wilkinson}, Mark I. and {Kleyna}, Jan T. and {Evans}, N. Wyn and {Gilmore}, Gerard F. and {Irwin}, Michael J. and {Grebel}, Eva K.},
        title = "{Kinematically Cold Populations at Large Radii in the Draco and Ursa Minor Dwarf Spheroidal Galaxies}",
      journal = {\apjl},
         year = 2004,
        month = aug,
       volume = {611},
       number = {1},
        pages = {L21-L24},
          doi = {10.1086/423619},
archivePrefix = {arXiv},
       eprint = {astro-ph/0406520},
 primaryClass = {astro-ph},
       adsurl = {https://ui.adsabs.harvard.edu/abs/2004ApJ...611L..21W}
}

@ARTICLE{Walker_2008,
       author = {{Walker}, Matthew G. and {Mateo}, Mario and {Olszewski}, Edward W.},
        title = "{Systemic Proper Motions of Milky Way Satellites from Stellar Redshifts: The Carina, Fornax, Sculptor, and Sextans Dwarf Spheroidals}",
      journal = {\apjl},
         year = 2008,
        month = dec,
       volume = {688},
       number = {2},
        pages = {L75},
          doi = {10.1086/595586},
archivePrefix = {arXiv},
       eprint = {0810.1511},
 primaryClass = {astro-ph},
       adsurl = {https://ui.adsabs.harvard.edu/abs/2008ApJ...688L..75W}
}

@ARTICLE{Carrera_2002,
       author = {{Carrera}, Ricardo and {Aparicio}, Antonio and {Mart{\'\i}nez-Delgado}, David and {Alonso-Garc{\'\i}a}, Javier},
        title = "{The Star Formation History and Spatial Distribution of Stellar Populations in the Ursa Minor Dwarf Spheroidal Galaxy}",
      journal = {\aj},
         year = 2002,
        month = jun,
       volume = {123},
       number = {6},
        pages = {3199-3209},
          doi = {10.1086/340702},
archivePrefix = {arXiv},
       eprint = {astro-ph/0203300},
 primaryClass = {astro-ph},
       adsurl = {https://ui.adsabs.harvard.edu/abs/2002AJ....123.3199C}
}

@ARTICLE{Lee_2009,
       author = {{Lee}, Myung Gyoon and {Yuk}, In-Soo and {Park}, Hong Soo and {Harris}, Jason and {Zaritsky}, Dennis},
        title = "{Star Formation History and Chemical Evolution of the Sextans Dwarf Spheroidal Galaxy}",
      journal = {\apj},
         year = 2009,
        month = sep,
       volume = {703},
       number = {1},
        pages = {692-701},
          doi = {10.1088/0004-637X/703/1/692},
archivePrefix = {arXiv},
       eprint = {0907.5102},
 primaryClass = {astro-ph.CO},
       adsurl = {https://ui.adsabs.harvard.edu/abs/2009ApJ...703..692L}
}

@ARTICLE{Pietrzynski_2009,
       author = {{Pietrzy{\'n}ski}, Grzegorz and {Gieren}, Wolfgang and {Szewczyk}, Olaf and {Walker}, Alistair and {Rizzi}, Luca and {Bresolin}, Fabio and {Kudritzki}, Rolf-Peter and {Nalewajko}, Krzysztof and {Storm}, Jesper and {Dall'Ora}, Massimo and {Ivanov}, Valentin},
        title = "{The Araucaria Project: the Distance to the Sculptor Dwarf Spheroidal Galaxy from Infrared Photometry of RR Lyrae Stars}",
      journal = {\aj},
         year = 2008,
        month = jun,
       volume = {135},
       number = {6},
        pages = {1993-1997},
          doi = {10.1088/0004-6256/135/6/1993},
archivePrefix = {arXiv},
       eprint = {0804.0347},
 primaryClass = {astro-ph},
       adsurl = {https://ui.adsabs.harvard.edu/abs/2008AJ....135.1993P}
}

@ARTICLE{Bonanos_2004,
       author = {{Bonanos}, A.~Z. and {Stanek}, K.~Z. and {Szentgyorgyi}, A.~H. and {Sasselov}, D.~D. and {Bakos}, G. {\'A}.},
        title = "{The RR Lyrae Distance to the Draco Dwarf Spheroidal Galaxy}",
      journal = {\aj},
         year = 2004,
        month = feb,
       volume = {127},
       number = {2},
        pages = {861-867},
          doi = {10.1086/381073},
archivePrefix = {arXiv},
       eprint = {astro-ph/0310477},
 primaryClass = {astro-ph},
       adsurl = {https://ui.adsabs.harvard.edu/abs/2004AJ....127..861B}
}

@ARTICLE{irwin_1995,
       author = {{Irwin}, M. and {Hatzidimitriou}, D.},
        title = "{Structural parameters for the Galactic dwarf spheroidals}",
      journal = {\mnras},
         year = 1995,
        month = dec,
       volume = {277},
       number = {4},
        pages = {1354-1378},
          doi = {10.1093/mnras/277.4.1354},
       adsurl = {https://ui.adsabs.harvard.edu/abs/1995MNRAS.277.1354I}
}

@ARTICLE{munoz_2018,
       author = {{Mu{\~n}oz}, Ricardo R. and {C{\^o}t{\'e}}, Patrick and {Santana}, Felipe A. and {Geha}, Marla and {Simon}, Joshua D. and {Oyarz{\'u}n}, Grecco A. and {Stetson}, Peter B. and {Djorgovski}, S.~G.},
        title = "{A MegaCam Survey of Outer Halo Satellites. III. Photometric and Structural Parameters}",
      journal = {\apj},
         year = 2018,
        month = jun,
       volume = {860},
       number = {1},
          eid = {66},
        pages = {66},
          doi = {10.3847/1538-4357/aac16b},
archivePrefix = {arXiv},
       eprint = {1806.06891},
 primaryClass = {astro-ph.GA},
       adsurl = {https://ui.adsabs.harvard.edu/abs/2018ApJ...860...66M}
}

@ARTICLE{Pietrzynski_2009b,
       author = {{Pietrzy{\'n}ski}, Grzegorz and {G{\'o}rski}, Marek and {Gieren}, Wolfgang and {Ivanov}, Valentin D. and {Bresolin}, Fabio and {Kudritzki}, Rolf-Peter},
        title = "{The Araucaria Project. Infrared Tip of the Red Giant Branch Distances to the Carina and Fornax Dwarf Spheroidal Galaxies}",
      journal = {\aj},
         year = 2009,
        month = aug,
       volume = {138},
       number = {2},
        pages = {459-465},
          doi = {10.1088/0004-6256/138/2/459},
archivePrefix = {arXiv},
       eprint = {0906.0082},
 primaryClass = {astro-ph.GA},
       adsurl = {https://ui.adsabs.harvard.edu/abs/2009AJ....138..459P}
}

@ARTICLE{Koposov_2011,
       author = {{Koposov}, Sergey E. and {Gilmore}, G. and {Walker}, M.~G. and {Belokurov}, V. and {Evans}, N. Wyn and {Fellhauer}, M. and {Gieren}, W. and {Geisler}, D. and {Monaco}, L. and {Norris}, J.~E. and {Okamoto}, S. and {Pe{\~n}arrubia}, J. and {Wilkinson}, M. and {Wyse}, R.~F.~G. and {Zucker}, D.~B.},
        title = "{Accurate Stellar Kinematics at Faint Magnitudes: Application to the Bo{\"o}tes I Dwarf Spheroidal Galaxy}",
      journal = {\apj},
         year = 2011,
        month = aug,
       volume = {736},
       number = {2},
          eid = {146},
        pages = {146},
          doi = {10.1088/0004-637X/736/2/146},
archivePrefix = {arXiv},
       eprint = {1105.4102},
 primaryClass = {astro-ph.GA},
       adsurl = {https://ui.adsabs.harvard.edu/abs/2011ApJ...736..146K}
}

@ARTICLE{Martin_2008,
       author = {{Martin}, Nicolas F. and {de Jong}, Jelte T.~A. and {Rix}, Hans-Walter},
        title = "{A Comprehensive Maximum Likelihood Analysis of the Structural Properties of Faint Milky Way Satellites}",
      journal = {\apj},
         year = 2008,
        month = sep,
       volume = {684},
       number = {2},
        pages = {1075-1092},
          doi = {10.1086/590336},
archivePrefix = {arXiv},
       eprint = {0805.2945},
 primaryClass = {astro-ph},
       adsurl = {https://ui.adsabs.harvard.edu/abs/2008ApJ...684.1075M}
}

@ARTICLE{DallOra_2006,
       author = {{Dall'Ora}, Massimo and {Clementini}, Gisella and {Kinemuchi}, Karen and {Ripepi}, Vincenzo and {Marconi}, Marcella and {Di Fabrizio}, Luca and {Greco}, Claudia and {Rodgers}, Christopher T. and {Kuehn}, Charles and {Smith}, Horace A.},
        title = "{Variable Stars in the Newly Discovered Milky Way Satellite in Bootes}",
      journal = {\apjl},
         year = 2006,
        month = dec,
       volume = {653},
       number = {2},
        pages = {L109-L112},
          doi = {10.1086/510665},
archivePrefix = {arXiv},
       eprint = {astro-ph/0611285},
 primaryClass = {astro-ph},
       adsurl = {https://ui.adsabs.harvard.edu/abs/2006ApJ...653L.109D}
}

@ARTICLE{Cicuendez2018,
       author = {{Cicu{\'e}ndez}, L. and {Battaglia}, G.},
        title = "{Appearances can be deceiving: clear signs of accretion in the seemingly ordinary Sextans dSph}",
      journal = {\mnras},
         year = 2018,
        month = oct,
       volume = {480},
       number = {1},
        pages = {251-260},
          doi = {10.1093/mnras/sty1748},
archivePrefix = {arXiv},
       eprint = {1804.02336},
 primaryClass = {astro-ph.GA},
       adsurl = {https://ui.adsabs.harvard.edu/abs/2018MNRAS.480..251C}
}

@ARTICLE{Okamoto2017,
       author = {{Okamoto}, S. and {Arimoto}, N. and {Tolstoy}, E. and {Jablonka}, P. and {Irwin}, M.~J. and {Komiyama}, Y. and {Yamada}, Y. and {Onodera}, M.},
        title = "{Population gradient in the Sextans dSph: comprehensive mapping of a dwarf galaxy by Suprime-Cam}",
      journal = {\mnras},
         year = 2017,
        month = may,
       volume = {467},
       number = {1},
        pages = {208-217},
          doi = {10.1093/mnras/stx086},
archivePrefix = {arXiv},
       eprint = {1701.04422},
 primaryClass = {astro-ph.GA},
       adsurl = {https://ui.adsabs.harvard.edu/abs/2017MNRAS.467..208O}
}

@ARTICLE{Kleyna2004,
       author = {{Kleyna}, Jan T. and {Wilkinson}, Mark I. and {Evans}, N. Wyn and {Gilmore}, Gerard},
        title = "{A photometrically and kinematically distinct core in the Sextans dwarf spheroidal galaxy}",
      journal = {\mnras},
         year = 2004,
        month = nov,
       volume = {354},
       number = {4},
        pages = {L66-L72},
          doi = {10.1111/j.1365-2966.2004.08434.x},
archivePrefix = {arXiv},
       eprint = {astro-ph/0409066},
 primaryClass = {astro-ph},
       adsurl = {https://ui.adsabs.harvard.edu/abs/2004MNRAS.354L..66K}
}

@ARTICLE{Bettinelli_2018,
       author = {{Bettinelli}, M. and {Hidalgo}, S.~L. and {Cassisi}, S. and {Aparicio}, A. and {Piotto}, G.},
        title = "{The star formation history of the Sextans dwarf spheroidal galaxy: a true fossil of the pre-reionization era}",
      journal = {\mnras},
         year = 2018,
        month = may,
       volume = {476},
       number = {1},
        pages = {71-79},
          doi = {10.1093/mnras/sty226},
archivePrefix = {arXiv},
       eprint = {1801.08145},
 primaryClass = {astro-ph.GA},
       adsurl = {https://ui.adsabs.harvard.edu/abs/2018MNRAS.476...71B}
}

@ARTICLE{Weisz_2015,
       author = {{Weisz}, Daniel R. and {Dolphin}, Andrew E. and {Skillman}, Evan D. and {Holtzman}, Jon and {Gilbert}, Karoline M. and {Dalcanton}, Julianne J. and {Williams}, Benjamin F.},
        title = "{The Star Formation Histories of Local Group Dwarf Galaxies. III. Characterizing Quenching in Low-mass Galaxies}",
      journal = {\apj},
         year = 2015,
        month = may,
       volume = {804},
       number = {2},
          eid = {136},
        pages = {136},
          doi = {10.1088/0004-637X/804/2/136},
archivePrefix = {arXiv},
       eprint = {1503.05195},
 primaryClass = {astro-ph.GA},
       adsurl = {https://ui.adsabs.harvard.edu/abs/2015ApJ...804..136W}
}

@ARTICLE{Queiroz2020,
       author = {{Queiroz}, A.~B.~A. and {Anders}, F. and {Chiappini}, C. and {Khalatyan}, A. and {Santiago}, B.~X. and {Steinmetz}, M. and {Valentini}, M. and {Miglio}, A. and {Bossini}, D. and {Barbuy}, B. and {Minchev}, I. and {Minniti}, D. and {Garc{\'\i}a Hern{\'a}ndez}, D.~A. and {Schultheis}, M. and {Beaton}, R.~L. and {Beers}, T.~C. and {Bizyaev}, D. and {Brownstein}, J.~R. and {Cunha}, K. and {Fern{\'a}ndez-Trincado}, J.~G. and {Frinchaboy}, P.~M. and {Lane}, R.~R. and {Majewski}, S.~R. and {Nataf}, D. and {Nitschelm}, C. and {Pan}, K. and {Roman-Lopes}, A. and {Sobeck}, J.~S. and {Stringfellow}, G. and {Zamora}, O.},
        title = "{From the bulge to the outer disc: StarHorse stellar parameters, distances, and extinctions for stars in APOGEE DR16 and other spectroscopic surveys}",
      journal = {\aap},
         year = 2020,
        month = jun,
       volume = {638},
          eid = {A76},
        pages = {A76},
          doi = {10.1051/0004-6361/201937364},
archivePrefix = {arXiv},
       eprint = {1912.09778},
 primaryClass = {astro-ph.GA},
       adsurl = {https://ui.adsabs.harvard.edu/abs/2020A&A...638A..76Q}
}

@ARTICLE{Anders2019,
       author = {{Anders}, F. and {Khalatyan}, A. and {Chiappini}, C. and {Queiroz}, A.~B. and {Santiago}, B.~X. and {Jordi}, C. and {Girardi}, L. and {Brown}, A.~G.~A. and {Matijevi{\v{c}}}, G. and {Monari}, G. and {Cantat-Gaudin}, T. and {Weiler}, M. and {Khan}, S. and {Miglio}, A. and {Carrillo}, I. and {Romero-G{\'o}mez}, M. and {Minchev}, I. and {de Jong}, R.~S. and {Antoja}, T. and {Ramos}, P. and {Steinmetz}, M. and {Enke}, H.},
        title = "{Photo-astrometric distances, extinctions, and astrophysical parameters for Gaia DR2 stars brighter than G = 18}",
      journal = {\aap},
         year = 2019,
        month = aug,
       volume = {628},
          eid = {A94},
        pages = {A94},
          doi = {10.1051/0004-6361/201935765},
archivePrefix = {arXiv},
       eprint = {1904.11302},
 primaryClass = {astro-ph.GA},
       adsurl = {https://ui.adsabs.harvard.edu/abs/2019A&A...628A..94A}
}

@ARTICLE{Queiroz2018,
       author = {{Queiroz}, A.~B.~A. and {Anders}, F. and {Santiago}, B.~X. and {Chiappini}, C. and {Steinmetz}, M. and {Dal Ponte}, M. and {Stassun}, K.~G. and {da Costa}, L.~N. and {Maia}, M.~A.~G. and {Crestani}, J. and {Beers}, T.~C. and {Fern{\'a}ndez-Trincado}, J.~G. and {Garc{\'\i}a-Hern{\'a}ndez}, D.~A. and {Roman-Lopes}, A. and {Zamora}, O.},
        title = "{StarHorse: a Bayesian tool for determining stellar masses, ages, distances, and extinctions for field stars}",
      journal = {\mnras},
         year = 2018,
        month = may,
       volume = {476},
       number = {2},
        pages = {2556-2583},
          doi = {10.1093/mnras/sty330},
archivePrefix = {arXiv},
       eprint = {1710.09970},
 primaryClass = {astro-ph.IM},
       adsurl = {https://ui.adsabs.harvard.edu/abs/2018MNRAS.476.2556Q}
}

@ARTICLE{Santiago2016,
       author = {{Santiago}, Bas{\'\i}lio X. and {Brauer}, Doroth{\'e}e E. and {Anders}, Friedrich and {Chiappini}, Cristina and {Queiroz}, Anna B. and {Girardi}, L{\'e}o and {Rocha-Pinto}, Helio J. and {Balbinot}, Eduardo and {da Costa}, Luiz N. and {Maia}, Marcio A.~G. and {Schultheis}, Mathias and {Steinmetz}, Matthias and {Miglio}, Andrea and {Montalb{\'a}n}, Josefina and {Schneider}, Donald P. and {Beers}, Timothy C. and {Frinchaboy}, Peter M. and {Lee}, Young Sun and {Zasowski}, Gail},
        title = "{Spectro-photometric distances to stars: A general purpose Bayesian approach}",
      journal = {\aap},
         year = 2016,
        month = jan,
       volume = {585},
          eid = {A42},
        pages = {A42},
          doi = {10.1051/0004-6361/201323177},
archivePrefix = {arXiv},
       eprint = {1501.05500},
 primaryClass = {astro-ph.IM},
       adsurl = {https://ui.adsabs.harvard.edu/abs/2016A&A...585A..42S}
}

@ARTICLE{gravity2018,
       author = {{Gravity Collaboration} and {Abuter}, R. and {Amorim}, A. and {Baub{\"o}ck}, M. and {Berger}, J.~P. and {Bonnet}, H. and {Brandner}, W. and {Cl{\'e}net}, Y. and {Coud{\'e} Du Foresto}, V. and {de Zeeuw}, P.~T. and {Deen}, C. and {Dexter}, J. and {Duvert}, G. and {Eckart}, A. and {Eisenhauer}, F. and {F{\"o}rster Schreiber}, N.~M. and {Garcia}, P. and {Gao}, F. and {Gendron}, E. and {Genzel}, R. and {Gillessen}, S. and {Guajardo}, P. and {Habibi}, M. and {Haubois}, X. and {Henning}, Th. and {Hippler}, S. and {Horrobin}, M. and {Huber}, A. and {Jim{\'e}nez-Rosales}, A. and {Jocou}, L. and {Kervella}, P. and {Lacour}, S. and {Lapeyr{\`e}re}, V. and {Lazareff}, B. and {Le Bouquin}, J. -B. and {L{\'e}na}, P. and {Lippa}, M. and {Ott}, T. and {Panduro}, J. and {Paumard}, T. and {Perraut}, K. and {Perrin}, G. and {Pfuhl}, O. and {Plewa}, P.~M. and {Rabien}, S. and {Rodr{\'\i}guez-Coira}, G. and {Rousset}, G. and {Sternberg}, A. and {Straub}, O. and {Straubmeier}, C. and {Sturm}, E. and {Tacconi}, L.~J. and {Vincent}, F. and {von Fellenberg}, S. and {Waisberg}, I. and {Widmann}, F. and {Wieprecht}, E. and {Wiezorrek}, E. and {Woillez}, J. and {Yazici}, S.},
        title = "{Detection of orbital motions near the last stable circular orbit of the massive black hole SgrA*}",
      journal = {\aap},
         year = 2018,
        month = oct,
       volume = {618},
          eid = {L10},
        pages = {L10},
          doi = {10.1051/0004-6361/201834294},
archivePrefix = {arXiv},
       eprint = {1810.12641},
 primaryClass = {astro-ph.GA},
       adsurl = {https://ui.adsabs.harvard.edu/abs/2018A&A...618L..10G}
}

@ARTICLE{schonrich2010,
       author = {{Sch{\"o}nrich}, Ralph and {Binney}, James and {Dehnen}, Walter},
        title = "{Local kinematics and the local standard of rest}",
      journal = {\mnras},
         year = 2010,
        month = apr,
       volume = {403},
       number = {4},
        pages = {1829-1833},
          doi = {10.1111/j.1365-2966.2010.16253.x},
archivePrefix = {arXiv},
       eprint = {0912.3693},
 primaryClass = {astro-ph.GA},
       adsurl = {https://ui.adsabs.harvard.edu/abs/2010MNRAS.403.1829S}
}

@ARTICLE{schonrich2012,
       author = {{Sch{\"o}nrich}, Ralph},
        title = "{Galactic rotation and solar motion from stellar kinematics}",
      journal = {\mnras},
         year = 2012,
        month = nov,
       volume = {427},
       number = {1},
        pages = {274-287},
          doi = {10.1111/j.1365-2966.2012.21631.x},
archivePrefix = {arXiv},
       eprint = {1207.3079},
 primaryClass = {astro-ph.GA},
       adsurl = {https://ui.adsabs.harvard.edu/abs/2012MNRAS.427..274S}
}

@ARTICLE{hayes2018c,
       author = {{Hayes}, Christian R. and {Law}, David R. and {Majewski}, Steven R.},
        title = "{Constraining the Solar Galactic Reflex Velocity using Gaia Observations of the Sagittarius Stream}",
      journal = {\apjl},
         year = 2018,
        month = nov,
       volume = {867},
       number = {2},
          eid = {L20},
        pages = {L20},
          doi = {10.3847/2041-8213/aae9dd},
archivePrefix = {arXiv},
       eprint = {1809.07654},
 primaryClass = {astro-ph.GA},
       adsurl = {https://ui.adsabs.harvard.edu/abs/2018ApJ...867L..20H}
}

@ARTICLE{js87,
       author = {{Johnson}, Dean R.~H. and {Soderblom}, David R.},
        title = "{Calculating Galactic Space Velocities and Their Uncertainties, with an Application to the Ursa Major Group}",
      journal = {\aj},
         year = 1987,
        month = apr,
       volume = {93},
        pages = {864},
          doi = {10.1086/114370},
       adsurl = {https://ui.adsabs.harvard.edu/abs/1987AJ.....93..864J}
}

@ARTICLE{Armandroff_1995,
       author = {{Armandroff}, Taft E. and {Olszewski}, Edward W. and {Pryor}, Carlton},
        title = "{The Mass-To-Light Ratios of the Draco and Ursa Minor Dwarf Spheroidal Galaxies.I. Radial Velocities from Multifiber Spectroscopy}",
      journal = {\aj},
         year = 1995,
        month = nov,
       volume = {110},
        pages = {2131},
          doi = {10.1086/117675},
       adsurl = {https://ui.adsabs.harvard.edu/abs/1995AJ....110.2131A}
}

@ARTICLE{Mateo_1991,
       author = {{Mateo}, Mario and {Olszewski}, Edward and {Welch}, Douglas L. and {Fischer}, Philippe and {Kunkel}, William},
        title = "{A Kinematic Study of the Fornax Dwarf Spheroid Galaxy}",
      journal = {\aj},
         year = 1991,
        month = sep,
       volume = {102},
        pages = {914},
          doi = {10.1086/115923},
       adsurl = {https://ui.adsabs.harvard.edu/abs/1991AJ....102..914M}
}

@ARTICLE{Walker_2009,
       author = {{Walker}, Matthew G. and {Mateo}, Mario and {Olszewski}, Edward W.},
        title = "{Stellar Velocities in the Carina, Fornax, Sculptor, and Sextans dSph Galaxies: Data From the Magellan/MMFS Survey}",
      journal = {\aj},
         year = 2009,
        month = feb,
       volume = {137},
       number = {2},
        pages = {3100-3108},
          doi = {10.1088/0004-6256/137/2/3100},
archivePrefix = {arXiv},
       eprint = {0811.0118},
 primaryClass = {astro-ph},
       adsurl = {https://ui.adsabs.harvard.edu/abs/2009AJ....137.3100W}
}

@ARTICLE{Walker_2015,
       author = {{Walker}, Matthew G. and {Olszewski}, Edward W. and {Mateo}, Mario},
        title = "{Bayesian analysis of resolved stellar spectra: application to MMT/Hectochelle observations of the Draco dwarf spheroidal}",
      journal = {\mnras},
         year = 2015,
        month = apr,
       volume = {448},
       number = {3},
        pages = {2717-2732},
          doi = {10.1093/mnras/stv099},
archivePrefix = {arXiv},
       eprint = {1503.02589},
 primaryClass = {astro-ph.GA},
       adsurl = {https://ui.adsabs.harvard.edu/abs/2015MNRAS.448.2717W}
}

@ARTICLE{McConnachie_2012,
       author = {{McConnachie}, Alan W.},
        title = "{The Observed Properties of Dwarf Galaxies in and around the Local Group}",
      journal = {\aj},
         year = 2012,
        month = jul,
       volume = {144},
       number = {1},
          eid = {4},
        pages = {4},
          doi = {10.1088/0004-6256/144/1/4},
archivePrefix = {arXiv},
       eprint = {1204.1562},
 primaryClass = {astro-ph.CO},
       adsurl = {https://ui.adsabs.harvard.edu/abs/2012AJ....144....4M}
}

@ARTICLE{Munoz_2005,
       author = {{Mu{\~n}oz}, Ricardo R. and {Frinchaboy}, Peter M. and {Majewski}, Steven R. and {Kuhn}, Jeffrey R. and {Chou}, Mei-Yin and {Palma}, Christopher and {Sohn}, Sangmo Tony and {Patterson}, Richard J. and {Siegel}, Michael H.},
        title = "{Exploring Halo Substructure with Giant Stars: The Velocity Dispersion Profiles of the Ursa Minor and Draco Dwarf Spheroidal Galaxies at Large Angular Separations}",
      journal = {\apjl},
         year = 2005,
        month = oct,
       volume = {631},
       number = {2},
        pages = {L137-L141},
          doi = {10.1086/497396},
archivePrefix = {arXiv},
       eprint = {astro-ph/0504035},
 primaryClass = {astro-ph},
       adsurl = {https://ui.adsabs.harvard.edu/abs/2005ApJ...631L.137M}
}

@ARTICLE{Palma_2003,
       author = {{Palma}, Christopher and {Majewski}, Steven R. and {Siegel}, Michael H. and {Patterson}, Richard J. and {Ostheimer}, James C. and {Link}, Robert},
        title = "{Exploring Halo Substructure with Giant Stars. IV. The Extended Structure of the Ursa Minor Dwarf Spheroidal Galaxy}",
      journal = {\aj},
         year = 2003,
        month = mar,
       volume = {125},
       number = {3},
        pages = {1352-1372},
          doi = {10.1086/367594},
archivePrefix = {arXiv},
       eprint = {astro-ph/0205194},
 primaryClass = {astro-ph},
       adsurl = {https://ui.adsabs.harvard.edu/abs/2003AJ....125.1352P}
}

@ARTICLE{Westfall_2006,
       author = {{Westfall}, Kyle B. and {Majewski}, Steven R. and {Ostheimer}, James C. and {Frinchaboy}, Peter M. and {Kunkel}, William E. and {Patterson}, Richard J. and {Link}, Robert},
        title = "{Exploring Halo Substructure with Giant Stars. VIII. The Extended Structure of the Sculptor Dwarf Spheroidal Galaxy}",
      journal = {\aj},
         year = 2006,
        month = jan,
       volume = {131},
       number = {1},
        pages = {375-406},
          doi = {10.1086/496975},
archivePrefix = {arXiv},
       eprint = {astro-ph/0508091},
 primaryClass = {astro-ph},
       adsurl = {https://ui.adsabs.harvard.edu/abs/2006AJ....131..375W}
}

@ARTICLE{majewski_2017,
   author = {{Majewski}, S.~R. and {Schiavon}, R.~P. and {Frinchaboy}, P.~M. and 
	{Allende Prieto}, C. and {Barkhouser}, R. and {Bizyaev}, D. and 
	{Blank}, B. and {Brunner}, S. and {Burton}, A. and {Carrera}, R. and 
	{Chojnowski}, S.~D. and {Cunha}, K. and {Epstein}, C. and {Fitzgerald}, G. and 
	{Garc{\'{\i}}a P{\'e}rez}, A.~E. and {Hearty}, F.~R. and {Henderson}, C. and 
	{Holtzman}, J.~A. and {Johnson}, J.~A. and {Lam}, C.~R. and 
	{Lawler}, J.~E. and {Maseman}, P. and {M{\'e}sz{\'a}ros}, S. and 
	{Nelson}, M. and {Nguyen}, D.~C. and {Nidever}, D.~L. and {Pinsonneault}, M. and 
	{Shetrone}, M. and {Smee}, S. and {Smith}, V.~V. and {Stolberg}, T. and 
	{Skrutskie}, M.~F. and {Walker}, E. and {Wilson}, J.~C. and 
	{Zasowski}, G. and {Anders}, F. and {Basu}, S. and {Beland}, S. and 
	{Blanton}, M.~R. and {Bovy}, J. and {Brownstein}, J.~R. and 
	{Carlberg}, J. and {Chaplin}, W. and {Chiappini}, C. and {Eisenstein}, D.~J. and 
	{Elsworth}, Y. and {Feuillet}, D. and {Fleming}, S.~W. and {Galbraith-Frew}, J. and 
	{Garc{\'{\i}}a}, R.~A. and {Garc{\'{\i}}a-Hern{\'a}ndez}, D.~A. and 
	{Gillespie}, B.~A. and {Girardi}, L. and {Gunn}, J.~E. and {Hasselquist}, S. and 
	{Hayden}, M.~R. and {Hekker}, S. and {Ivans}, I. and {Kinemuchi}, K. and 
	{Klaene}, M. and {Mahadevan}, S. and {Mathur}, S. and {Mosser}, B. and 
	{Muna}, D. and {Munn}, J.~A. and {Nichol}, R.~C. and {O'Connell}, R.~W. and 
	{Parejko}, J.~K. and {Robin}, A.~C. and {Rocha-Pinto}, H. and 
	{Schultheis}, M. and {Serenelli}, A.~M. and {Shane}, N. and 
	{Silva Aguirre}, V. and {Sobeck}, J.~S. and {Thompson}, B. and 
	{Troup}, N.~W. and {Weinberg}, D.~H. and {Zamora}, O.},
    title = "{The Apache Point Observatory Galactic Evolution Experiment (APOGEE)}",
  journal = {\aj},
archivePrefix = "arXiv",
   eprint = {1509.05420},
 primaryClass = "astro-ph.IM",
     year = 2017,
    month = sep,
   volume = 154,
      eid = {94},
    pages = {94},
      doi = {10.3847/1538-3881/aa784d},
   adsurl = {http://adsabs.harvard.edu/abs/2017AJ....154...94M}
}

@ARTICLE{majewski_2000,
       author = {{Majewski}, Steven R. and {Ostheimer}, James C. and {Kunkel}, William E.
        and {Patterson}, Richard J.},
        title = "{Exploring Halo Substructure with Giant Stars. I. Survey Description and Calibration of the Photometric Search Technique}",
      journal = {\aj},
         year = 2000,
        month = Nov,
       volume = {120},
        pages = {2550-2568},
          doi = {10.1086/316836},
archivePrefix = {arXiv},
       eprint = {astro-ph/0006411},
 primaryClass = {astro-ph},
       adsurl = {https://ui.adsabs.harvard.edu/\#abs/2000AJ....120.2550M}
}

@ARTICLE{blanton_2017,
   author = {{Blanton}, M.~R. and {Bershady}, M.~A. and {Abolfathi}, B. and 
	{Albareti}, F.~D. and {Allende Prieto}, C. and {Almeida}, A. and 
	{Alonso-Garc{\'{\i}}a}, J. and {Anders}, F. and {Anderson}, S.~F. and 
	{Andrews}, B. and et al.},
    title = "{Sloan Digital Sky Survey IV: Mapping the Milky Way, Nearby Galaxies, and the Distant Universe}",
  journal = {\aj},
archivePrefix = "arXiv",
   eprint = {1703.00052},
     year = 2017,
    month = jul,
   volume = 154,
      eid = {28},
    pages = {28},
      doi = {10.3847/1538-3881/aa7567},
   adsurl = {http://adsabs.harvard.edu/abs/2017AJ....154...28B}
}

@ARTICLE{WISE, author = {{Wright}, E.~L. and {Eisenhardt}, P.~R.~M. and {Mainzer}, A.~K. and {Ressler}, M.~E. and {Cutri}, R.~M. and {Jarrett}, T. and {Kirkpatrick}, J.~D. and {Padgett}, D. and {McMillan}, R.~S. and {Skrutskie}, M. and {Stanford}, S.~A. and {Cohen}, M. and {Walker}, R.~G. and {Mather}, J.~C. and {Leisawitz}, D. and {Gautier}, III, T.~N. and {McLean}, I. and {Benford}, D. and {Lonsdale}, C.~J. and {Blain}, A. and {Mendez}, B. and {Irace}, W.~R. and {Duval}, V. and {Liu}, F. and {Royer}, D. and {Heinrichsen}, I. and {Howard}, J. and {Shannon}, M. and {Kendall}, M. and {Walsh}, A.~L. and {Larsen}, M. and {Cardon}, J.~G. and {Schick}, S. and {Schwalm}, M. and {Abid}, M. and {Fabinsky}, B. and {Naes}, L. and {Tsai}, C.-W.}, title = "{The Wide-field Infrared Survey Explorer (WISE): Mission Description and Initial On-orbit Performance}", journal = {\aj}, archivePrefix = "arXiv", eprint = {1008.0031}, primaryClass = "astro-ph.IM", year = 2010, month = dec, volume = 140, eid = {1868}, pages = {1868-1881}, doi = {10.1088/0004-6256/140/6/1868}, adsurl = {http://adsabs.harvard.edu/abs/2010AJ....140.1868W} }

@ARTICLE{Zasowski_2017,
       author = {{Zasowski}, G. and {Cohen}, R.~E. and {Chojnowski}, S.~D. and {Santana}, F. and {Oelkers}, R.~J. and {Andrews}, B. and {Beaton}, R.~L. and {Bender}, C. and {Bird}, J.~C. and {Bovy}, J. and {Carlberg}, J.~K. and {Covey}, K. and {Cunha}, K. and {Dell'Agli}, F. and {Fleming}, Scott W. and {Frinchaboy}, P.~M. and {Garc{\'\i}a-Hern{\'a}ndez}, D.~A. and {Harding}, P. and {Holtzman}, J. and {Johnson}, J.~A. and {Kollmeier}, J.~A. and {Majewski}, S.~R. and {M{\'e}sz{\'a}ros}, Sz. and {Munn}, J. and {Mu{\~n}oz}, R.~R. and {Ness}, M.~K. and {Nidever}, D.~L. and {Poleski}, R. and {Rom{\'a}n-Z{\'u}{\~n}iga}, C. and {Shetrone}, M. and {Simon}, J.~D. and {Smith}, V.~V. and {Sobeck}, J.~S. and {Stringfellow}, G.~S. and {Szigeti{\'a}ros}, L. and {Tayar}, J. and {Troup}, N.},
        title = "{Target Selection for the SDSS-IV APOGEE-2 Survey}",
      journal = {\aj},
         year = 2017,
        month = nov,
       volume = {154},
       number = {5},
          eid = {198},
        pages = {198},
          doi = {10.3847/1538-3881/aa8df9},
archivePrefix = {arXiv},
       eprint = {1708.00155},
 primaryClass = {astro-ph.GA},
       adsurl = {https://ui.adsabs.harvard.edu/abs/2017AJ....154..198Z}
}

@ARTICLE{vizier2000,
       author = {{Ochsenbein}, F. and {Bauer}, P. and {Marcout}, J.},
        title = "{The VizieR database of astronomical catalogues}",
      journal = {\aaps},
         year = "2000",
        month = "Apr",
       volume = {143},
        pages = {23-32},
          doi = {10.1051/aas:2000169},
archivePrefix = {arXiv},
       eprint = {astro-ph/0002122},
 primaryClass = {astro-ph},
       adsurl = {https://ui.adsabs.harvard.edu/abs/2000A&AS..143...23O}
}

@ARTICLE{simbad_2000,
       author = {{Wenger}, M. and {Ochsenbein}, F. and {Egret}, D. and {Dubois}, P. and
         {Bonnarel}, F. and {Borde}, S. and {Genova}, F. and {Jasniewicz}, G. and
         {Lalo{\"e}}, S. and {Lesteven}, S. and {Monier}, R.},
        title = "{The SIMBAD astronomical database. The CDS reference database for astronomical objects}",
      journal = {\aaps},
         year = "2000",
        month = "Apr",
       volume = {143},
        pages = {9-22},
          doi = {10.1051/aas:2000332},
archivePrefix = {arXiv},
       eprint = {astro-ph/0002110},
 primaryClass = {astro-ph},
       adsurl = {https://ui.adsabs.harvard.edu/abs/2000A&AS..143....9W}
}

@ARTICLE{Gratton_2000,
       author = {{Gratton}, R.~G. and {Sneden}, C. and {Carretta}, E. and {Bragaglia}, A.},
        title = "{Mixing along the red giant branch in metal-poor field stars}",
      journal = {\aap},
         year = 2000,
        month = feb,
       volume = {354},
        pages = {169-187},
       adsurl = {https://ui.adsabs.harvard.edu/abs/2000A&A...354..169G}
}

@ARTICLE{Pace_2022,
       author = {{Pace}, Andrew B. and {Erkal}, Denis and {Li}, Ting S.},
        title = "{Proper Motions, Orbits, and Tidal Influences of Milky Way Dwarf Spheroidal Galaxies}",
      journal = {\apj},
         year = 2022,
        month = dec,
       volume = {940},
       number = {2},
          eid = {136},
        pages = {136},
          doi = {10.3847/1538-4357/ac997b},
archivePrefix = {arXiv},
       eprint = {2205.05699},
 primaryClass = {astro-ph.GA},
       adsurl = {https://ui.adsabs.harvard.edu/abs/2022ApJ...940..136P}
}

@ARTICLE{Kirby_2019,
       author = {{Kirby}, Evan N. and {Xie}, Justin L. and {Guo}, Rachel and {de los Reyes}, Mithi A.~C. and {Bergemann}, Maria and {Kovalev}, Mikhail and {Shen}, Ken J. and {Piro}, Anthony L. and {McWilliam}, Andrew},
        title = "{Evidence for Sub-Chandrasekhar Type Ia Supernovae from Stellar Abundances in Dwarf Galaxies}",
      journal = {\apj},
         year = 2019,
        month = aug,
       volume = {881},
       number = {1},
          eid = {45},
        pages = {45},
          doi = {10.3847/1538-4357/ab2c02},
archivePrefix = {arXiv},
       eprint = {1906.10126},
 primaryClass = {astro-ph.SR},
       adsurl = {https://ui.adsabs.harvard.edu/abs/2019ApJ...881...45K}
}

@ARTICLE{de_los_Reyes_2022,
       author = {{de los Reyes}, Mithi A.~C. and {Kirby}, Evan N. and {Ji}, Alexander P. and {Nu{\~n}ez}, Evan H.},
        title = "{Simultaneous Constraints on the Star Formation History and Nucleosynthesis of Sculptor dSph}",
      journal = {\apj},
         year = 2022,
        month = jan,
       volume = {925},
       number = {1},
          eid = {66},
        pages = {66},
          doi = {10.3847/1538-4357/ac332b},
archivePrefix = {arXiv},
       eprint = {2110.01690},
 primaryClass = {astro-ph.GA},
       adsurl = {https://ui.adsabs.harvard.edu/abs/2022ApJ...925...66D}
}

@ARTICLE{Geisler_2021,
       author = {{Geisler}, D. and {Villanova}, S. and {O'Connell}, J.~E. and {Cohen}, R.~E. and {Moni Bidin}, C. and {Fern{\'a}ndez-Trincado}, J.~G. and {Mu{\~n}oz}, C. and {Minniti}, D. and {Zoccali}, M. and {Rojas-Arriagada}, A. and {Contreras Ramos}, R. and {Catelan}, M. and {Mauro}, F. and {Cort{\'e}s}, C. and {Ferreira Lopes}, C.~E. and {Arentsen}, A. and {Starkenburg}, E. and {Martin}, N.~F. and {Tang}, B. and {Parisi}, C. and {Alonso-Garc{\'\i}a}, J. and {Gran}, F. and {Cunha}, K. and {Smith}, V. and {Majewski}, S.~R. and {J{\"o}nsson}, H. and {Garc{\'\i}a-Hern{\'a}ndez}, D.~A. and {Horta}, D. and {M{\'e}sz{\'a}ros}, S. and {Monaco}, L. and {Monachesi}, A. and {Mu{\~n}oz}, R.~R. and {Brownstein}, J. and {Beers}, T.~C. and {Lane}, R.~R. and {Barbuy}, B. and {Sobeck}, J. and {Henao}, L. and {Gonz{\'a}lez-D{\'\i}az}, D. and {Miranda}, R.~E. and {Reinarz}, Y. and {Santander}, T.~A.},
        title = "{CAPOS: The bulge Cluster APOgee Survey. I. Overview and initial ASPCAP results}",
      journal = {\aap},
         year = 2021,
        month = aug,
       volume = {652},
          eid = {A157},
        pages = {A157},
          doi = {10.1051/0004-6361/202140436},
archivePrefix = {arXiv},
       eprint = {2106.00024},
 primaryClass = {astro-ph.GA},
       adsurl = {https://ui.adsabs.harvard.edu/abs/2021A&A...652A.157G}
}

@ARTICLE{Lucchesi_2020,
       author = {{Lucchesi}, R. and {Lardo}, C. and {Primas}, F. and {Jablonka}, P. and {North}, P. and {Battaglia}, G. and {Starkenburg}, E. and {Hill}, V. and {Irwin}, M. and {Francois}, P. and {Shetrone}, M. and {Tolstoy}, E. and {Venn}, K.},
        title = "{Homogeneity in the early chemical evolution of the Sextans dwarf spheroidal galaxy}",
      journal = {\aap},
         year = 2020,
        month = dec,
       volume = {644},
          eid = {A75},
        pages = {A75},
          doi = {10.1051/0004-6361/202037534},
archivePrefix = {arXiv},
       eprint = {2001.11033},
 primaryClass = {astro-ph.GA},
       adsurl = {https://ui.adsabs.harvard.edu/abs/2020A&A...644A..75L}
}

@ARTICLE{Starkenburg_2013,
       author = {{Starkenburg}, E. and {Hill}, V. and {Tolstoy}, E. and {Fran{\c{c}}ois}, P. and {Irwin}, M.~J. and {Boschman}, L. and {Venn}, K.~A. and {de Boer}, T.~J.~L. and {Lemasle}, B. and {Jablonka}, P. and {Battaglia}, G. and {Groot}, P. and {Kaper}, L.},
        title = "{The extremely low-metallicity tail of the Sculptor dwarf spheroidal galaxy}",
      journal = {\aap},
         year = 2013,
        month = jan,
       volume = {549},
          eid = {A88},
        pages = {A88},
          doi = {10.1051/0004-6361/201220349},
archivePrefix = {arXiv},
       eprint = {1211.4592},
 primaryClass = {astro-ph.GA},
       adsurl = {https://ui.adsabs.harvard.edu/abs/2013A&A...549A..88S}
}

@ARTICLE{Cohen_2009,
       author = {{Cohen}, Judith G. and {Huang}, Wenjin},
        title = "{The Chemical Evolution of the Draco Dwarf Spheroidal Galaxy}",
      journal = {\apj},
         year = 2009,
        month = aug,
       volume = {701},
       number = {2},
        pages = {1053-1075},
          doi = {10.1088/0004-637X/701/2/1053},
archivePrefix = {arXiv},
       eprint = {0906.1006},
 primaryClass = {astro-ph.GA},
       adsurl = {https://ui.adsabs.harvard.edu/abs/2009ApJ...701.1053C}
}

@ARTICLE{Shetrone_2003,
       author = {{Shetrone}, Matthew and {Venn}, Kim A. and {Tolstoy}, Eline and {Primas}, Francesca and {Hill}, Vanessa and {Kaufer}, Andreas},
        title = "{VLT/UVES Abundances in Four Nearby Dwarf Spheroidal Galaxies. I. Nucleosynthesis and Abundance Ratios}",
      journal = {\aj},
         year = 2003,
        month = feb,
       volume = {125},
       number = {2},
        pages = {684-706},
          doi = {10.1086/345966},
archivePrefix = {arXiv},
       eprint = {astro-ph/0211167},
 primaryClass = {astro-ph},
       adsurl = {https://ui.adsabs.harvard.edu/abs/2003AJ....125..684S}
}

@ARTICLE{Shetrone_2001,
       author = {{Shetrone}, Matthew D. and {C{\^o}t{\'e}}, Patrick and {Sargent}, W.~L.~W.},
        title = "{Abundance Patterns in the Draco, Sextans, and Ursa Minor Dwarf Spheroidal Galaxies}",
      journal = {\apj},
         year = 2001,
        month = feb,
       volume = {548},
       number = {2},
        pages = {592-608},
          doi = {10.1086/319022},
archivePrefix = {arXiv},
       eprint = {astro-ph/0009505},
 primaryClass = {astro-ph},
       adsurl = {https://ui.adsabs.harvard.edu/abs/2001ApJ...548..592S}
}

@ARTICLE{Frebel_2016,
       author = {{Frebel}, Anna and {Norris}, John E. and {Gilmore}, Gerard and {Wyse}, Rosemary F.~G.},
        title = "{The Chemical Evolution of the Bootes I Ultra-faint Dwarf Galaxy}",
      journal = {\apj},
         year = 2016,
        month = aug,
       volume = {826},
       number = {2},
          eid = {110},
        pages = {110},
          doi = {10.3847/0004-637X/826/2/110},
archivePrefix = {arXiv},
       eprint = {1605.05732},
 primaryClass = {astro-ph.GA},
       adsurl = {https://ui.adsabs.harvard.edu/abs/2016ApJ...826..110F}
}

@ARTICLE{Feltzing_2009,
       author = {{Feltzing}, S. and {Eriksson}, K. and {Kleyna}, J. and {Wilkinson}, M.~I.},
        title = "{Evidence of enrichment by individual SN from elemental abundance ratios in the very metal-poor dSph galaxy Bo{\"o}tes I}",
      journal = {\aap},
         year = 2009,
        month = dec,
       volume = {508},
       number = {1},
        pages = {L1-L4},
          doi = {10.1051/0004-6361/200912833},
archivePrefix = {arXiv},
       eprint = {0910.1557},
 primaryClass = {astro-ph.GA},
       adsurl = {https://ui.adsabs.harvard.edu/abs/2009A&A...508L...1F}
}

@ARTICLE{Gilmore_2013,
       author = {{Gilmore}, Gerard and {Norris}, John E. and {Monaco}, Lorenzo and {Yong}, David and {Wyse}, Rosemary F.~G. and {Geisler}, D.},
        title = "{Elemental Abundances and their Implications for the Chemical Enrichment of the Bo{\"o}tes I Ultrafaint Galaxy}",
      journal = {\apj},
         year = 2013,
        month = jan,
       volume = {763},
       number = {1},
          eid = {61},
        pages = {61},
          doi = {10.1088/0004-637X/763/1/61},
archivePrefix = {arXiv},
       eprint = {1212.0598},
 primaryClass = {astro-ph.GA},
       adsurl = {https://ui.adsabs.harvard.edu/abs/2013ApJ...763...61G}
}

@ARTICLE{Norris_2010b,
       author = {{Norris}, John E. and {Wyse}, Rosemary F.~G. and {Gilmore}, Gerard and {Yong}, David and {Frebel}, Anna and {Wilkinson}, Mark I. and {Belokurov}, V. and {Zucker}, Daniel B.},
        title = "{Chemical Enrichment in the Faintest Galaxies: The Carbon and Iron Abundance Spreads in the Bo{\"o}tes I Dwarf Spheroidal Galaxy and the Segue 1 System}",
      journal = {\apj},
         year = 2010,
        month = nov,
       volume = {723},
       number = {2},
        pages = {1632-1650},
          doi = {10.1088/0004-637X/723/2/1632},
archivePrefix = {arXiv},
       eprint = {1008.0137},
 primaryClass = {astro-ph.GA},
       adsurl = {https://ui.adsabs.harvard.edu/abs/2010ApJ...723.1632N}
}

@ARTICLE{Ishigaki_2014,
       author = {{Ishigaki}, M.~N. and {Aoki}, W. and {Arimoto}, N. and {Okamoto}, S.},
        title = "{Chemical compositions of six metal-poor stars in the ultra-faint dwarf spheroidal galaxy Bo{\"o}tes I}",
      journal = {\aap},
         year = 2014,
        month = feb,
       volume = {562},
          eid = {A146},
        pages = {A146},
          doi = {10.1051/0004-6361/201322796},
archivePrefix = {arXiv},
       eprint = {1401.1265},
 primaryClass = {astro-ph.GA},
       adsurl = {https://ui.adsabs.harvard.edu/abs/2014A&A...562A.146I}
}

@ARTICLE{Cohen_2010,
       author = {{Cohen}, Judith G. and {Huang}, Wenjin},
        title = "{The Chemical Evolution of the Ursa Minor Dwarf Spheroidal Galaxy}",
      journal = {\apj},
         year = 2010,
        month = aug,
       volume = {719},
       number = {1},
        pages = {931-949},
          doi = {10.1088/0004-637X/719/1/931},
archivePrefix = {arXiv},
       eprint = {1006.3538},
 primaryClass = {astro-ph.CO},
       adsurl = {https://ui.adsabs.harvard.edu/abs/2010ApJ...719..931C}
}

@ARTICLE{Letarte_2010,
       author = {{Letarte}, B. and {Hill}, V. and {Tolstoy}, E. and {Jablonka}, P. and {Shetrone}, M. and {Venn}, K.~A. and {Spite}, M. and {Irwin}, M.~J. and {Battaglia}, G. and {Helmi}, A. and {Primas}, F. and {Fran{\c{c}}ois}, P. and {Kaufer}, A. and {Szeifert}, T. and {Arimoto}, N. and {Sadakane}, K.},
        title = "{A high-resolution VLT/FLAMES study of individual stars in the centre of the Fornax dwarf spheroidal galaxy}",
      journal = {\aap},
         year = 2010,
        month = nov,
       volume = {523},
          eid = {A17},
        pages = {A17},
          doi = {10.1051/0004-6361/200913413},
archivePrefix = {arXiv},
       eprint = {1007.1007},
 primaryClass = {astro-ph.GA},
       adsurl = {https://ui.adsabs.harvard.edu/abs/2010A&A...523A..17L}
}

@ARTICLE{Lemasle_2014,
       author = {{Lemasle}, B. and {de Boer}, T.~J.~L. and {Hill}, V. and {Tolstoy}, E. and {Irwin}, M.~J. and {Jablonka}, P. and {Venn}, K. and {Battaglia}, G. and {Starkenburg}, E. and {Shetrone}, M. and {Letarte}, B. and {Fran{\c{c}}ois}, P. and {Helmi}, A. and {Primas}, F. and {Kaufer}, A. and {Szeifert}, T.},
        title = "{VLT/FLAMES spectroscopy of red giant branch stars in the Fornax dwarf spheroidal galaxy}",
      journal = {\aap},
         year = 2014,
        month = dec,
       volume = {572},
          eid = {A88},
        pages = {A88},
          doi = {10.1051/0004-6361/201423919},
archivePrefix = {arXiv},
       eprint = {1409.7703},
 primaryClass = {astro-ph.GA},
       adsurl = {https://ui.adsabs.harvard.edu/abs/2014A&A...572A..88L}
}

@ARTICLE{Hill_2019,
       author = {{Hill}, V. and {Sk{\'u}lad{\'o}ttir}, {\'A}. and {Tolstoy}, E. and {Venn}, K.~A. and {Shetrone}, M.~D. and {Jablonka}, P. and {Primas}, F. and {Battaglia}, G. and {de Boer}, T.~J.~L. and {Fran{\c{c}}ois}, P. and {Helmi}, A. and {Kaufer}, A. and {Letarte}, B. and {Starkenburg}, E. and {Spite}, M.},
        title = "{VLT/FLAMES high-resolution chemical abundances in Sculptor: a textbook dwarf spheroidal galaxy}",
      journal = {\aap},
         year = 2019,
        month = jun,
       volume = {626},
          eid = {A15},
        pages = {A15},
          doi = {10.1051/0004-6361/201833950},
archivePrefix = {arXiv},
       eprint = {1812.01486},
 primaryClass = {astro-ph.GA},
       adsurl = {https://ui.adsabs.harvard.edu/abs/2019A&A...626A..15H}
}

@ARTICLE{Tafelmeyer_2010,
       author = {{Tafelmeyer}, M. and {Jablonka}, P. and {Hill}, V. and {Shetrone}, M. and {Tolstoy}, E. and {Irwin}, M.~J. and {Battaglia}, G. and {Helmi}, A. and {Starkenburg}, E. and {Venn}, K.~A. and {Abel}, T. and {Francois}, P. and {Kaufer}, A. and {North}, P. and {Primas}, F. and {Szeifert}, T.},
        title = "{Extremely metal-poor stars in classical dwarf spheroidal galaxies: Fornax, Sculptor, and Sextans}",
      journal = {\aap},
         year = 2010,
        month = dec,
       volume = {524},
          eid = {A58},
        pages = {A58},
          doi = {10.1051/0004-6361/201014733},
archivePrefix = {arXiv},
       eprint = {1008.3721},
 primaryClass = {astro-ph.SR},
       adsurl = {https://ui.adsabs.harvard.edu/abs/2010A&A...524A..58T}
}

@ARTICLE{Simon_2015,
       author = {{Simon}, Joshua D. and {Jacobson}, Heather R. and {Frebel}, Anna and {Thompson}, Ian B. and {Adams}, Joshua J. and {Shectman}, Stephen A.},
        title = "{Chemical Signatures of the First Supernovae in the Sculptor Dwarf Spheroidal Galaxy}",
      journal = {\apj},
         year = 2015,
        month = apr,
       volume = {802},
       number = {2},
          eid = {93},
        pages = {93},
          doi = {10.1088/0004-637X/802/2/93},
archivePrefix = {arXiv},
       eprint = {1412.5176},
 primaryClass = {astro-ph.GA},
       adsurl = {https://ui.adsabs.harvard.edu/abs/2015ApJ...802...93S}
}

@ARTICLE{Aoki_2009,
       author = {{Aoki}, W. and {Arimoto}, N. and {Sadakane}, K. and {Tolstoy}, E. and {Battaglia}, G. and {Jablonka}, P. and {Shetrone}, M. and {Letarte}, B. and {Irwin}, M. and {Hill}, V. and {Francois}, P. and {Venn}, K. and {Primas}, F. and {Helmi}, A. and {Kaufer}, A. and {Tafelmeyer}, M. and {Szeifert}, T. and {Babusiaux}, C.},
        title = "{Chemical composition of extremely metal-poor stars in the Sextans dwarf spheroidal galaxy}",
      journal = {\aap},
         year = 2009,
        month = aug,
       volume = {502},
       number = {2},
        pages = {569-578},
          doi = {10.1051/0004-6361/200911959},
archivePrefix = {arXiv},
       eprint = {0904.4307},
 primaryClass = {astro-ph.GA},
       adsurl = {https://ui.adsabs.harvard.edu/abs/2009A&A...502..569A}
}

@ARTICLE{theler_2020,
       author = {{Theler}, R. and {Jablonka}, P. and {Lucchesi}, R. and {Lardo}, C. and {North}, P. and {Irwin}, M. and {Battaglia}, G. and {Hill}, V. and {Tolstoy}, E. and {Venn}, K. and {Helmi}, A. and {Kaufer}, A. and {Primas}, F. and {Shetrone}, M.},
        title = "{The chemical evolution of the dwarf spheroidal galaxy Sextans}",
      journal = {\aap},
         year = 2020,
        month = oct,
       volume = {642},
          eid = {A176},
        pages = {A176},
          doi = {10.1051/0004-6361/201937146},
archivePrefix = {arXiv},
       eprint = {1911.08627},
 primaryClass = {astro-ph.GA},
       adsurl = {https://ui.adsabs.harvard.edu/abs/2020A&A...642A.176T}
}

@ARTICLE{Jablonka_2015,
       author = {{Jablonka}, P. and {North}, P. and {Mashonkina}, L. and {Hill}, V. and {Revaz}, Y. and {Shetrone}, M. and {Starkenburg}, E. and {Irwin}, M. and {Tolstoy}, E. and {Battaglia}, G. and {Venn}, K. and {Helmi}, A. and {Primas}, F. and {Fran{\c{c}}ois}, P.},
        title = "{The early days of the Sculptor dwarf spheroidal galaxy}",
      journal = {\aap},
         year = 2015,
        month = nov,
       volume = {583},
          eid = {A67},
        pages = {A67},
          doi = {10.1051/0004-6361/201525661},
archivePrefix = {arXiv},
       eprint = {1506.08636},
 primaryClass = {astro-ph.GA},
       adsurl = {https://ui.adsabs.harvard.edu/abs/2015A&A...583A..67J}
}

@ARTICLE{Koch_2008,
       author = {{Koch}, Andreas and {Grebel}, Eva K. and {Gilmore}, Gerard F. and {Wyse}, Rosemary F.~G. and {Kleyna}, Jan T. and {Harbeck}, Daniel R. and {Wilkinson}, Mark I. and {Evans}, N. Wyn},
        title = "{Complexity on Small Scales. III. Iron and {\ensuremath{\alpha}} Element Abundances in the Carina Dwarf Spheroidal Galaxy}",
      journal = {\aj},
         year = 2008,
        month = apr,
       volume = {135},
       number = {4},
        pages = {1580-1597},
          doi = {10.1088/0004-6256/135/4/1580},
archivePrefix = {arXiv},
       eprint = {0802.2104},
 primaryClass = {astro-ph},
       adsurl = {https://ui.adsabs.harvard.edu/abs/2008AJ....135.1580K}
}

@ARTICLE{Norris_2010c,
       author = {{Norris}, John E. and {Yong}, David and {Gilmore}, Gerard and {Wyse}, Rosemary F.~G.},
        title = "{Boo-1137{\textemdash}an Extremely Metal-Poor Star in the Ultra-Faint Dwarf Spheroidal Galaxy Bo{\"o}tes I}",
      journal = {\apj},
         year = 2010,
        month = mar,
       volume = {711},
       number = {1},
        pages = {350-360},
          doi = {10.1088/0004-637X/711/1/350},
archivePrefix = {arXiv},
       eprint = {0911.5350},
 primaryClass = {astro-ph.GA},
       adsurl = {https://ui.adsabs.harvard.edu/abs/2010ApJ...711..350N}
}

@ARTICLE{Kirby_2010,
       author = {{Kirby}, Evan N. and {Guhathakurta}, Puragra and {Simon}, Joshua D. and {Geha}, Marla C. and {Rockosi}, Constance M. and {Sneden}, Christopher and {Cohen}, Judith G. and {Sohn}, Sangmo Tony and {Majewski}, Steven R. and {Siegel}, Michael},
        title = "{Multi-element Abundance Measurements from Medium-resolution Spectra. II. Catalog of Stars in Milky Way Dwarf Satellite Galaxies}",
      journal = {\apjs},
         year = 2010,
        month = dec,
       volume = {191},
       number = {2},
        pages = {352-375},
          doi = {10.1088/0067-0049/191/2/352},
archivePrefix = {arXiv},
       eprint = {1011.4516},
 primaryClass = {astro-ph.GA},
       adsurl = {https://ui.adsabs.harvard.edu/abs/2010ApJS..191..352K}
}

@ARTICLE{Kirby_2009,
       author = {{Kirby}, Evan N. and {Guhathakurta}, Puragra and {Bolte}, Michael and {Sneden}, Christopher and {Geha}, Marla C.},
        title = "{Multi-element Abundance Measurements from Medium-resolution Spectra. I. The Sculptor Dwarf Spheroidal Galaxy}",
      journal = {\apj},
         year = 2009,
        month = nov,
       volume = {705},
       number = {1},
        pages = {328-346},
          doi = {10.1088/0004-637X/705/1/328},
archivePrefix = {arXiv},
       eprint = {0909.3092},
 primaryClass = {astro-ph.GA},
       adsurl = {https://ui.adsabs.harvard.edu/abs/2009ApJ...705..328K}
}

@ARTICLE{Saga_1,
       author = {{Suda}, Takuma and {Katsuta}, Yutaka and {Yamada}, Shimako and {Suwa}, Tamon and {Ishizuka}, Chikako and {Komiya}, Yutaka and {Sorai}, Kazuo and {Aikawa}, Masayuki and {Fujimoto}, Masayuki Y.},
        title = "{Stellar Abundances for the Galactic Archeology (SAGA) Database --- Compilation of the Characteristics of Known Extremely Metal-Poor Stars}",
      journal = {\pasj},
         year = 2008,
        month = oct,
       volume = {60},
        pages = {1159},
          doi = {10.1093/pasj/60.5.1159},
archivePrefix = {arXiv},
       eprint = {0806.3697},
 primaryClass = {astro-ph},
       adsurl = {https://ui.adsabs.harvard.edu/abs/2008PASJ...60.1159S}
}

@ARTICLE{Saga_2,
       author = {{Suda}, Takuma and {Yamada}, Shimako and {Katsuta}, Yutaka and {Komiya}, Yutaka and {Ishizuka}, Chikako and {Aoki}, Wako and {Fujimoto}, Masayuki Y.},
        title = "{The Stellar Abundances for Galactic Archaeology (SAGA) data base - II. Implications for mixing and nucleosynthesis in extremely metal-poor stars and chemical enrichment of the Galaxy}",
      journal = {\mnras},
         year = 2011,
        month = apr,
       volume = {412},
       number = {2},
        pages = {843-874},
          doi = {10.1111/j.1365-2966.2011.17943.x},
archivePrefix = {arXiv},
       eprint = {1010.6272},
 primaryClass = {astro-ph.GA},
       adsurl = {https://ui.adsabs.harvard.edu/abs/2011MNRAS.412..843S}
}

@ARTICLE{saga_3,
       author = {{Yamada}, Shimako and {Suda}, Takuma and {Komiya}, Yutaka and {Aoki}, Wako and {Fujimoto}, Masayuki Y.},
        title = "{The Stellar Abundances for Galactic Archaeology (SAGA) Database - III. Analysis of enrichment histories for elements and two modes of star formation during the early evolution of the Milky Way}",
      journal = {\mnras},
         year = 2013,
        month = dec,
       volume = {436},
       number = {2},
        pages = {1362-1380},
          doi = {10.1093/mnras/stt1652},
archivePrefix = {arXiv},
       eprint = {1309.3430},
 primaryClass = {astro-ph.GA},
       adsurl = {https://ui.adsabs.harvard.edu/abs/2013MNRAS.436.1362Y}
}

@ARTICLE{saga_4,
       author = {{Suda}, Takuma and {Hidaka}, Jun and {Aoki}, Wako and {Katsuta}, Yutaka and {Yamada}, Shimako and {Fujimoto}, Masayuki Y. and {Ohtani}, Yukari and {Masuyama}, Miyu and {Noda}, Kazuhiro and {Wada}, Kentaro},
        title = "{Stellar Abundances for Galactic Archaeology Database. IV. Compilation of stars in dwarf galaxies}",
      journal = {\pasj},
         year = 2017,
        month = oct,
       volume = {69},
       number = {5},
          eid = {76},
        pages = {76},
          doi = {10.1093/pasj/psx059},
archivePrefix = {arXiv},
       eprint = {1703.10009},
 primaryClass = {astro-ph.GA},
       adsurl = {https://ui.adsabs.harvard.edu/abs/2017PASJ...69...76S}
}

@ARTICLE{Skul_2015,
       author = {{Sk{\'u}lad{\'o}ttir}, {\'A}. and {Andrievsky}, S.~M. and {Tolstoy}, E. and {Hill}, V. and {Salvadori}, S. and {Korotin}, S.~A. and {Pettini}, M.},
        title = "{Sulphur in the Sculptor dwarf spheroidal galaxy. Including NLTE corrections}",
      journal = {\aap},
         year = 2015,
        month = aug,
       volume = {580},
          eid = {A129},
        pages = {A129},
          doi = {10.1051/0004-6361/201525956},
archivePrefix = {arXiv},
       eprint = {1505.03155},
 primaryClass = {astro-ph.GA},
       adsurl = {https://ui.adsabs.harvard.edu/abs/2015A&A...580A.129S}
}

@ARTICLE{Skul_2017,
       author = {{Sk{\'u}lad{\'o}ttir}, {\'A}. and {Tolstoy}, E. and {Salvadori}, S. and {Hill}, V. and {Pettini}, M.},
        title = "{Zinc abundances in the Sculptor dwarf spheroidal galaxy}",
      journal = {\aap},
         year = 2017,
        month = oct,
       volume = {606},
          eid = {A71},
        pages = {A71},
          doi = {10.1051/0004-6361/201731158},
archivePrefix = {arXiv},
       eprint = {1708.00511},
 primaryClass = {astro-ph.GA},
       adsurl = {https://ui.adsabs.harvard.edu/abs/2017A&A...606A..71S}
}

@ARTICLE{Lemasle_2012,
       author = {{Lemasle}, B. and {Hill}, V. and {Tolstoy}, E. and {Venn}, K.~A. and {Shetrone}, M.~D. and {Irwin}, M.~J. and {de Boer}, T.~J.~L. and {Starkenburg}, E. and {Salvadori}, S.},
        title = "{VLT/FLAMES spectroscopy of red giant branch stars in the Carina dwarf spheroidal galaxy}",
      journal = {\aap},
         year = 2012,
        month = feb,
       volume = {538},
          eid = {A100},
        pages = {A100},
          doi = {10.1051/0004-6361/201118132},
archivePrefix = {arXiv},
       eprint = {1112.0431},
 primaryClass = {astro-ph.GA},
       adsurl = {https://ui.adsabs.harvard.edu/abs/2012A&A...538A.100L}
}

@ARTICLE{Battaglia_2011,
       author = {{Battaglia}, G. and {Tolstoy}, E. and {Helmi}, A. and {Irwin}, M. and {Parisi}, P. and {Hill}, V. and {Jablonka}, P.},
        title = "{Study of the Sextans dwarf spheroidal galaxy from the DART Ca II triplet survey}",
      journal = {\mnras},
         year = 2011,
        month = feb,
       volume = {411},
       number = {2},
        pages = {1013-1034},
          doi = {10.1111/j.1365-2966.2010.17745.x},
archivePrefix = {arXiv},
       eprint = {1009.4857},
 primaryClass = {astro-ph.CO},
       adsurl = {https://ui.adsabs.harvard.edu/abs/2011MNRAS.411.1013B}
}

@ARTICLE{Norris_2017,
       author = {{Norris}, John E. and {Yong}, David and {Venn}, Kim A. and {Gilmore}, Gerard and {Casagrande}, Luca and {Dotter}, Aaron},
        title = "{The Populations of Carina. II. Chemical Enrichment}",
      journal = {\apjs},
         year = 2017,
        month = jun,
       volume = {230},
       number = {2},
          eid = {28},
        pages = {28},
          doi = {10.3847/1538-4365/aa755e},
archivePrefix = {arXiv},
       eprint = {1705.09023},
 primaryClass = {astro-ph.GA},
       adsurl = {https://ui.adsabs.harvard.edu/abs/2017ApJS..230...28N}
}

@ARTICLE{Venn_2012,
       author = {{Venn}, Kim A. and {Shetrone}, Matthew D. and {Irwin}, Mike J. and {Hill}, Vanessa and {Jablonka}, Pascale and {Tolstoy}, Eline and {Lemasle}, Bertrand and {Divell}, Mike and {Starkenburg}, Else and {Letarte}, Bruno and {Baldner}, Charles and {Battaglia}, Giuseppina and {Helmi}, Amina and {Kaufer}, Andreas and {Primas}, Francesca},
        title = "{Nucleosynthesis and the Inhomogeneous Chemical Evolution of the Carina Dwarf Galaxy}",
      journal = {\apj},
         year = 2012,
        month = jun,
       volume = {751},
       number = {2},
          eid = {102},
        pages = {102},
          doi = {10.1088/0004-637X/751/2/102},
archivePrefix = {arXiv},
       eprint = {1204.0787},
 primaryClass = {astro-ph.GA},
       adsurl = {https://ui.adsabs.harvard.edu/abs/2012ApJ...751..102V}
}

@ARTICLE{Ural_2015,
       author = {{Ural}, U{\u{g}}ur and {Cescutti}, Gabriele and {Koch}, Andreas and {Kleyna}, Jan and {Feltzing}, Sofia and {Wilkinson}, Mark I.},
        title = "{An inefficient dwarf: chemical abundances and the evolution of the Ursa Minor dwarf spheroidal galaxy}",
      journal = {\mnras},
         year = 2015,
        month = may,
       volume = {449},
       number = {1},
        pages = {761-770},
          doi = {10.1093/mnras/stv294},
archivePrefix = {arXiv},
       eprint = {1502.04133},
 primaryClass = {astro-ph.GA},
       adsurl = {https://ui.adsabs.harvard.edu/abs/2015MNRAS.449..761U}
}

@ARTICLE{Pont_2004,
       author = {{Pont}, Fr{\'e}d{\'e}ric and {Zinn}, Robert and {Gallart}, Carme and {Hardy}, Eduardo and {Winnick}, Rebeccah},
        title = "{The Chemical Enrichment History of the Fornax Dwarf Spheroidal Galaxy from the Infrared Calcium Triplet}",
      journal = {\aj},
         year = 2004,
        month = feb,
       volume = {127},
       number = {2},
        pages = {840-860},
          doi = {10.1086/380608},
archivePrefix = {arXiv},
       eprint = {astro-ph/0310870},
 primaryClass = {astro-ph},
       adsurl = {https://ui.adsabs.harvard.edu/abs/2004AJ....127..840P}
}
\bibliographystyle{aasjournal}

\begin{appendix}
    \section{Additional dSph Membership Discussion}\label{app:memb}

\subsection{Non-Members targeted as Members} \label{app:members_nonmembers}
There are 46 targets that were targeted based on literature data but are not classified as members in this work (or in DR17). 
Some of these do not have ASPCAP results because the total SNR of the spectrum is $<~5$. 
Although the final SNR is between 5 and 32, another subset of targets have very poor spectral quality typified with no identifiable features, poor RV templates, and, most likely, spurious ASPCAP results.
Some of these, however, do have radial velocities consistent with the targeted dSph, but other aspects of the RV measurements are concerning (for example, the visit-level cross correlation peaks are marginal). 
These low SNR objects are indicated in \autoref{fig:targ} as targeted members (orange) not classified as members in APOGEE/this work (open) and with a black ``x'' to indicate the poor data quality.

As given in \autoref{tab:confirmed_nonmembers}, there are 13 targets with reliable spectra (based on visual inspection) that we reclassify as non-members and 1 target that is a one of the Fornax GCs that had poor Gaia EDR3 results.
Seven of these have dwarf-like stellar parameters from the APOGEE spectra, but all thirteen fail the astrometric criteria. 
Because the astrometry is subject to change with future Gaia releases, the five stars that are spectroscopic giants may be reclassified in the future.


\begin{table}
    \centering
    \caption{Additional dSph Members in the 6-sigma Sample}
    \label{tab:shetronemembers}    
    \begin{tabular}{c|c}
    \hline \hline
    \texttt{APOGEE\_ID}     &  dSph \\
    \hline
    2M14001049+1431454     & Bootes I \\
    2M14011845+1415250     & Bootes I \\
    2M14035451+1453285     & Bootes I \\
    2M06394720-5057439     & Carina \\  
    2M06415434-5102460     & Carina \\  
    2M17184293+5738497     & Draco \\  
    2M17193185+5733347     & Draco \\           
    2M02394528-3431581     & Fornax \\ 
    2M02400555-3428374     & Fornax \\  
    2M02400770-3432103     & Fornax \\  
    2M02410076-3419005     & Fornax \\  
    2M02410892-3434232     & Fornax \\  
    2M02415773-3437015     & Fornax \\  
    2M01000007-3338343     & Sculptor \\  
    2M01004017-3334353     & Sculptor \\     
    2M10121623-0144455     & Sextans \\  
    2M10122620-0138301     & Sextans \\    
    2M10122701-0130214     & Sextans \\    
    AP10123769-0211303     & Sextans \\    
    2M10123857-0129189     & Sextans \\    
    2M10134122-0206086     & Sextans \\    
    2M10152592-0204318     & Sextans \\    
    2M15063038+6656411     & Ursa Minor \\  
    \hline \hline
    \end{tabular}
\end{table}

\begin{table}
\centering
\movetabledown=1.7in
\begin{rotatetable}
    \centering
    \caption{APOGEE Non-Members classified as Members in the Literature}
    \label{tab:confirmed_nonmembers}    
    \begin{tabular}{c|c c c l}
    \hline \hline
    \texttt{APOGEE\_ID}   &  dSph & Lit. Membership & \texttt{SNREV} & Notes \\
    \hline
    2M06405987-5100423 & Carina   & 2MASS Blend, fiber target non-member          & 178 & dwarf stellar parameters; Astrometric non-member \\
    2M06410358-5057048 & Carina   & \cite{Walker_2009}   & 56  & Astrometric non-member \\
    2M06420906-5058365 & Carina   & \cite{Walker_2009}   & 37  & dwarf stellar parameters; Astrometric non-member \\
    2M06391906-5106303 & Carina   & 2MASS Blend, fiber target non-member  & 34  & dwarf stellar parameters; Astrometric non-member; \\
    2M17210799+5720370 & Draco    & \cite{Walker_2015}   & 93  & Astrometric non-member;  \\
    2M17212624+5752066 & Draco    & \cite{Armandroff_1995} & 71  & dwarf stellar parameters; Astrometric non-member \\
    2M02400793-3437162 & Fornax   & \cite{Walker_2009}   & 59  & Astrometric non-member; \\
    2M02401001-3429589 & Fornax   & \cite{Mateo_1991}    & 49  & Astrometric non-member; \\
    2M02395022-3435595 & Fornax   & \cite{Walker_2009}   & 55  & Astrometric non-member; \\
    2M10130818-0144502 & Sextans  & \cite{Walker_2009}   & 37  &  dwarf FERRE results; Astrometric non-member \\
    2M15125534+6748381 & Ursa Minor & \cite{Palma_2003} (photometry only) & 62 & Astrometric non-member; Binary? \\
    \hline
    2M02394817-3415285 & Fornax   & & 217 &  GC Fornax H3; Poor Gaia EDR3 results \\
    \hline
    2M01014472-3338305 & Sculptor & Targeting Flag Error & 59  &  dwarf stellar parameters; Astrometric non-member \\
    2M00590484-3342548 & Sculptor & Targeting Flag Error & 57  &  dwarf stellar parameters; Astrometric non-member \\ 
    \hline \hline
    \end{tabular}
\end{rotatetable}    
\end{table}
\subsection{Detailed Star by Star Assessments} \label{app:star_by_star}

There are 23 additional stars among the 7 dSphs that are classified as members in this work but not in the \texttt{MEMBER} convenience flag.
The members not indicated in DR17 via the \texttt{MEMBER} tag are listed in \autoref{tab:shetronemembers}.

There are 12 stars that were excluded from this analysis because their ASPCAP solutions were unrealistic, e.g. dwarf surface gravity, unrealistically hot or cool.   We list these stars in \autoref{tab:bad}.   In many cases the SNR is low, so a poor APOGEE solution is not surprising; however, for a few with very good SNR they are luminous AGB stars and likely variables, which means that the combination of many epochs of spectra probably results in a nonsense solution.

\begin{deluxetable*}{llcl}[h]
\tablecaption{Stars Excluded based on Param Range \label{tab:bad}}
\tablehead{
   \colhead{\texttt{APOGEE\_ID}} & \colhead{Galaxy} & \colhead{\texttt{SNREV}} & \colhead{Notes}
}
\startdata
2M17203116+5758060 & Draco      &  22  & Note 1 \\
2M15030656+6723166 & Ursa\_Minor &  41  & Note 1 \\
2M10115110-0154426 & Sextans    &  44  & Note 1 \\
AP10123769-0211303 & Sextans    &  16  & Note 1 \\
2M10130429-0134589 & Sextans    &  19  & Note 1 \\
2M10152592-0204318 & Sextans    &  135  & Note 1, Carbon Star \\
2M10153545-0131108 & Sextans    &  20  & Note 1 \\
2M00590406-3340317 & Sculptor   &  26  & Note 1 \\
2M00593045-3336050 & Sculptor   &  29  & Note 1 \\
2M01012084-3353047 & Sculptor   &  127 & Note 1, Mira \\
2M01021830-3337358 & Sculptor   &  16  & Note 1 \\
2M01024977-3340378 & Sculptor   &  14  & Note 1 \\
2M06404244-5100428 & CARINA     &  33  & Note 1 \\
2M02384457-3448255 & FORNAX     &  90  & Note 1 \\
\enddata
\tablenotetext{1}{{ASPCAP stellar parameters out of reasonable range}}
\end{deluxetable*}

\subsection{Carbon Stars in High Quality Sample} 

\subsubsection{Literature Classifications, but No Carbon enhancement} 

The following stars in our main sample are listed as C* in SIMBAD but the ASPCAP solution shows little carbon enhancement. 
CARINA:2M06403082-5059153, CARINA:2M06404775-5106033, CARINA:2M06411470-5051099, CARINA:2M06414823-5055016, FORNAX:2M02395861-3425279.

\subsection{ASPCAP-identified Carbon Stars} 

We identified carbon stars on the basis of their carbon abundance, stellar parameters, and spectral identification. 
The details of these identifications will be presented in a companion
paper, but for completeness in our data file and sample, we include the identifications here. 
More specifically, candidate carbon stars are flagged. 
We further note stars previously identified as carbon stars that have no clear carbon enhancement in our spectra. 



\section{Derivation of Upper Limit Formula} \label{app:upper_limits}

Our upper limit methodology is driven by the following idea:  for a given abundance measurement to be considered a detection, the average line depth of the relevant lines of that element should exceed the noise on that average.  So by measuring the average line depth of the lines that APOGEE uses to measure abundances of each element, and by setting some threshold $\chi$, such that the average line depth is $\chi$ times larger than the spectral noise averaged over those lines, we can provide an estimate of what signal-to-noise ratio we need to measure that abundance.

Since it begins to become cost prohibitive to estimate this S/N for every abundance measurement, we instead estimate these S/N thresholds at a range of relevant abundances and temperatures and use this to produce the upper limit relations we report in this paper.  Our procedure to do this is as follows.

\begin{enumerate}
    \item Produce synthetic spectra with the element of interest, and spectra where the lines of that element are missing.
    \item From the difference spectra measure the weighted average line depth, using APOGEE's window weights for that element.
    \item Calculate the S/N threshold needed to measure the abundance of that element.
    \item Repeat this process at a range of temperatures and abundances.
    \item Fit a polynomial model for the abundance upper limit as a function of temperature and S/N for each element.
    \item Use these polynomial models to flag APOGEE measurements that are likely upper limits according to our choice of threshold.
\end{enumerate}

\subsection{Synthetic Spectra Grids}

For the non-CNO elements that we analyze we calculate synthetic spectra at a range of temperatures from \teff{} $= 4200 - 4800$ K in steps of 200 K, and at a range of abundances from \xh{X} $= -2.5 - -0.5$ in steps of 0.5 dex, with \feh{} $= -1.0$, $\log g = 1.0$, v$_{\rm micro} = 2$, and hold all other abundances solar-scaled.  And then repeat these calculations decreasing the \xfe{X} abundance by -5 to simulate removing the lines of that element.

Most elements are measured from atomic lines, however, C, N, and O are primarily measured from molecular lines (in particular CO, CN, and OH), whose line strengths are interdependent on the abundances of each element, so we use two slightly different temperature and abundance grids to calculate their abundances.

For C, N, and O, we again calculate synthetic spectra at a range of temperatures from \teff{} $= 4200 - 4800$ K in steps of 200 K, but then at a range of metallicities from \xh{Fe} $= -2.5 - -0.5$ in steps of 0.5 dex.  We do this while holding \xfe{C} $ = -0.75$, \xfe{N} $ = 0.25$ and O at solar-scaled, which are typical values seen in our dSph sample.  And again we use $\log g = 1.0$, v$_{\rm micro} = 2$, and hold all other abundances solar-scaled.  To simulate the lack of C, N, and O features, instead of reducing the abundances of any of these elements, which are important to the molecular equilibrium in stellar atmospheres, we instead repeat these syntheses but with line lists that have the relevant lines removed for each element (CO, CN, and C I lines removed for C, CN removed for N, and CO and OH removed for O).

Using the above parameters and abundances we synthesize spectra using the 1D LTE Turbospectrum (v19.1.4) radiative transfer code \citep{turbospectrum1,turbospectrum2}, with spherical radiative transfer and using the MARCS model atmosphere grid \citep{marcs}.  For consistency with ASPCAP's abundance determinations, we use the APOGEE DR17 Turbospectrum linelist, version \emph{180901t20}.  We synthesize the relevant wavelength range for each element in steps of 0.05 \AA{} and then convolve these spectra with a Gaussian profile with a full-width half-maximum of 800 m\AA{} to account for typical instrumental resolution, and then convolve them with a Gaussian velocity profile of 5 km s$^{-1}$, to account for typical macroturbulent velocity broadening.  Finally we interpolate the synthetic spectra and resample them to APOGEE's wavelength scale.

\subsection{Measuring S/N Thresholds}

\begin{figure*} 
    \centering
    \includegraphics[width=0.9\textwidth]{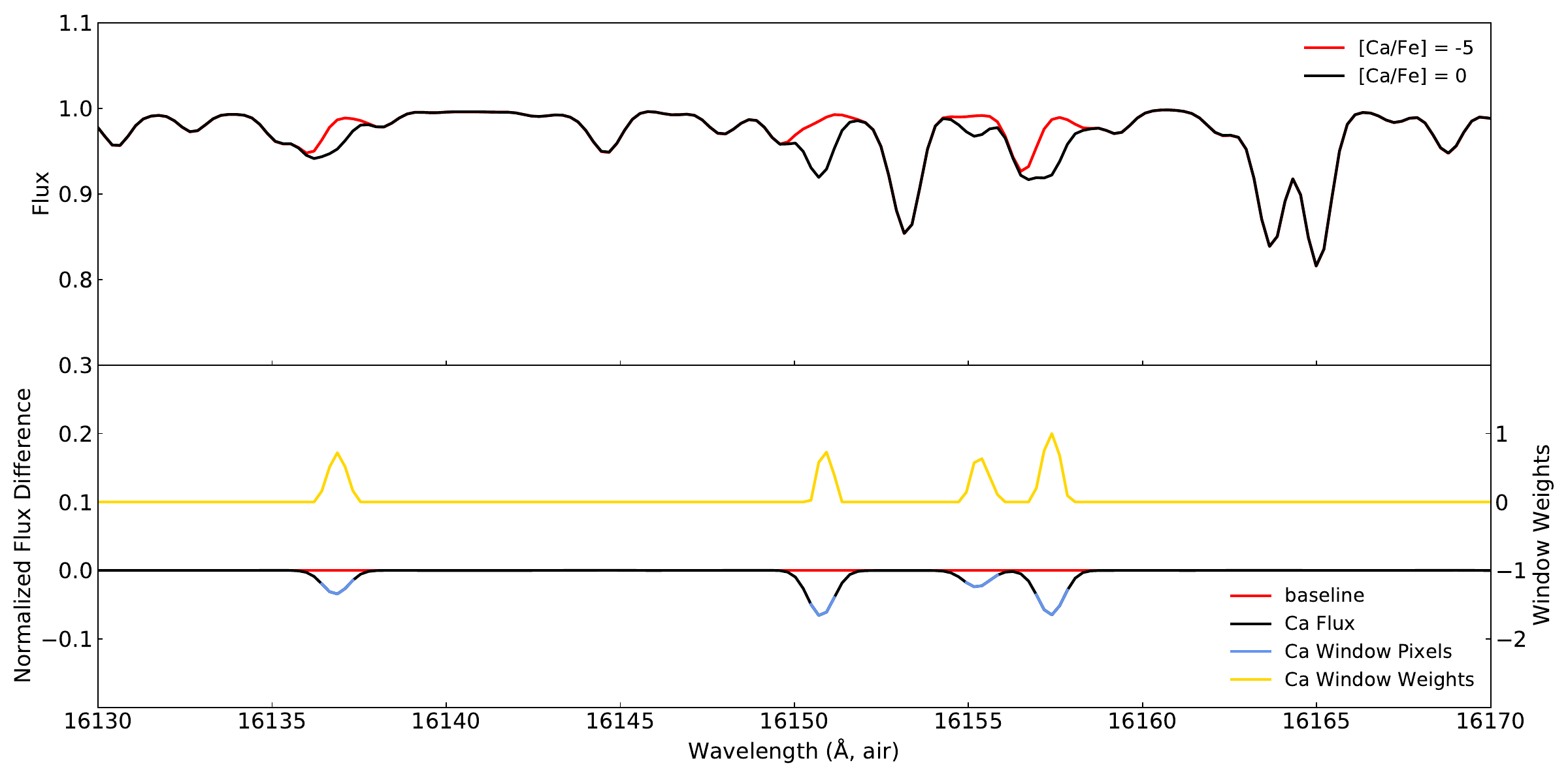}
    \caption{(Top panel) Normalized synthetic spectra at a temperature of 4200 K, metallicity of \feh{} $-1$, and \xfe{Ca} $= 0$ (black) and $-5$ (red). (Bottom panel) Difference of the solar Ca abundance spectrum and no Ca spectrum (black), the null difference (red), along with the APOGEE window strengths (yellow). Pixels in the difference spectra where the window strengths are non-zero are highlighted (blue).
    }
    \label{fig:synth_spec_ca}
\end{figure*} 

From our calculations, we have pairs of synthetic spectra with and without the element, X, in question.  We used these spectra, combined with APOGEE's window weights for that element to measure the S/N threshold needed to detect this abundance-temperature combination.  In the following, we provide a case study of this procedure with Ca as our example element.

In the top panel of Figure \ref{fig:synth_spec_ca}, we show our pair of synthetic spectra at T$_{\rm eff} = 4200$ K and \feh{} $= -1$ with \xfe{Ca} $= 0$ and \xfe{Ca} $= -5$, around the region that APOGEE uses to measure Ca.  We then take the difference between these two synthetic spectra which is shown in the bottom panel of Figure \ref{fig:synth_spec_ca}.  The difference spectrum indicates the flux from the Ca lines at an abundance of \xh{Ca} $= -1$.  We then measure a weighted average Ca line depth by taking a weighted average of the difference spectrum, weighted by the APOGEE Ca window weights for each pixel, which are shown as the gold line in Figure \ref{fig:synth_spec_ca}.  Numerically this is:
\begin{displaymath}
D_{\rm wavg} = \frac{\sum_i w_i D_i}{\sum_i w_i}
\end{displaymath}

Where $w_i$ is the window weight, and $D_i$ is the Ca flux from the difference spectrum, for pixels $i$.  For this example, the weighted average line depth is 0.0416, across a weighted number of pixels, $N_{\rm wavg} = \sum_i w_i = 8.35$.  We then use this weighted average line depth to estimate the S/N threshold needed to confidently detect Ca in the \xfe{Ca} $=0$ spectrum.  For a given S/N, the noise per pixel in a normalized spectrum would be $\sigma_{\rm pix} = 1/({\rm S/N})$.  When averaging this noise over the $N_{\rm wavg}$ the error we would expect on measuring $D_{\rm wavg}$ would be 
\begin{displaymath}
\sigma_{\rm wavg} = \frac{\sigma_{\rm pix}}{\sqrt{N_{\rm wavg}}} = \frac{1}{{\rm S/N} \cdot \sqrt{N_{\rm wavg}}}
\end{displaymath}

We therefore have an estimate of our Ca ``signal,'' $D_{\rm wavg}$, and the expected error on this measurement for a given S/N, $\sigma_{\rm wavg}$, all that remains is to set a threshold, $\chi$, above which the measurement would be considered a confident detection.  Traditionally, studies will consider a $(\chi = ) \ 3-5\sigma$ signal for detection depending on the circumstances.  Since our procedure has made a number of assumptions and is likely a fairly simple estimate, we have chosen to use $\chi = 4$ as our threshold, which we have verified is reasonable by comparing with previous literature measurements and through a visual inspection of cases near these thresholds.  Stating this another way, we use a detection threshold of $4\sigma$.

This choice allows us to work backward and determine the equivalent S/N threshold needed to measure a given $D_{\rm wavg}$ that we are able to measure from synthetic spectra.  Specifically this S/N threshold can by found with some rearrangement:
\begin{displaymath}
{\rm S/N}_{\rm limit} = \frac{4}{D_{\rm wavg} \sqrt{N_{\rm wavg}}}
\end{displaymath}

For our example of \xh{Ca} $= -1$, with $D_{\rm wavg}$ measured over $N_{\rm wavg} = 8.35$ pixels, the signal-to-noise threshold for a measurement would be ${\rm S/N}_{\rm limit} = 33.3$.

\subsection{Deriving Upper Limit Relations}

\begin{figure*} 
    \centering
    \includegraphics[width=0.5\textwidth]{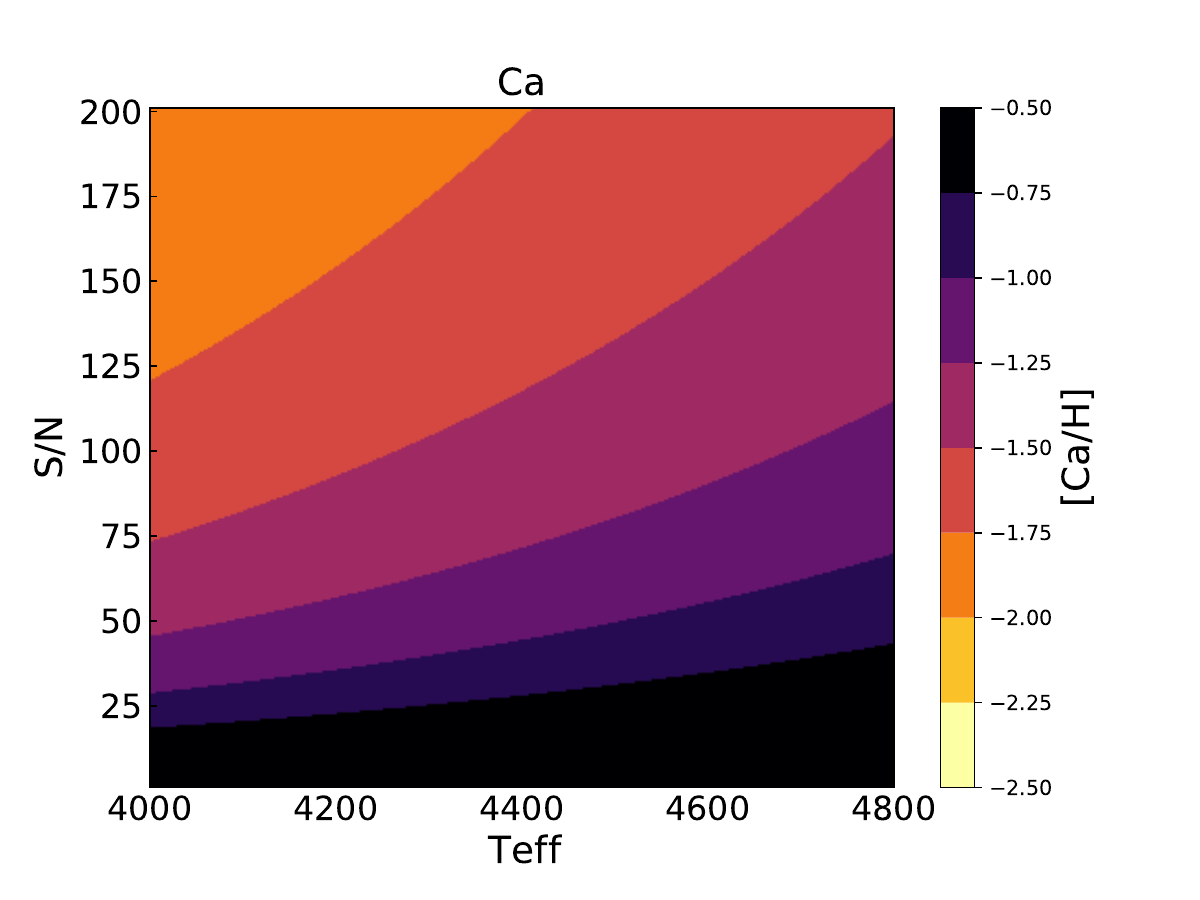}
    \caption{Derived upper limits for Ca as a function of temperature and S/N.
    }
    \label{fig:upper_limit_relation_ex}
\end{figure*} 

In order to turn individual S/N limit measurements into more general upper limit relations, we first measure S/N thresholds for each element, temperature, and abundance combination.  We then fit these with a polynomial surface to provide a parametrized upper limit threshold using the following functional form:

    \begin{equation}\label{eq:upper_limit_2}
        {\rm [X/H]}_{\rm limit} = \alpha + \beta \bigg(\frac{T_{\rm eff}}{1000\, K}\bigg) + \gamma \log_{10}(S/N) + \delta \bigg[(\log_{10}(S/N))\bigg]^{2}, 
    \end{equation}

The residuals between the calculated thresholds and the polynomial fits are below 0.1 dex, and therefore should provide a reasonable estimate of the detection threshold, though we note that these estimates are designed to remove the most egregious cases of non-detections and may be less discriminatory for marginal detections.

An example of one of these limit thresholds is shown for Ca in Figure \ref{fig:upper_limit_relation_ex}.  We have quantized the color-map for simplicity of visual interpretation, but note that the actual thresholds as implemented are continuous functions of S/N and temperature.  This shows a typical case, where the upper limits are higher for lower S/N stars and warmer stars.

\subsubsection{Notes on Individual Elements}  

\paragraph{Sodium (Na)}
On the basis of our upper limit analysis for Na, we determined that Na is not detectable in the dSph sample except for one cool, metal-rich Fornax star. The upper limits we compute are typically greater than 0.2 dex and thus do not constrain chemical evolution in these dSphs. 

\paragraph{Sulfur (S)}
Although the formula we derive for upper limits does reject 93\% of the sample, the remaining 7\% are extremely super-Solar (mean [S/Fe] $> +0.73$).  As mentioned above, the nature of our methodology can result in a detection being generated from very high [S/Fe] scattered false positives.    Inspection of the spectra does not reveal any believable detections.  The upper limits are typically greater than 0.3 dex, so they do not provide interesting constraints on dSph nucleosynthesis.   

\paragraph{Vanadium (V)}
Based on our upper limit analysis for V, we determine that V is not detectable in the dSph sample except for six cool, metal-rich Fornax stars and two of these have unrealistically high detections and are likely false positives.   The remaining 4 points have a mean [V/Fe] of -0.46 at a mean [Fe/H] = -0.75.   The upper limits we compute are typically greater than 1.0 dex and thus do not constrain chemical evolution in these dSph.

\subsubsection{Notes on Metallicity}  

\paragraph{Iron (Fe)}
As a primary metallicity indicator, [Fe/H] is critical for the analysis and interpretation of the dSph abundance patterns.    Based on our analysis, there are 25 stars that should be removed from the dSph sample because [Fe/H] should not be detectable.   In the \citet{Hill_2019} and \citet{Kirby_2010} cross-matched samples, only 1 star is removed from each sample based on the upper limits of [Fe/H].  The agreement with Hill is excellent, while there is a large miss-match with Kirby.   This suggests that some stars which have been removed from our sample based on our upper limits could be marginally detectable.    We will exclude the 25 Fe-upper limit stars from discussions and figures that involved [Fe/H].  

\paragraph{Magnesium (Mg)}
[Mg/H] is an alternative metallicity indicator that will be used in this analysis.    Based on our formula, there are 12 stars with undetectable Mg lines.  Visual inspection of these 12 star's spectra reveals that the strongest lines within the APOGEE windows remain detectable in some of the stars while they are not detectable in others.  We take a consistent and conservative approach and remove the 12 Mg-upper limit stars from discussions and figures that use [Mg/H] as a metallicity indicator.

\section{Comparison Samples} \label{app:comparison}

When showing the dSph abundance patterns in Figures \ref{fig:cn_grid} - \ref{fig:misc_grid} of \autoref{sec:chem_patterns}, we include a MW comparison sample also observed by APOGEE to help put the dSph abundance patterns in a greater galactic context. Starting from the full DR17 APOGEE catalog, we perform the following restrictions:  

\begin{itemize}
    \item remove stars associated with the LMC and SMC using the spatial and kinematic cuts given in \citet{Hasselquist2021_sats};
    \item remove the Sgr dSph and stream stars that are presented in \citet{Hayes_2020} and \citet{Hasselquist2021_sats};
    \item remove any stars labeled as a cluster or dSph member using the \texttt{MEMBER} flag in the allStar file;
    \item remove stars with \texttt{STAR\_BAD}, \texttt{M\_H\_BAD}, or \texttt{ALPHA\_M\_BAD};
    \item remove any duplicate observations;
    \item and finally, we use only the stars that should provide the highest quality abundance measurements, namely those with \texttt{SNREV} $>$ 100, \logg\ $<$ 4.0, \teff\ $<$ 5500 K, metallicities, \feh{} $<$ -0.5.
\end{itemize}

Once these cuts were performed, the remaining stars had their positions and velocities transformed into a Galactocentric frame.  
To do so, we used \texttt{StarHorse} spectro-photometric distances \citep{Santiago2016, Queiroz2018, Anders2019, Queiroz2020} assuming $R_{\rm GC, \odot} = 8.122$ kpc \citep{gravity2018}.  
We additionally used the \citet{js87} method to transform the APOGEE radial velocities and {\it Gaia} EDR3 proper motions to heliocentric 3D spatial velocities, and convert these to Galactocentric space velocities assuming a total solar motion of $(V_{r}, \ V_{\phi} \ V_{z})_{\odot} = (14, 253, 7)$ km s$^{-1}$ in the right-handed velocity sense \citep{schonrich2010,schonrich2012,hayes2018c}.

With these Galactocentric positions and velocities, we define four MW comparison samples as follows: 
\begin{itemize}
    \item {\bf High-$\alpha$ disk sample:}  150 $<$ v$_{\rm \phi}$ $<$ 290 \kms{}, \xfe{Mg} $>$ 0.25, $|\rm z| < 4$ kpc, and $\log g < 1.3$ (to limit the sample to a similar range as the dSph stars)
    \item {\bf Low-$\alpha$ disk sample:} 150 $<$ v$_{\rm \phi}$ $<$ 290 \kms{}, \xfe{Mg} $<$ 0.2, $|\rm z| < 4$ kpc, and $\log g < 1.3$ 
    \item {\bf Bulge-like sample:} $-$100 $<$ v$_{\rm \phi}$ $<$ 100 \kms{}, $|\rm R_{GC, \, sph}| < 3$ kpc, and $\log g < 1.3$
    \item {\bf Halo sample:} $-$100 $<$ v$_{\rm \phi}$ $<$ 100 \kms{}, \xfe{Mg} $<$ $-0.333 \cdot$ \feh{} $- 0.067$, $|\rm z| > 4$ kpc, and $\log g < 1.3$
\end{itemize}

\begin{figure*}
     \centering
     \includegraphics[width=0.9\textwidth]{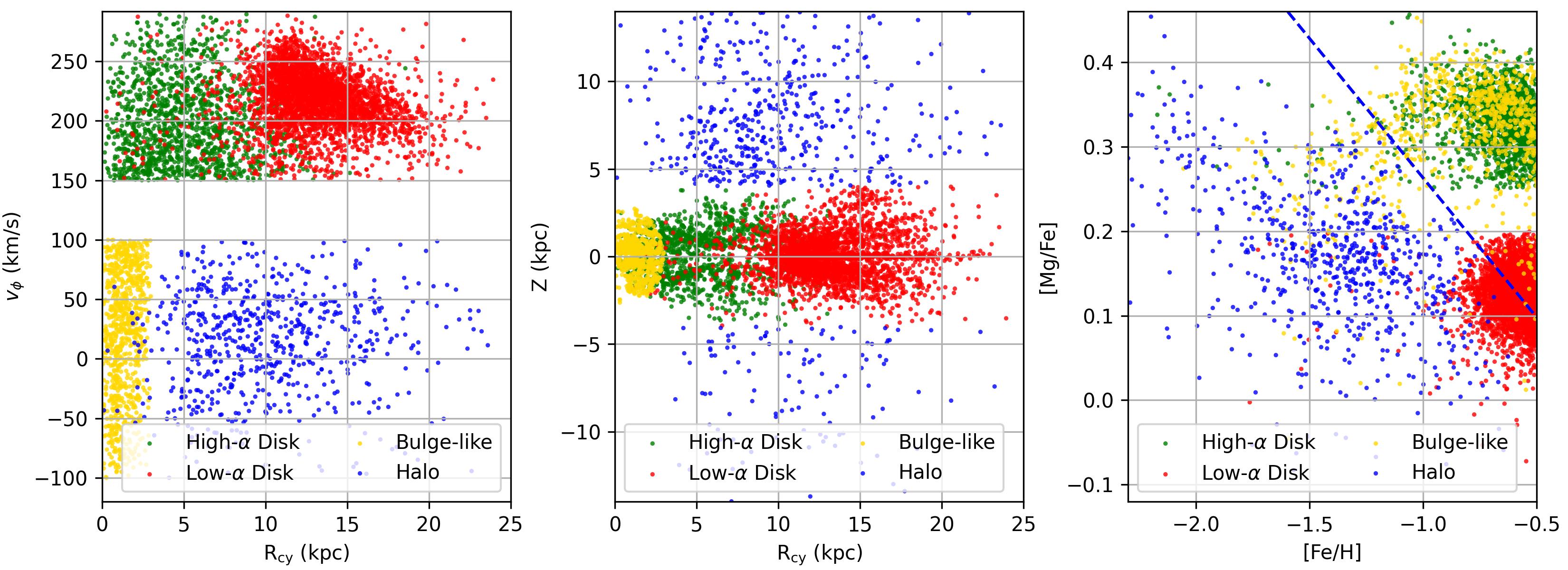}
     \caption{Visualizations of the MW Comparison sample used in various figures throughout the main text. The left panel shows Galactic rotational velocity, v$_{\rm \phi}$, as a function of Galactic cylindrical radius ($\rm R_{GC, \, cy}$), and the middle panel shows distance from Galactic plane, $Z$, plotted as a function of $\rm R_{GC, \, cy}$. The right panel shows the [Mg/Fe]-[Fe/H] abundance plane. All parameters shown here are used to divide the MW comparison sample into the sub samples listed in the legend, as described in the text. }
     \label{fig:comparison_sample}
\end{figure*}

These MW comparison samples are shown in Figure \ref{fig:comparison_sample}, where we plot the parameters used to separate these samples. We find that the halo sample is primarily comprised of GSE stars, so we refer to this sample as ``GSE+Halo'' in the text for clarity.

\end{appendix}
\end{document}